\documentclass[aip,pop,reprint,numerical,twocolumn]{revtex4-1}

\usepackage{circuitikz}
\usepackage{amsmath}
\usepackage{amssymb}
\usepackage{graphicx}
\usepackage{xspace}
\usepackage{upgreek}
\usepackage{accents}
\usepackage{subfigure} 
\usepackage{colortbl}
\usepackage{xfrac}

\newcommand{\eg}{{e.g.\/}\xspace}
\newcommand{\ie}{{i.e.,\/}\xspace}

\newcommand{\eq}[1]{(\ref{#1})}
\newcommand{\Eq}[1]{Eq.~(\ref{#1})}
\newcommand{\Eqs}[1]{Eqs.~(\ref{#1})}
\newcommand{\Fig}[1]{Fig.~\ref{#1}}
\newcommand{\Figs}[1]{Figs.~\ref{#1}} 
\newcommand{\Sec}[1]{Sec.~\ref{#1}}

\newcommand{\Refa}[1]{Ref.~\onlinecite{#1}}
\newcommand{\Refs}[1]{Refs.~\onlinecite{#1}}

\newcommand{\const}{\text{const}}

\usepackage{hyperref}
\hypersetup{
  colorlinks   	= true, 	%Colours links instead of ugly boxes
  urlcolor     	= blue, 	%Colour for external hyperlinks
  linkcolor    	= blue, 	%Colour of internal links
  citecolor   	= blue 	%Colour of citations
}
\newcommand{\I}{I_{\rm max}}

\graphicspath{{Figures/}} %Setting the graphicspath

\begin{document}

\title{Exploring the parameter space of MagLIF implosions using similarity scaling.~~III.~Rise-time scaling}
\date{\today}
\author{D.~ E.~Ruiz}
\email{deruiz@sandia.gov}
\affiliation{Sandia National Laboratories, P.O. Box 5800, Albuquerque, NM 87185, USA}
\author{P.~F.~Schmit}
\affiliation{Lawrence Livermore National Laboratory, Livermore, CA 94550, USA}
\author{M.~R.~Weis}
\affiliation{Sandia National Laboratories, P.O. Box 5800, Albuquerque, NM 87185, USA}
\author{K.~J.~Peterson}
\affiliation{Sandia National Laboratories, P.O. Box 5800, Albuquerque, NM 87185, USA}
\author{M.~K.~Matzen}
\affiliation{Sandia National Laboratories, P.O. Box 5800, Albuquerque, NM 87185, USA}

%%%%%%%%%%%%%%%%%%%%%%%%%%%%%%%%%%%%%%%%%%%%%%%%%
%%%%%%%%%%%%%%%%%%%%%%%%%%%%%%%%%%%%%%%%%%%%%%%%%
%%%%%%%%%%%%%%%%%%%%%%%%%%%%%%%%%%%%%%%%%%%%%%%%%

\begin{abstract}

Magnetized Liner Inertial Fusion (MagLIF) is a z-pinch magneto-inertial-fusion (MIF) concept studied on the Z Pulsed Power Facility at Sandia National Laboratories.  Two important metrics characterizing current delivery to a z-pinch load are the peak current and the current-rise time, which is roughly the time interval to reach peak current.  It is known that, when driving a z-pinch load with a longer current-rise time, the performance of the z-pinch decreases.  However, a theory to understand and quantify this effect is still lacking.  In this paper, we utilize a framework based on similarity scaling to analytically investigate the variations in performance of MagLIF loads when varying the current-rise time, or equivalently, the implosion timescale.  To maintain similarity between the implosions, we provide the scaling prescriptions of the experimental input parameters defining a MagLIF load and derive the expected scaling laws for the stagnation conditions and for various performance metrics.  We compare predictions of the theory to 2D numerical simulations using the radiation, magneto-hydrodynamic code \textsc{hydra}.  For several metrics, we find acceptable agreement between the theory and simulations.  Our results show that the voltage $\varphi_{\rm load}$ near the MagLIF load follows a weak scaling law $\smash{\varphi_{\rm load} \propto t_\varphi^{-0.12}}$ with respect to the characteristic timescale $t_\varphi$ of the voltage source, instead of the ideal $\smash{\varphi_{\rm load} \propto t_\varphi^{-1}}$ scaling.  This occurs because the imploding height of the MagLIF load must increase to preserve end losses.  As a consequence of the longer imploding liners, the required total laser preheat energy and delivered electric energy increase.  Overall, this study may help understand the trade-offs of the MagLIF design space when considering future pulsed-power generators with shorter and longer current-rise times.

\end{abstract}

%\pacs{52.35.Fp, 47.10.ab, 52.25.-b, 52.65.Rr}

\maketitle

%%%%%%%%%%%%%%%%%%%%%%%%%%%%%%%%%%%%%%%%%%%%%%%%%
%%%%%%%%%%%%%%%%%%%%%%%%%%%%%%%%%%%%%%%%%%%%%%%%%
%%%%%%%%%%%%%%%%%%%%%%%%%%%%%%%%%%%%%%%%%%%%%%%%%
\section{Introduction}

In a nutshell, inertial-confinement fusion (ICF) is the pursuit of compressing energy in both time and space to assemble a hot and dense DT plasma fuel to thermonuclear conditions.\cite{Atzeni:2009phys,Lindl:1998bq}  One of the main three approaches to ICF is magneto-inertial fusion (MIF), which introduces strong magnetic fields in the fuel in order to relax the stringent requirements on high pressures and high implosion velocities in laser-driven ICF.\cite{Widner_1977,Lindemuth_1981,Lindemuth:1983aa,Lindemuth:2015fu}  In particular, the Magnetized Liner Inertial Fusion (MagLIF) platform\cite{Slutz:2010hd,Gomez:2014eta} is a MIF concept currently being studied at the pulsed-power Z facility at Sandia National Laboratories.\cite{Gomez:2019bg,Sefkow:2014ik,Gomez:2014eta,Knapp:2019gf,Sinars:2020bv,YagerElorriaga:2022cp}  The Z facility delivers a 20-MA electrical current pulse to the cylindrical MagLIF z-pinch, which then implodes under the action of the Lorentz force.\citep{Gomez:2020cd}  MagLIF uses a metallic cylindrical tamper, or liner, to compress the plasma fuel.  The implosions are considerably slower (on the order of 100 ns) than those in laser-driven ICF, so the fuel must be axially premagnetized to mitigate thermal-conduction losses.  Although a shock is produced in the liner during the current rise, it does not appreciably shock heat the fuel.  To reach high temperatures at reasonable convergence ratios, the fuel is therefore preheated by a 2--4-kJ, 1-TW laser to raise the initial adiabat.\cite{HarveyThompson:2019ff,HarveyThompson:2018dd,Weis:2021id}  MagLIF experiments in the laboratory have demonstrated significant thermonuclear yield production\cite{Gomez:2014eta,Knapp:2019gf,Gomez:2019bg,Gomez:2020cd,YagerElorriaga:2022cp} and high plasma magnetization.\cite{Schmit:2014fg,Knapp:2015kc,Lewis:2021kz}

It is known that z-pinch loads driven by electrical currents with longer rise times tend to show reduced performance. Intuitively, this occurs because energy is not sufficiently compressed in the temporal direction; in other words, the characteristic power delivered to the load decreases.  Therefore, slower imploding z-pinch loads are less effective for attaining high pressures and temperatures.  This behavior was observed in previous simulation work scoping the performance of the MagLIF concept on future, pulsed-power drivers.  In \Refa{Slutz:2018iq}, thousands of 1D \textsc{lasnex} simulations and subsequent 2D \textsc{lasnex} simulations were performed to find optimized MagLIF configurations that maximized fusion yield or fusion gain for a given peak current and current-rise time.  These \textit{optimized-scaling} studies showed that the peak current and energy required to obtain a given fusion gain increases monotonically with the current-rise time.  In \Refa{Slutz:2018iq}, it was also noted that the required preheat energy for MagLIF scales unfavorably with the rise time.

In this paper, we provide a framework to understand and \textit{analytically} quantify this effect.  We apply the similarity-scaling framework developed in \Refa{foot:Ruiz_framework} (hereby referred as Paper I) to investigate the performance of MagLIF loads when varying the characteristic rise time of the electrical current, or equivalently, the timescale of the implosions.  It is shown that we can analytically estimate the stagnation conditions achieved by the similarity-scaled implosions.  We compare the theoretical predictions to 2D numerical simulation results using the radiation, magneto-hydrodynamic code \textsc{hydra},\cite{Marinak:1996fs,Koning:2009} which is one of the main design codes for MagLIF experiments.\cite{Sefkow:2014ik,HarveyThompson:2018dd,Weis:2021id}  For a wide range of metrics, we find acceptable agreement between the theory predictions and the simulation results.

This paper is organized as follows.  In \Sec{sec:prescriptions}, we summarize the scaling rules of MagLIF input parameters when varying the characteristic timescale of the implosions.  In \Sec{sec:numerical}, we give the specific input parameters for the baseline MagLIF load and discuss some general notions on how MagLIF loads are scaled when varying the timescale of the implosions.  In \Sec{sec:implosion}, we compare the implosion dynamics of the similarity-scaled MagLIF liners.  In \Sec{sec:stagnation}, we derive the scaling laws for the thermodynamic quantities describing the fuel at stagnation and compare the scaling laws to simulation results.  In \Sec{sec:loss}, we discuss the scaling of the burn-width time and of the energy-loss mechanisms.  In \Sec{sec:performance}, we test predictions of the theory for the neutron yield.  In \Sec{sec:discussion}, we discuss advantages and disadvantages of increasing the current-rise time with a specific emphasis on the electrical-engineering requirements.  In \Sec{sec:conclusions}, we summarize our main results.

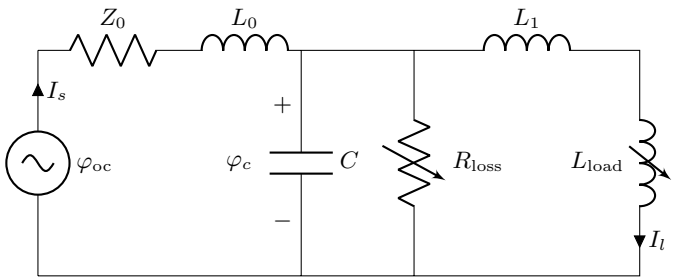
\begin{figure}
	\begin{circuitikz}[american,scale=1] \draw
	(0,3) to[sinusoidal voltage source,l=$\varphi_{\rm oc}$,i=$I_s$] (0,0)
	(0,3) to[resistor,l=$Z_0$] (2,3)
  			to[inductor,l=$L_0$] (3.5,3) --(5,3)
  		    to[inductor,l=$L_1$] (8,3)
  		    to[vL,l_=$L_{\rm load}$,i=$I_l$] (8,0) -- (0,0) 
    (3.5,3) to[capacitor,l^=$C$,v=$\varphi_c$] (3.5,0)
    (5,0) to[vR,l_=$R_{\rm loss}$] (5,3);
	\end{circuitikz}
	\caption{Representative circuit diagram of the Z generator.}
	\label{fig:scaling:circuit}
\end{figure}

%%%%%%%%%%%%%%%%%%%%%%%%%%%%%%%%%%%%%%%%%%%%%%%%%
%%%%%%%%%%%%%%%%%%%%%%%%%%%%%%%%%%%%%%%%%%%%%%%%%
%%%%%%%%%%%%%%%%%%%%%%%%%%%%%%%%%%%%%%%%%%%%%%%%%
\section{Rise-time--scaling prescriptions}
\label{sec:prescriptions}

\begin{figure}
 	\includegraphics[scale=.43]{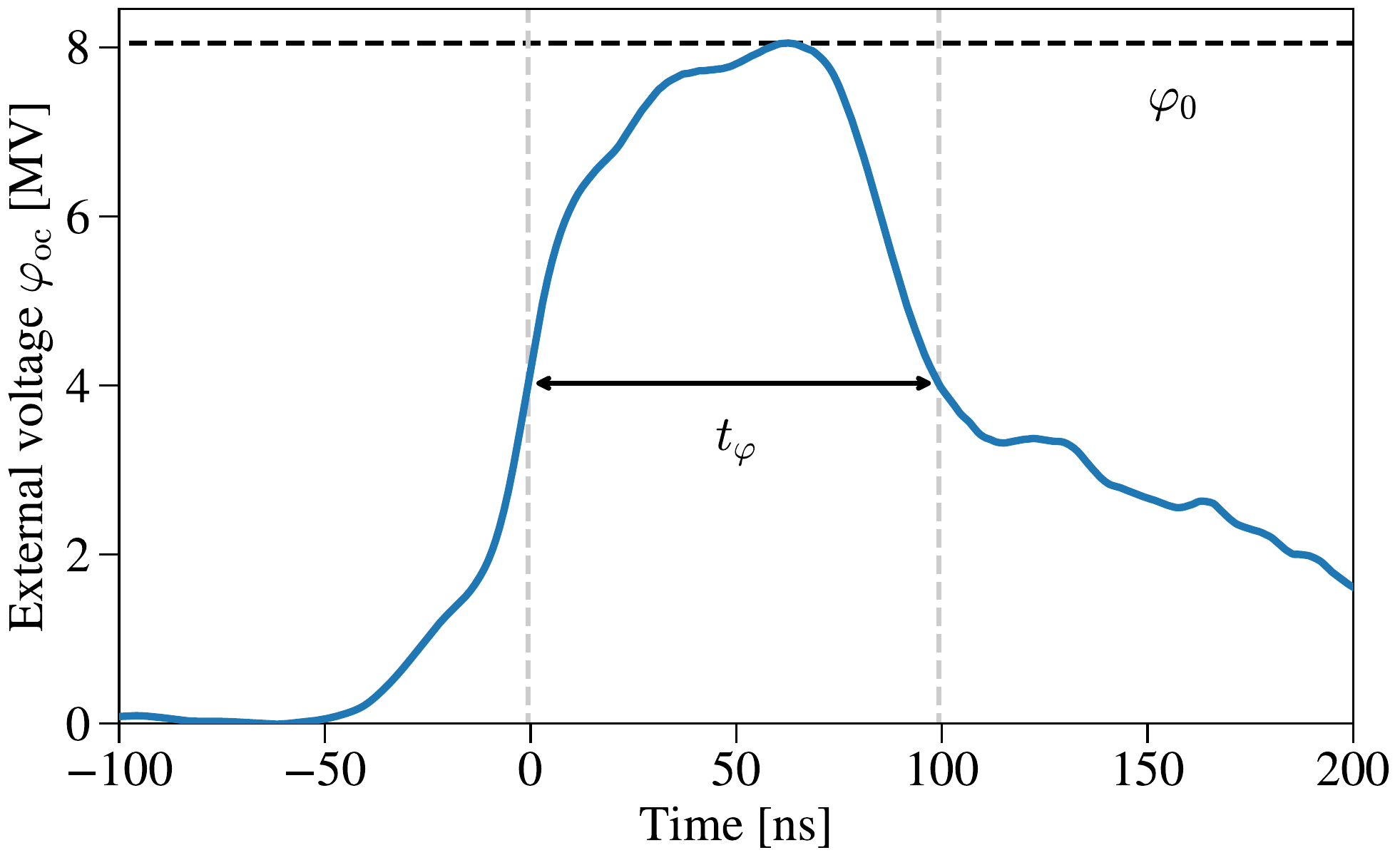}
	\caption{Example of a voltage source $\varphi_{\rm oc}$ as a function of time.  The characteristic voltage $\varphi_0$ is defined as the maximum value of $\varphi_{\rm oc}$.  The characteristic time $t_\varphi$ is defined as the full-width half-maximum (FWHM) of the voltage curve.}
	\label{fig:scaling:Voc}
\end{figure}

In Paper I,\cite{foot:Ruiz_framework} a general framework was introduced for similarity scaling MagLIF loads.  In contrast to Paper II,\cite{foot:Ruiz_current} where the independent scaling variable was the characteristic current $I_\star$, here we shall focus on the scaling of MagLIF loads with respect to the characteristic time $t_\varphi$ of the external voltage drive $\varphi_{\rm oc}(t)$ appearing in \Fig{fig:scaling:circuit}.  The characteristic time $t_\varphi$ is defined as the full-width, half-maximum (FWHM) of the external voltage driving the z-pinch circuit (see \Fig{fig:scaling:Voc}).  If similarity-scaling arguments hold, all timescales present in MagLIF implosions should scale linearly with $t_\varphi$; therefore, varying $t_\varphi$ is equivalent to changing the current-rise time $t_{\rm rise}$.  The assumption of the linear scaling of all timescales with $t_\varphi$ will be tested throughout this paper.

%%%%%%%%%%%%%%%%%%%%%%%%%%%%%%%%%%%%%%%%%%%%%%%%%
\subsection{Scaling prescriptions for a MagLIF load}

Let us write the scaling rules for the input parameters describing a MagLIF load.  In Paper~I, two dimensionless parameters were identified that characterize the magnetic drive of the z-pinch implosion and the liner susceptibility towards the magneto-Rayleigh--Taylor (MRT) instability.\cite{Harris:1962hu,Weis:2015hk,Velikovich:2015jl,Sinars:2010de,McBride:2012db,McBride:2013gda,Awe:2014gba,Ruiz:2022aa}  The first parameter
\begin{equation}
	\Pi  	\doteq \frac{\mu_0 I_\star^2 }{4 \pi \widehat{m} R_{\rm out,0}^2 /  t_\varphi^2}
	\label{eq:scaling:Pi}
\end{equation}
characterizes how strongly the magnetic drive accelerates the liner.\cite{Ryutov:2014hr}  Here $I_\star$ is a characteristic current driving the implosion [explicitly defined in \Eq{eq:scaling:Istar}], $R_{\rm out,0}$ is the initial outer radius of the liner, $\widehat{m}$ is the liner mass per-unit-length, and $\mu_0$ is the magnetic permeability of free space.  The second parameter 
\begin{equation}
	\Psi 	\doteq 2 \pi \frac{R_{\rm out,0}^2 \rho_{\rm ref}}{\widehat{m}} 
				\left(\frac{p_{\rm mag}}{2 p_{\rm ref}} \right)^{2/\gamma}.
	\label{eq:scaling:Psi}	
\end{equation}
serves as a measure of the liner susceptibility towards instabilities.\cite{Schmit:2020jd}  In \Eq{eq:scaling:Psi}, $p_{\rm mag,ext} \doteq (\mu_0 I_\star^2)/(16 \pi^2 R_{\rm out,0}^2)$ is the characteristic external magnetic pressure driving the implosion.  $p_{\rm ref} $, $\rho_{\rm ref}$, and $\gamma$ are respectively the reference pressure, mass density, and polytropic index that describe the equation-of-state (EOS) [$p_{\rm liner}(\rho) = p_{\rm ref} (\rho/\rho_{\rm ref})^\gamma]$ of an adiabatically compressed liner material.  To understand the dependencies entering $\Psi$, we note that the term $R_{\rm out,0}^2/\widehat{m}$ is proportional to the initial aspect ratio (AR$\doteq R_{\rm out,0}/\delta R_0$) for the case of thin-shell liners.  $\Psi$ also depends on the ratio $p_{\rm mag,ext}/p_{\rm ref}$ since the liner becomes more compressed and unstable with higher magnetic pressures.  Hence, $\Psi$ increases with AR and $p_{\rm mag,ext}$.  When $\Psi$ is large, the liner is more susceptible to instabilities.  (See Paper I for more details.)

In this study, the characteristic current $I_\star$ is held constant, and we consider the liner parameters $p_{\rm ref} $, $\rho_{\rm ref}$, and $\gamma$ fixed.  When varying the characteristic time $t_\varphi$ and enforcing conservation of the dimensionless parameters $\Pi$ and $\Psi$, we obtain two scaling laws for $R_{\rm out,0}$ and $\widehat{m}$, which are given by
\begin{align}
	\frac{R_{\rm out,0}'}{R_{\rm out,0}}
			&=	\left( \frac{t_\varphi'}{t_\varphi} \right)^{\frac{\gamma}{2\gamma-1}}  , 
		\label{eq:scaling:Rout} \\
	\frac{\widehat{m}'}{\widehat{m}}
			&=\left( \frac{t_\varphi'}{t_\varphi} \right)^{\frac{2\gamma-2}{2\gamma-1}}  .
		\label{eq:scaling:mhat} 	
\end{align}
As in Papers I and II, for an arbitrary quantity $Q$ corresponding to the \emph{baseline} MagLIF load, the quantity $Q'$ corresponds to the value of the \textit{scaled} MagLIF configuration.  From \Eqs{eq:scaling:Rout} and \eq{eq:scaling:mhat}, we find that the liner outer radius $R_{\rm out,0}$ and the mass per-unit-length $\widehat{m}$ generally increase for longer characteristic implosion times.

The scaled liner inner radius $R_{\rm in,0}'$ is obtained from $R_{\rm out,0}'$ and $\widehat{m}'$ via the formula:
\begin{equation}
	\widehat{m}' \doteq \pi \rho_{\rm liner,0} \left( R_{\rm out,0}'^2 - R_{\rm in,0}'^2 \right),
	\label{eq:scaling:Rin}
\end{equation}
where $\rho_{\rm liner,0}$ is the initial density of the liner.  Along with the specification of the dimensions for the baseline liner and the polytropic index $\gamma$, \Eqs{eq:scaling:Rout}--\eq{eq:scaling:Rin} determine the scaling of the liner radial dimensions.

As discussed in Paper I, the deceleration of the liner by the compressed fuel pressure is characterized by the dimensionless parameter $\Phi$ which is given by
\begin{equation}
	\Phi 	\doteq \frac{4}{3} \frac{E_{\rm preheat}  }{\widehat{m} h R_{\rm out,0}^2 /  t_\varphi^2},
			\label{eq:scaling:Phi}
\end{equation}
where $E_{\rm preheat}$ is the preheat energy delivered to the fuel and $h$ is the imploding axial length of the liner.  The dimensionless parameter $\Phi$ measures the relative importance of the fuel preheat energy to the characteristic liner kinetic energy.  It characterizes the pushback done by the fuel on the imploding liner.\cite{Schmit:2020jd} When conserving $\Phi$, we find that the preheat energy per-unit-length $\smash{\widehat{E}_{\rm preheat}\doteq E_{\rm preheat}/h}$ remains constant when varying the characteristic time of the voltage drive:
\begin{equation}
	\widehat{E}_{\rm preheat}' = \widehat{E}_{\rm preheat}.
	\label{eq:scaling:Epreheathat}
\end{equation}
We emphasize that $\smash{\widehat{E}_{\rm preheat}}$ remains constant, but the \textit{total} preheat energy $E_{\rm preheat}$ scales with the imploding length $h$ of the liner:
\begin{equation}
	\frac{E_{\rm preheat}'}{E_{\rm preheat}}
		=	\frac{\widehat{E}_{\rm preheat}'}{\widehat{E}_{\rm preheat}}
			\frac{h'}{h}
		=	\frac{h'}{h}.
	\label{eq:scaling:Epreheat}
\end{equation}
As shown below, when increasing the characteristic time $t_\varphi$, $h$ needs to increase to conserve end-flow energy losses.  Due to similarity, all timescales appearing in the problem should scale with $t_\varphi$.  Therefore, the time $t_{\rm preheat}$ at which preheat occurs obeys the following scaling rule:
\begin{equation}
		\frac{t_{\rm preheat}'}{t_{\rm preheat}}
			= \frac{t_\varphi'}{t_\varphi} .
	\label{eq:scaling:tpreheat}
\end{equation}
This ensures that preheat occurs at the same stage, or equivalently same convergence ratio, of the implosions.

Following Paper I, we scale the initial fuel density $\rho_0$, the external magnetic field $B_{z,0}$, and the liner height $h$ such that the relative radiation, thermal-conduction, and end-flow energy losses are conserved.  Within the framework of similarity scaling, the dimensionless parameters characterizing these energy-loss processes have the following dependencies on the MagLIF input parameters:
\begin{align}
	\Upsilon_{\rm rad} & \propto \frac{\rho_0^2 T_{\rm preheat}^{1/2} R_{\rm in,0}^2 h }{E_{\rm preheat}} \, t_\varphi, 	
		\label{eq:scaling:Upsilonrad} \\
	\Upsilon_c  & \propto \frac{\rho_0 T_{\rm preheat}^2 h }{E_{\rm preheat} B_{z,0} } \, t_\varphi, 	
		\label{eq:scaling:Upsilonc} \\
	\Upsilon_{\rm end} & \propto \frac{T_{\rm preheat}^{1/2} }{h}\, t_\varphi, 
		\label{eq:scaling:Upsilonend}
\end{align}
where $\smash{T_{\rm preheat}}$ is the characteristic fuel temperature achieved during preheat and depends as  $\smash{T_{\rm preheat}\propto E_{\rm preheat}/(\rho_0 R_{\rm in,0}^2 h)}$.  The dimensionless parameter \eq{eq:scaling:Upsilonc} considers thermal-conduction losses in a Bohm-like regime.  This is a result of internal advection flows arising in the isobaric, hot plasma core as the plasma is cooled by the cold liner walls.\cite{Vekshtein:1983aa,Vekshtein:1986aa}  Such advection flows enhance thermal-conduction losses, specially in the high-magnetization regime.\cite{Velikovich:2015gs}  When varying $t_\varphi$ and enforcing conservation of \Eqs{eq:scaling:Upsilonrad}--\eq{eq:scaling:Upsilonend}, we obtain the scaling rules for $\rho_0$, $B_{z,0}$, and $h$:
\begin{align}
	\frac{\rho_0'}{\rho_0}&
			=	\left( \frac{R_{\rm in,0}'}{R_{\rm in,0} } \right)^{-2/3}
				\left( \frac{t_\varphi'}{t_\varphi} \right)^{-2/3}   ,
		\label{eq:scaling:rho}  \\
	\frac{B_{z,0}'}{B_{z,0}} &
			=	\left( \frac{R_{\rm in,0}'}{R_{\rm in,0} } \right)^{-10/3}
				\left( \frac{t_\varphi'}{t_\varphi} \right)^{5/3} ,
		\label{eq:scaling:Bz} \\
	\frac{h'}{h}&
			=	\left( \frac{R_{\rm in,0}'}{R_{\rm in,0} } \right)^{-2/3}
				\left( \frac{t_\varphi'}{t_\varphi} \right)^{4/3} .
		\label{eq:scaling:h} 
\end{align}

\begin{figure}
	\includegraphics[scale=0.43]{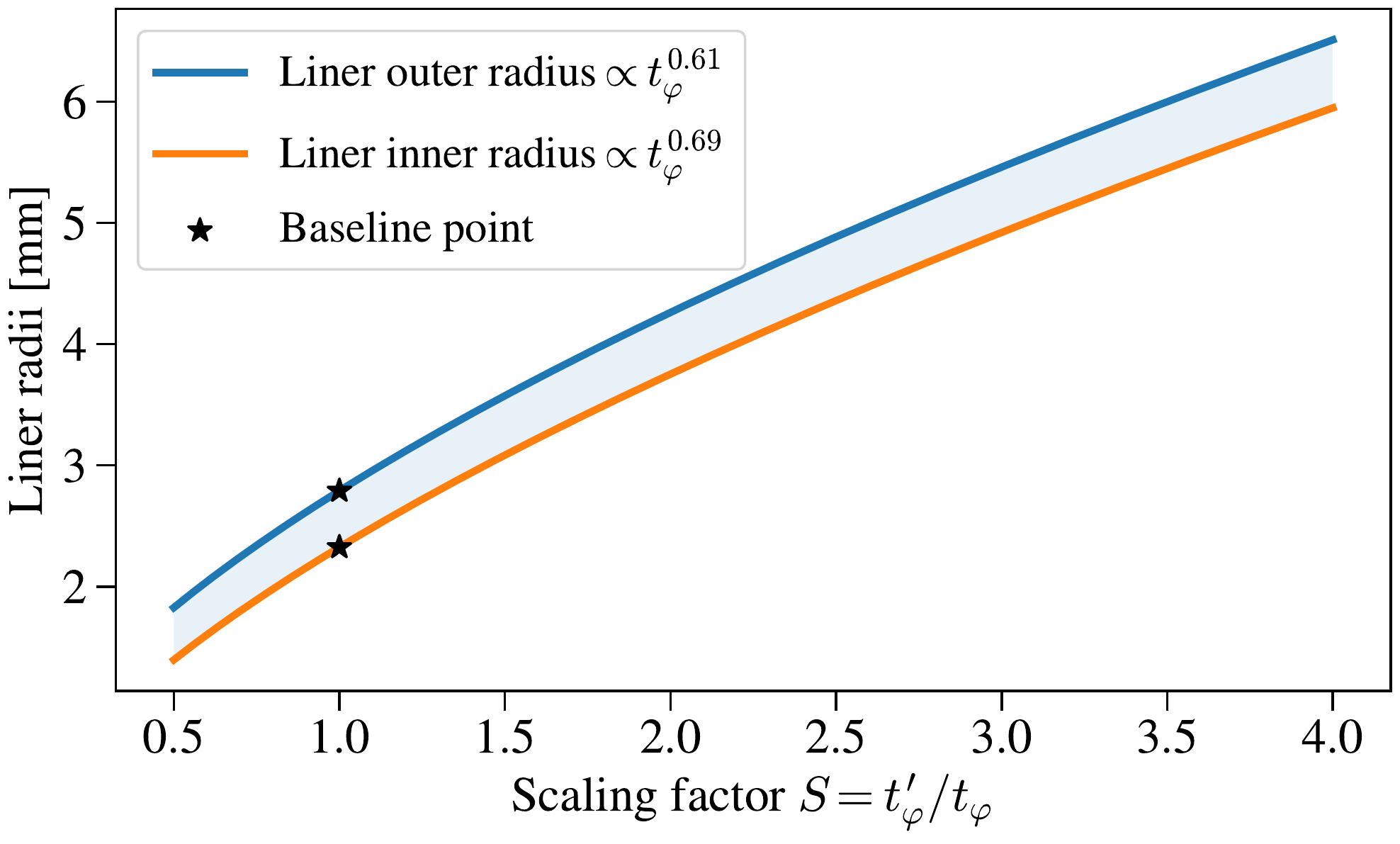}
	\caption{Scaling curves for the initial inner and outer radii of a MagLIF liner.  These curves are based off a typically fielded MagLIF target with $R_{\rm out,0} =2.79$~mm and $R_{\rm in,0}=2.325$~mm.\cite{Gomez:2020cd}  When the rise time increases, the liners become larger in radius.  Since magnetic pressure decreases when increasing the radius at fixed current, the thickness of the scaled liners does not change substantially when varying $t_\varphi$.}
	\label{fig:numerical:radii}
\end{figure}

Equations \eq{eq:scaling:Rout}--\eq{eq:scaling:Rin}, \eq{eq:scaling:Epreheathat}--\eq{eq:scaling:tpreheat}, and \eq{eq:scaling:rho}--\eq{eq:scaling:h} represent the scaling rules for the most important input parameters characterizing a MagLIF load.  Following the discussion in Paper II,\cite{foot:Ruiz_current} geometric similarity is invoked for the scaling of other secondary parameters describing a MagLIF load.  For the sake of completeness, we rewrite these scaling rules below.  The laser-spot size $R_{\rm spot}$ and the inner radius $R_{\rm cushion}$ of the cushions\cite{foot:cushions} are scaled proportionally to the initial inner radius $R_{\rm in,0}$ of the liner:
\begin{equation}
	\frac{R_{\rm spot}'}{R_{\rm spot}} 
		= \frac{R_{\rm cushion}'}{R_{\rm cushion}} 
		=	\frac{R_{\rm in,0}'}{R_{\rm in,0}}.
\end{equation}
Parameters describing other axial dimensions of the load, such as the height of the cushions, the anode--cathode gap, and the axial location of the laser-entrance-hole window, are scaled proportionally to the liner height $h$.  Upon denoting these quantities by $H$, we have
\begin{equation}
	\frac{H'}{H} = \frac{h'}{h}.
	\label{eq:scaling:H} 
\end{equation}
These additional scaling rules complete the scaling prescriptions for the input parameters defining a MagLIF load.

%%%%%%%%%%%%%%%%%%%%%%%%%%%%%%%%%%%%%%%%%%%%%%%%%
\subsection{Scaling prescriptions for the circuit parameters}

Now, let us discuss the scaling prescriptions required to maintain the electrical coupling between the circuit and the imploding MagLIF load.  In this work, the circuit model for the Z generator shown in \Fig{fig:scaling:circuit} is used to drive the simulated MagLIF implosions.  In \Fig{fig:scaling:circuit}, $\varphi_{\rm oc}(t)$ is the external time-varying drive voltage and is twice the forward-going voltage at the vacuum-insulator stack on Z.  (An example time-trace for $\varphi_{\rm oc}$ is shown in \Fig{fig:scaling:Voc}.)  $Z_0$ is the effective impedance of the pulsed-power generator, $L_0$ is the inductance of the outer magnetically-insulated transmission lines, $\varphi_c(t)$ is the corresponding voltage across the capacitor $C$ associated to the MITLs, $L_1$ is the initial inductance of post-convolute feed region, $R_{\rm loss}(t)$ is a shunt resistor using a prescribed time-dependent model, and $L_{\rm load}(t)$ is the time-varying inductance of the imploding MagLIF load.  The model used for the shunt resistor is given by
\begin{equation}
	R_{\rm loss} (t)  \doteq R_{\rm loss,i} + (R_{\rm loss,f}- R_{\rm loss,i}) f(t),
	\label{eq:electric:Rloss}
\end{equation}
where
\begin{equation}
	f(t) = \frac{1}{1+ \exp\left( - \frac{t-t_{\rm loss}}{\Delta t_{\rm loss}} \right) } .
	\label{eq:electrical:f}
\end{equation}
is a function describing the transition from the initial loss resistance $R_{\rm loss,i}$ of the circuit early in time to the final loss resistance $R_{\rm loss,f}$ at later times. Specific values for the circuit components are given in \Sec{sec:numerical}.

In Section~II of Paper I, the governing equations for the circuit in \Fig{fig:scaling:circuit} are given.  When rewriting the equations in dimensionless form, we obtain six dimensionless parameters describing the circuit inductance matching, the LR-circuit drive efficiency, the LC-circuit resonance, relative current losses, and the load--circuit coupling:
\begin{equation}
	\begin{aligned}
	c_1 &\doteq \frac{L_0}{L_0+L_1},  							& c_2 & \doteq \frac{Z_0 t_\varphi}{L_0+L_1},  \\
	c_3 &\doteq (L_0+L_1)^{1/2} \frac{C^{1/2}}{t_\varphi}, 	& c_4 &	\doteq \frac{R_{\rm loss,i} t_\varphi}{L_0+L_1},   \\
	c_5 &\doteq \frac{R_{\rm loss,f} t_\varphi}{L_0+L_1}, 		& c_6 & \doteq \frac{\mu_0 h}{2\pi(L_0+L_1)}.
	\end{aligned}
	\label{eq:electrical:variables}
\end{equation}
When varying $t_\varphi$ and enforcing conservation of the six dimensionless parameters above, we obtain the following scaling rules for the circuit components:
\begin{gather}
	\frac{Z_0'}{Z_0}  
		=	\frac{R_{\rm loss,i}'}{R_{\rm loss,i}} 
		=	\frac{R_{\rm loss,f}'}{R_{\rm loss,f}} 
		= 	\frac{h'}{h}  \frac{t_\varphi}{t_\varphi'},
	\label{eq:scaling:Z} \\
	\frac{L_0'}{L_0} 
		=	\frac{L_1'}{L_1} 
		= 	\frac{h'}{h} ,
	\label{eq:scaling:L} \\	
	\frac{C'}{C} 
		 = \frac{h}{h'} \left( \frac{t_\varphi'}{t_\varphi} \right)^2.
	\label{eq:scaling:C} 
\end{gather}
As shown, all resistances and impedances in the circuit scale proportionally to the target height and inversely proportional to $t_\varphi$.  Inductances scale proportionally to the load height $h$. The capacitance $C$ scales inversely proportionally to $h$ and proportionally to $t_\varphi^2$.  Finally, the time parameters describing the shunt resistor in \Eqs{eq:electric:Rloss} and \eq{eq:electrical:f} are scaled proportionally to $t_\varphi$ so that
\begin{equation}
	\frac{t_{\rm loss}'}{t_{\rm loss}}
		=	\frac{\Delta t_{\rm loss}' }{\Delta t_{\rm loss}}
		=	\frac{t_\varphi'}{t_\varphi} .
	\label{eq:scaling:tloss} 
\end{equation}

\begin{figure*}
	\includegraphics[scale=0.42]{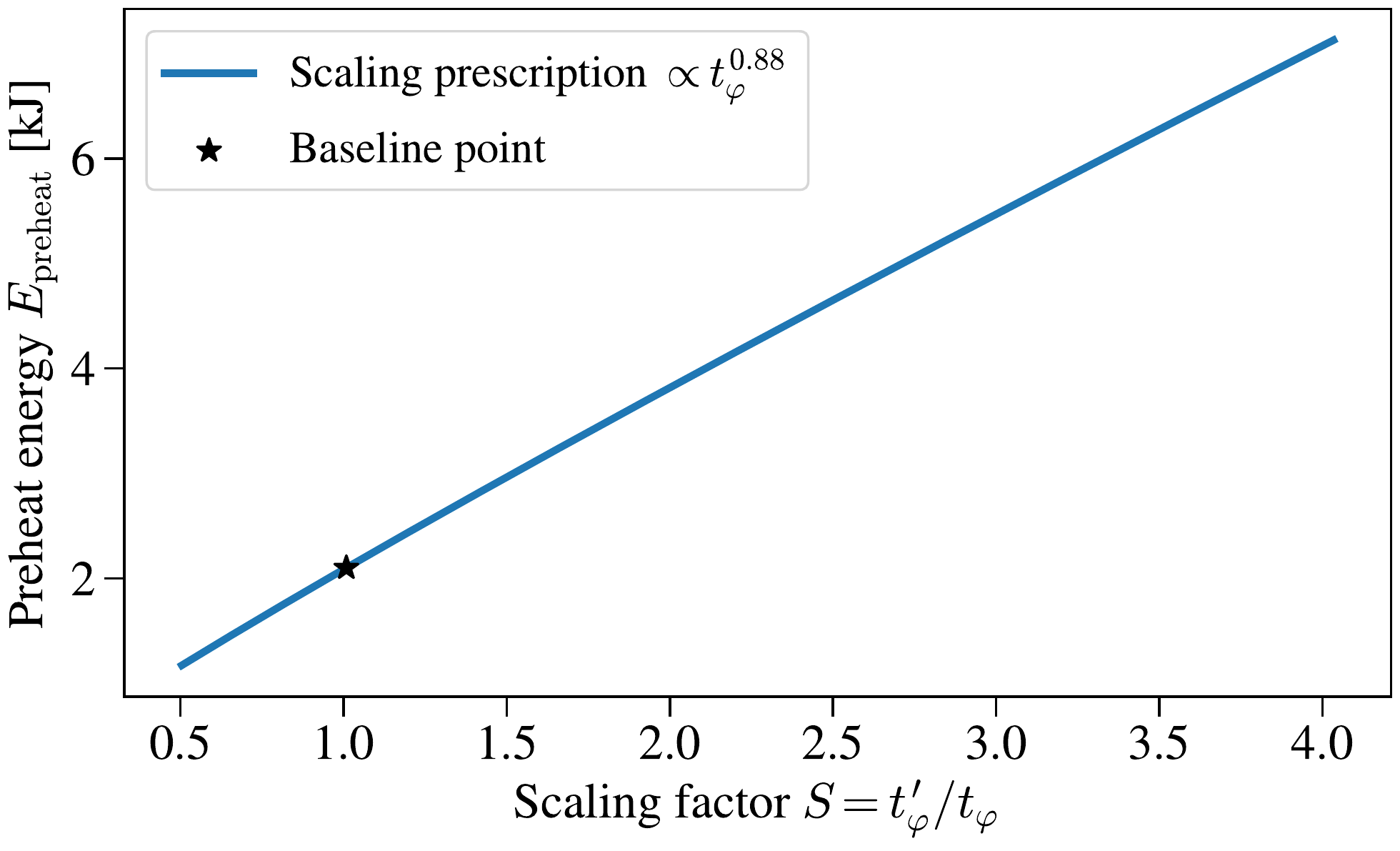}
	\hspace{0.3cm}
	\includegraphics[scale=0.42]{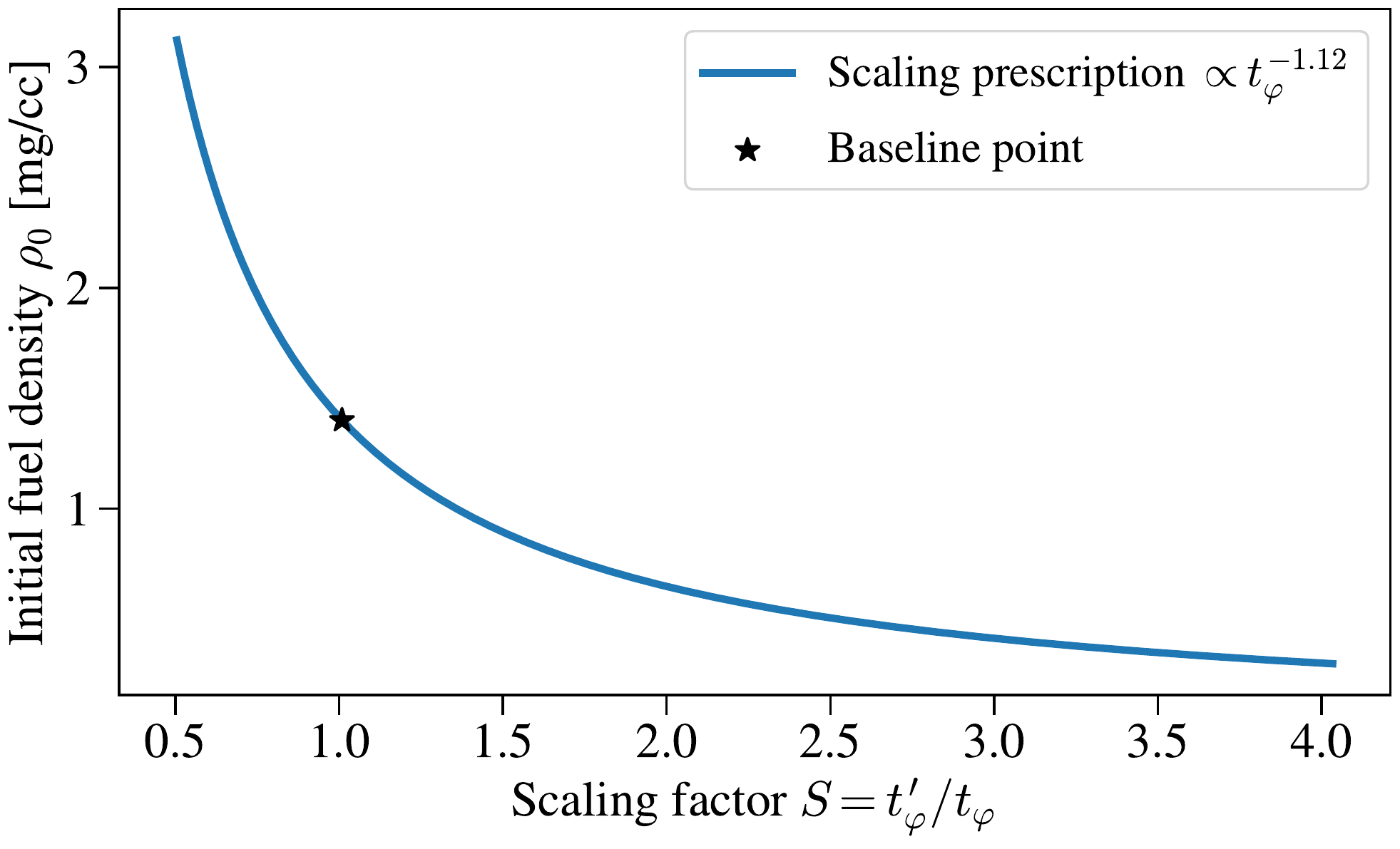}
	\includegraphics[scale=0.42]{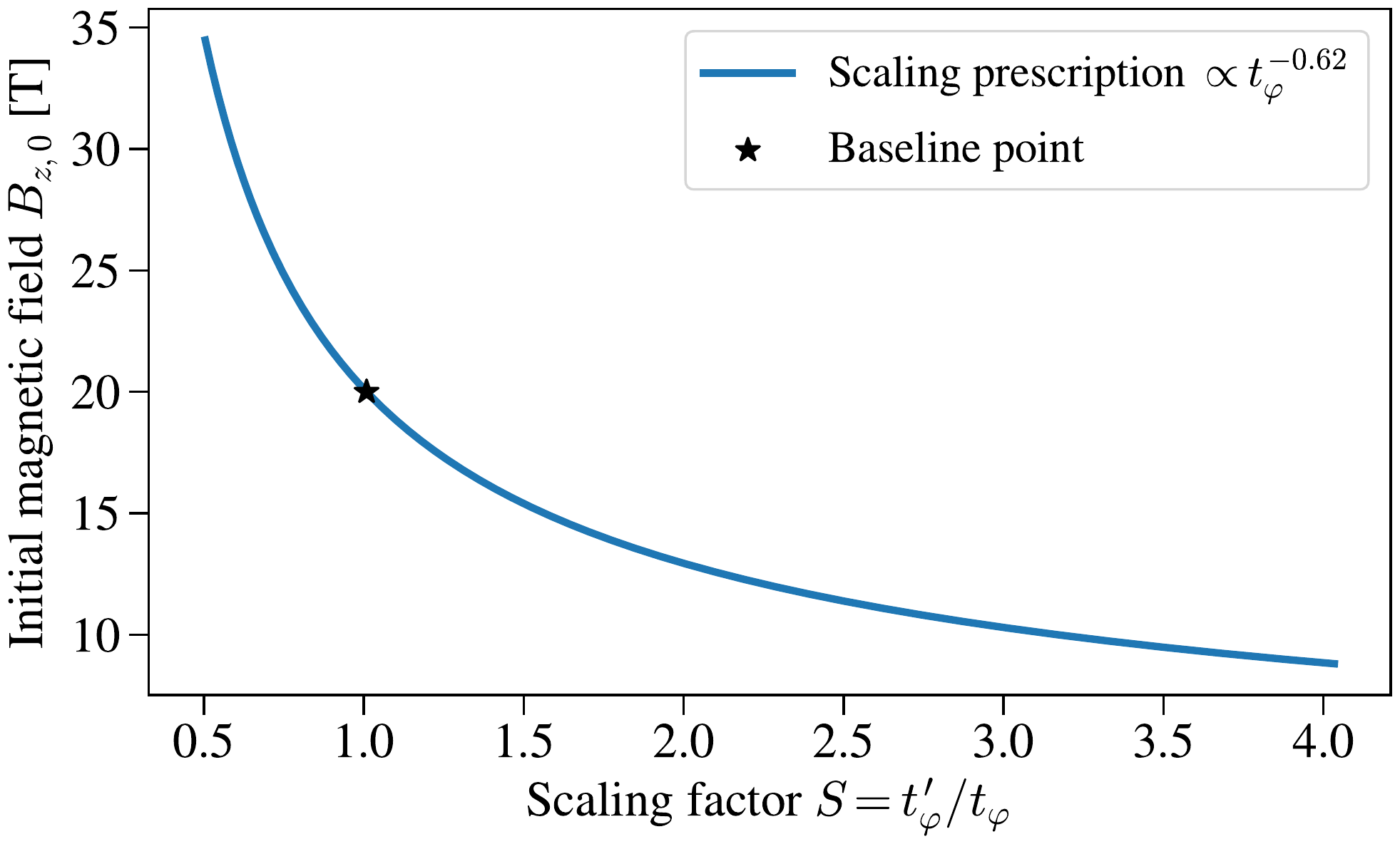}
	\hspace{0.3cm}
	\includegraphics[scale=0.42]{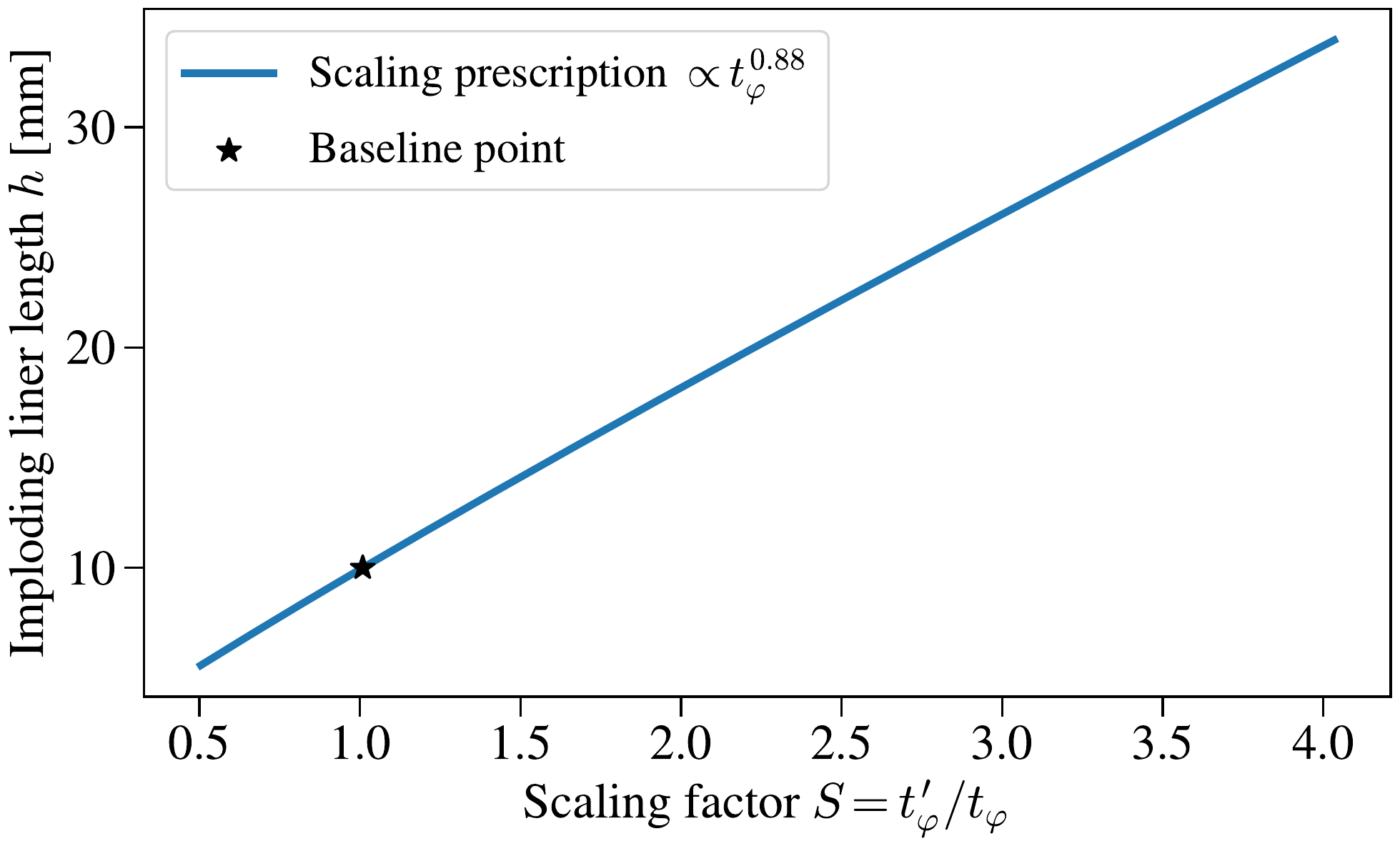}
	\caption{Scaling curves for the preheat energy $E_{\rm preheat}$, initial fuel density $\rho_0$, initial magnetic field $B_{z,0}$, and imploding height $h$ of the liner.  The legends show the approximate power-law fits to the scaling prescriptions.}
	\label{fig:numerical:parameters}
\end{figure*}

The external voltage source $\varphi_{\rm oc}$ can be written as $\varphi_{\rm oc}(t) = \varphi_0 \bar{\varphi}_{\rm oc}(t/t_{\rm \varphi})$, where $\bar{\varphi}_{\rm oc}(\bar{t})$ is the normalized voltage trace, $\bar{t}\doteq t/t_\varphi$ is the normalized time variable, and $\varphi_0$ is defined as the maximum value of $\varphi_{\rm oc}$.)  We require a scaling rule for $\varphi_0$.  To obtain such scaling rule, we recall that the characteristic current $I_\star$ driving the implosion can be defined as\cite{foot:Ruiz_framework}
\begin{equation}
	I_\star \doteq \frac{\varphi_0}{Z_0 + L_{\rm tot}/t_\varphi },
	\label{eq:scaling:Istar} 
\end{equation}
where $L_{\rm tot}\doteq L_0 + L_1$ is the total initial inductance in the circuit.  [Intuitively, $I_\star$ represents the current corresponding to a short-circuit configuration with no imploding load $(L_{\rm load}=0)$, no MITL capacitance, and no current losses.]  In this study, the peak current $\I\doteq \max(I_l)$ is intended to remain fixed for the scaled MagLIF configurations.  Therefore, the characteristic current $I_\star$ must also remain constant when varying $t_\varphi$.  Upon substituting \Eqs{eq:scaling:Z} and \eq{eq:scaling:L} into \Eq{eq:scaling:Istar}, we obtain the scaling law for the characteristic voltage $\varphi_0$:
\begin{equation}
	\frac{\varphi_0'}{\varphi_0}
		=	\frac{h'}{h} \frac{t_\varphi}{t_\varphi'} .
	\label{eq:scaling:varphi} 
\end{equation}
As expected, the characteristic voltage $\varphi_0$ and therefore all voltages in the circuit scale inversely proportionally to the characteristic time $t_\varphi$.  However, $\varphi_0$ also scales linearly with the liner height $h$, which increases for longer characteristic timescales $t_\varphi$.  Therefore, the voltages in the circuit may actually have a weaker scaling than the commonly assumed $t_\varphi^{-1}$ scaling.  We shall discuss this point further in \Sec{sec:discussion}.

%%%%%%%%%%%%%%%%%%%%%%%%%%%%%%%%%%%%%%%%%%%%%%%%%
%%%%%%%%%%%%%%%%%%%%%%%%%%%%%%%%%%%%%%%%%%%%%%%%%
%%%%%%%%%%%%%%%%%%%%%%%%%%%%%%%%%%%%%%%%%%%%%%%%%
\section{Numerical simulations and baseline load parameters}
\label{sec:numerical}

In the following sections, we compare predictions of the scaling theory against simulation results obtained from 2D \textsc{hydra} simulations.  \textsc{hydra} is a massively parallel arbitrary Lagrangian--Eulerian (ALE) radiation, resistive-diffusion, magneto-hydrodynamics code and is one of the main design tools for MagLIF experiments.\cite{Sefkow:2014ik,HarveyThompson:2018dd,Weis:2021id}  Further details on \textsc{hydra} are given in \Refs{Marinak:1996fs,Koning:2009}.

The 2D \textsc{hydra} simulations were externally driven using the circuit model shown in \Fig{fig:scaling:circuit}.  The circuit-model parameters for the baseline configuration are kept the same as those in Paper II.  For the sake of completeness, these parameters are $Z_0=0.18~\Omega$, $L_0=9.58$~nH, $C=0.1$~nF, and $L_{\rm load}=5$~nH.  For the shunt resistor, we used $R_{\rm loss,i}=80$~Ohm and  $R_{\rm loss,f}=0.25$~Ohm.  The time parameters for the shunt resistor are $t_{\rm loss}=3000$~ns and  $\Delta t_{\rm loss}=5$~ns.  The open-source voltage used is shown in \Fig{fig:scaling:Voc}.  The circuit parameters and the voltage drive are scaled according to \Eqs{eq:scaling:Z}--\eq{eq:scaling:varphi}.  

The baseline MagLIF configuration considered in this paper has an initial inner radius of $R_{\rm in,0}=2.325$~mm and outer radius of $R_{\rm out,0}=2.79$~mm.  Thus, the initial aspect ratio AR$\doteq R_{\rm out,0}/(R_{\rm out,0}-R_{\rm in,0})$ of this liner is six.   In these simulations, the liner is made of Be with initial density 1.858~g/cm$^3$, so the baseline mass per-unit-length is approximately 139~mg/cm.  Regarding the fuel parameters, we considered a pure deuterium D$_2$ gas fill at $\rho_0 = 1.4$~mg/cm$^3$ density.  The preimposed initial axial magnetic field is $B_{z,0}=$20~T.  For the preheat energy deposition, the fuel is heated uniformly by adding 2.1~kJ of energy into a plasma column of radius $R_{\rm spot} = $0.75~mm coaxial to the liner.  The deposition of energy begins approximately 70~ns before burn time and lasts for 10 ns.\cite{foot:preheat}  The imploding height of the liner is 10 mm.  With exception to the slightly higher density and initial magnetic field, these input parameters are representative of typically fielded MagLIF loads in present-day experiments.\cite{Gomez:2020cd}

In this study, we use a polytropic index $\gamma = 2.75$ for the Be liner, which is slightly larger than the $\gamma=2.25$ value considered in Paper II.\cite{foot:Ruiz_current}  Although using $\gamma=2.25$ leads to reasonable agreement between the theory and simulation results, considering $\gamma = 2.75$ corresponds to modeling the liner as a slightly more incompressible fluid.  When compared to current-scaling results shown in Paper II, scaling with respect to $t_\varphi$ leads to smaller variations of the magnetic shock in the liner material.  Therefore, we believe that a larger value for $\gamma$ is more appropriate when scaling with respect to $t_\varphi$.

It is important to note that, in the present work, $\gamma$ serves as a semi-empirical parameter that represents an ``effective" polytropic index characterizing the liner compressibility for the ensemble of similarity-scaled implosions studied in this paper.  As noted in Section III of Paper I, self-consistently obtaining an adiabatic polytropic index $\gamma$ based on tabular EOS for Be is difficult since the trajectories in the EOS phase-space are highly dependent on the liner-implosion dynamics, and in most cases, MagLIF liners are shocked by the magnetic pressure drive. Future work may include improving the present scaling study by replacing the simple adiabatic EOS model with constant $\gamma$ with a more sophisticated and complete material EOS model.

With the scaling prescriptions in \Eqs{eq:scaling:Rout}--\eq{eq:scaling:Rin} and the parameters given in the preceding paragraphs, we show in \Fig{fig:numerical:radii} the initial inner and outer radii of the  scaled MagLIF liners with respect to the temporal scaling factor $S\doteq t_\varphi'/t_\varphi$.  When increasing $S$, the liner becomes larger in radius.  This is mainly a consequence of constraining the liner to implode in a similar fashion by conserving the $\Pi$ parameter in \Eq{eq:scaling:Pi}.  Both radial dimensions follow almost identical scaling laws with the liner inner radius increasing at a slightly higher rate.  When comparing \Fig{fig:numerical:radii} to the corresponding Fig.~3 of Paper II, scaling MagLIF loads to longer timescales $t_\varphi$ does not substantially change the liner thickness.  Thus, the initial aspect ratio of the liner increases when increasing $t_\varphi$.  From a physics perspective, this occurs because the characteristic magnetic pressure acting on the liner decreases due to the larger outer radius.  Therefore, less magnetic compression of the liner occurs, so there is no need to substantially change the liner thickness.

The scaling law \eq{eq:scaling:Rin} for $R_{\rm in,0}$ has a complex dependency on the scaling prescriptions for $R_{\rm out,0}$ and $\widehat{m}$.  To simplify the upcoming analysis, we fit the scaling law for $R_{\rm in,0}$ within the range shown in \Fig{fig:numerical:radii} using a power law.  The resulting approximate power-law scaling rules for the liner radial dimensions are the following:
\begin{equation}
	\frac{R_{\rm out,0}'}{R_{\rm out,0}} 
		\simeq \left( \frac{t_\varphi'}{t_\varphi}  \right)^{0.61} , \qquad
	\frac{R_{\rm in,0}'}{R_{\rm in,0}} 
		\simeq \left( \frac{t_\varphi'}{t_\varphi}  \right)^{0.69}.
	\label{eq:numerical:R}
\end{equation}

\begin{figure}
	\includegraphics[scale=0.6]{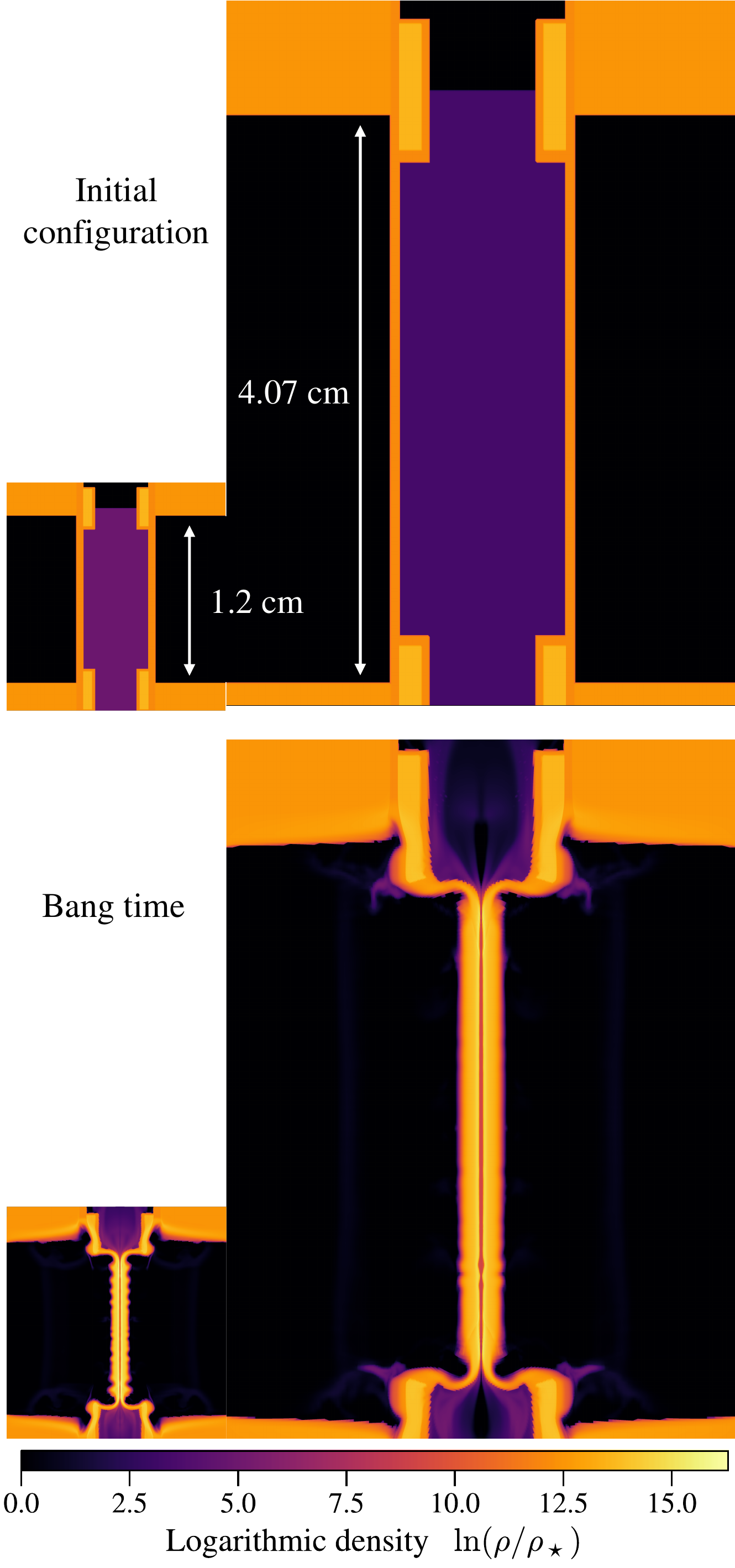}
	\caption{Top left: Logarithmic density plot for the anchor MagLIF load driven at a 20-MA peak with a nominal rise time.  Top right:  Logarithmic density plot of a similarity-scaled MagLIF load driven at 20-MA peak current with a 4x nominal rise time.  Bottom:  Corresponding logarithmic density plots near stagnation calculated using \textsc{hydra}.  Density maps are normalized to $\rho_\star = 10^{-5}$~g/cc.}
	\label{fig:liners}
\end{figure}

\begin{figure}
	\includegraphics[scale=.43]{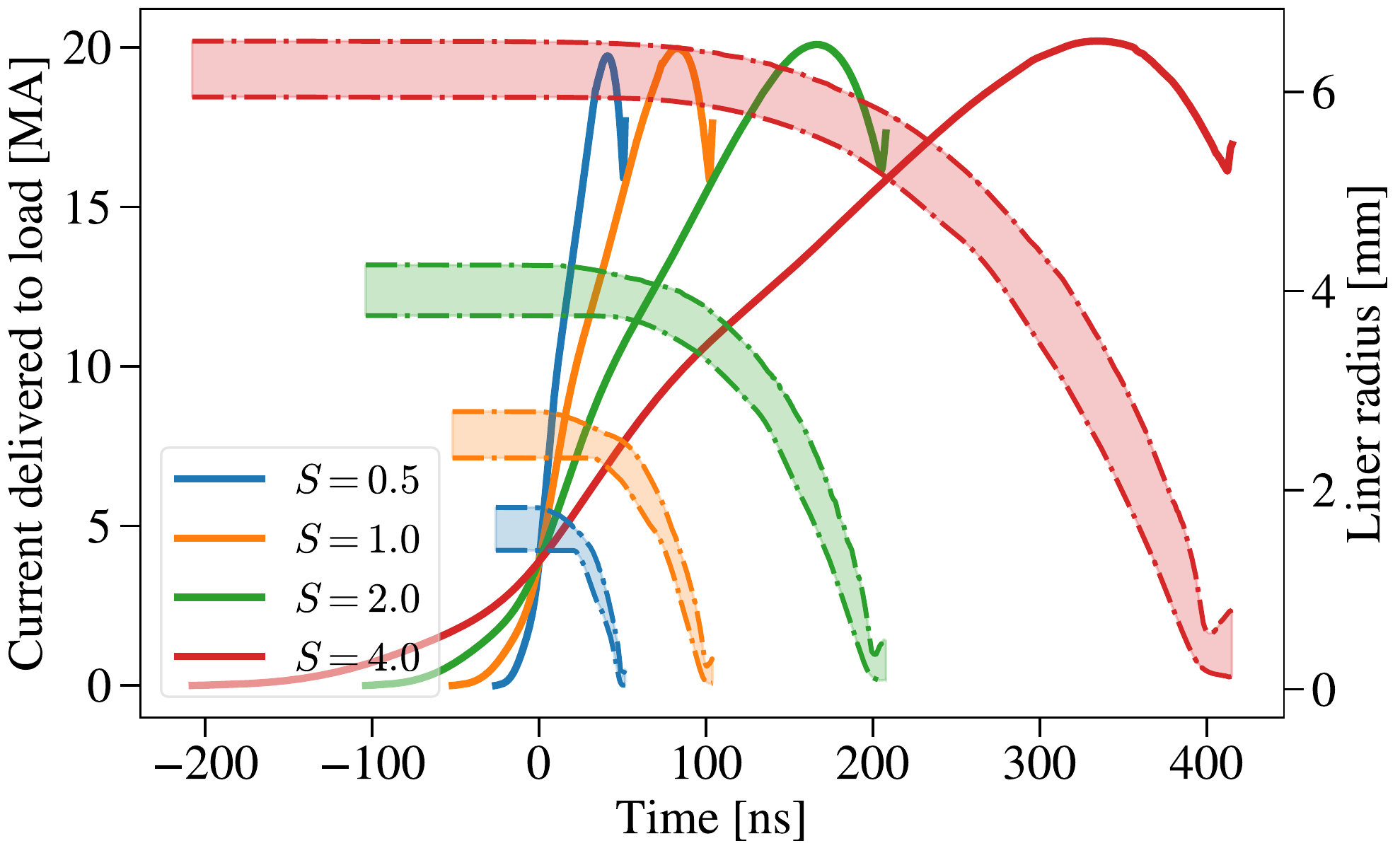}
	\caption{Liner position (shaded regions) and current delivered to the load (solid lines) versus time for different characteristic time scales using \textsc{hydra}.  Increasing the characteristic time $t_\varphi$ of the voltage drive dilates the corresponding time traces of the current delivered to the load and dilates the implosions themselves.  As expected, the peak currents are approximately constant.  In this work, the outer boundary of the liners is tracked used a $1/e\simeq37\%$ threshold of the maximum density (similar to \Refa{Bose:2017jf}), and the inner boundary is tracked using a Lagrangian marker.}
	\label{fig:radius_vs_time}
\end{figure}

In \Fig{fig:numerical:parameters}, we plot the scaling prescriptions for the preheat energy $E_{\rm preheat}$, the initial fuel density $\rho_0$, the applied axial magnetic field $B_{z,0}$, and the imploding target height $h$.  As done for \Eq{eq:numerical:R}, when fitting the exact scaling prescriptions \eq{eq:scaling:Epreheat} and \eq{eq:scaling:rho}--\eq{eq:scaling:h} to power laws, we find the following scaling relations:
\begin{gather}
	\frac{E_{\rm preheat}'}{E_{\rm preheat}} 
		= \frac{h'}{h}
		\simeq \left( \frac{t_\varphi'}{t_\varphi}  \right)^{0.88}  ,
		\label{eq:numerical:Epreheat_h}
\end{gather}
\begin{gather}
	\frac{\rho_0'}{\rho_0}   \simeq \left( \frac{t_\varphi'}{t_\varphi} \right)^{-1.12}, 
		\label{eq:numerical:rho}\\
	\frac{B_{z,0}'}{B_{z,0}}   \simeq \left(\frac{t_\varphi'}{t_\varphi} \right)^{-0.62} .
		\label{eq:numerical:Bz}
\end{gather}
When considering longer implosion timescales, there is more time for the fuel to escape the liner cavity. Therefore, the MagLIF liner height must increase.  As shown in \Eq{eq:numerical:Epreheat_h}, $h$ increases almost linearly with $t_\varphi$, but due to a slight decrease in the characteristic sound speed of the fuel, the scaling exponent for $h$ becomes less than unity.  Likewise, the preheat energy delivered to the target will scale proportionally to the target height.  From \Fig{fig:numerical:parameters}, we see that, when dilating $t_\varphi$ by a factor of 4, a 10-mm long target becomes 33.4-mm long, and the preheat energy increases from 2.1~kJ to 7.0~kJ.  Thus, for a MagLIF load, the required preheat energy scales unfavorably with respect to $t_\varphi$, or equivalently  the implosion time, which is in agreement with the results presented in \Refa{Slutz:2018iq}.  In \Eq{eq:numerical:rho}, the initial fuel density decreases for longer $t_\varphi$ because more slowly imploding MagLIF liners deliver a smaller pdV work rate to the fuel. [From Eq.~(53) of Paper I, one can immediately see that the pdV work rate per-unit-length scales inversely proportional to $t_\varphi$ when similarity scaling.]  Thus, to avoid unsustainable radiation losses, the initial fuel density must decrease for longer $t_\varphi$.  As shown in \Eq{eq:numerical:rho}, the initial fuel density can decrease almost five-fold when dilating the rise time by a factor of 4.  Notably, these lower initial fuel densities can potentially make delivery of the required preheat energy into the fuel more difficult via inverse Bremsstrahlung.  This problem is not addressed in this paper.  Finally, when carefully inspecting the conditions of the preheated plasma, one can note that thermal-conduction losses decrease because the plasma becomes more collisionless as the initial density decreases.  Since relative thermal losses decrease, the required magnetic field needed to mitigate them is smaller.  Thus, MagLIF loads scaled to longer implosion times require weaker externally applied magnetic fields.  According to \Eq{eq:numerical:Bz}, increasing the characteristic timescale $t_\varphi$ by a factor of four reduces the magnetic field from 20~T to 6.4~T.  In summary, increasing the timescale of a MagLIF implosion leads to higher requirements for the laser preheat, liner length, and load inductance.  In contrast, longer implosions can only support smaller initial density fills and require smaller externally applied magnetic fields.

\begin{figure}
	\includegraphics[scale=.43]{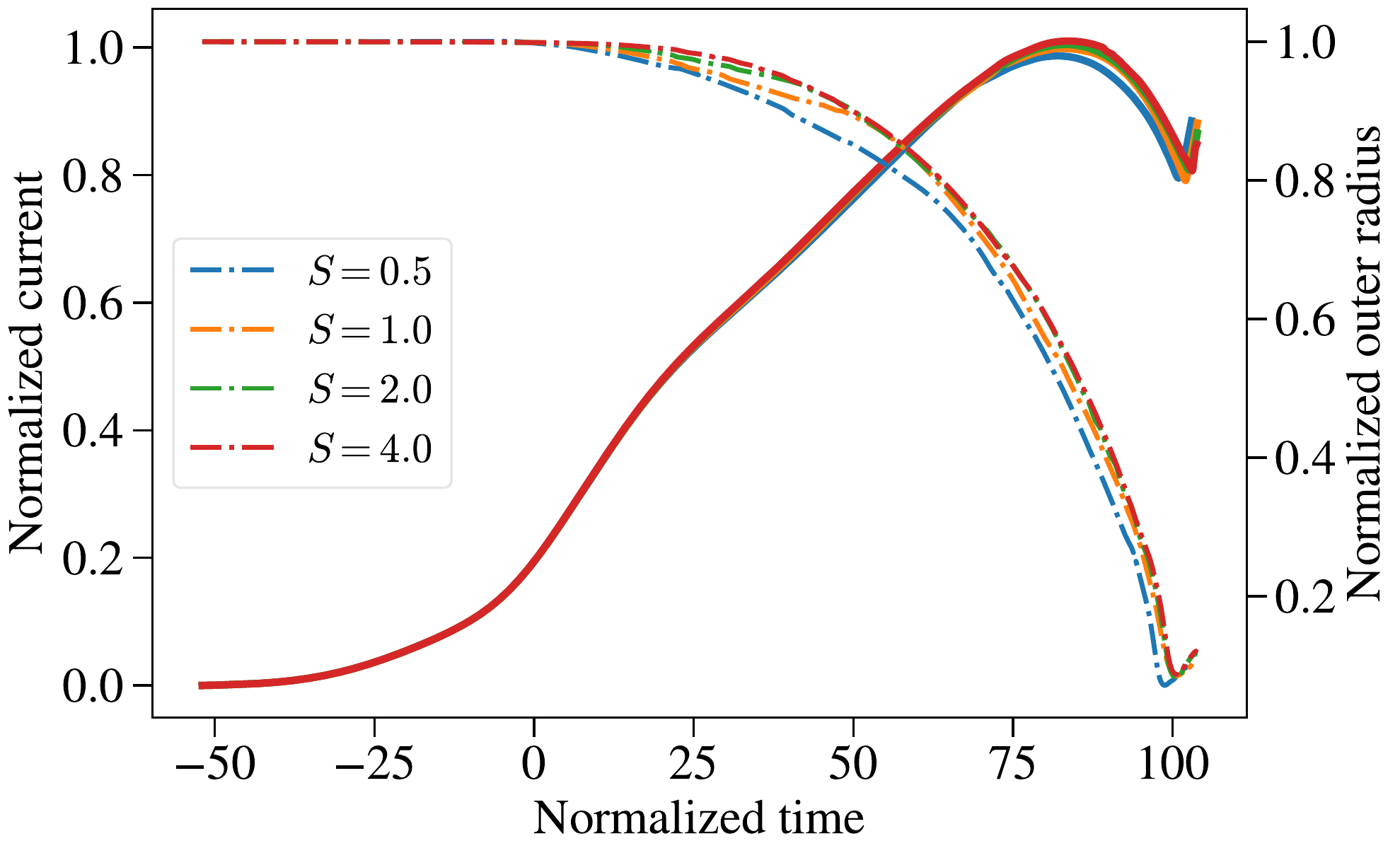}
	\caption{Normalized implosion trajectory of the outer liner radius (dashed lines) and normalized current traces (solid lines) calculated using \textsc{hydra}.  When following the scaling prescriptions provided in \Sec{sec:prescriptions}, the normalized trajectories of the outer liner radius are almost identical.  Likewise, the normalized current traces are also in very close agreement, as expected from the scaling theory.}
	\label{fig:radius_norm_vs_time}
\end{figure}

In \Fig{fig:liners} (top), we show logarithmic density plots of the initial configurations of the baseline MagLIF load and its similarity-scaled configuration driven by a 4x longer current pulse.  As shown, when increasing the characteristic timescale of the implosions, the liners become significantly longer and their radial dimensions increase.  The bottom part of \Fig{fig:liners} illustrates that these similarity-scaled loads look qualitatively similar near peak burn.

%%%%%%%%%%%%%%%%%%%%%%%%%%%%%%%%%%%%%%%%%%%%%%%%%
%%%%%%%%%%%%%%%%%%%%%%%%%%%%%%%%%%%%%%%%%%%%%%%%%
%%%%%%%%%%%%%%%%%%%%%%%%%%%%%%%%%%%%%%%%%%%%%%%%%
\section{Liner-implosion dynamics}
\label{sec:implosion}

In \Fig{fig:radius_vs_time}, we show radius-versus-time and current-versus-time plots calculated using $\textsc{hydra}$ for four similarity-scaled MagLIF liners.  The temporal scaling factors $S$ are 0.5, 1, 2, and 4.  The shaded areas represent the liner regions.  The solid curves represent the time traces of the current delivered to the loads.  As observed, the temporal duration of the implosions of these similarity-scaled MagLIF configurations qualitatively scale linearly with $S$.

Figure~\ref{fig:radius_norm_vs_time} plots the current traces normalized to 20~MA and the normalized liner outer radius $\bar{R}_{\rm out} \doteq R_{\rm out}/R_{\rm out,0}$.  We normalize the time axis according to the scaling factor so that $t \to t/S$.  As shown in the figure, the normalized outer radii of the scaled liners (shown in dashed lines) implode in a similar fashion with some minor  differences.  For example, one can note that the normalized outer radius for the smaller faster-imploding $S=0.5$ load (blue dashed curve) follows a slightly earlier implosion trajectory in normalized time.  In this case, the smaller initial outer radius of the liner is subject to a larger magnetic-field pressure, which scales as
\begin{equation}
	\frac{p_{\rm mag,ext}'}{p_{\rm mag,ext}}
		= \left( \frac{I_\star'}{I_\star} \frac{R_{\rm out,0}}{R_{\rm out,0'}} \right)^2
		\simeq \left( \frac{t_\varphi'}{t_\varphi} \right)^{-1.22},
\end{equation}
where we used \Eq{eq:numerical:R} and $I_\star = \const$.  Compared to the baseline case with $S=1$, the scaled $S=0.5$ liner is subject to a 2.3x stronger magnetic pressure which leads to stronger initial shock compression and moves the outer liner surface further inwards.  Another unintended consequence of this effect is that the inductance history of the $S=0.5$ liner becomes slightly larger.  This additional inductance slightly reduces the current delivery to the target as shown by the small offset of the blue solid line. The other scaled implosions in \Fig{fig:radius_vs_time} follow similar trajectories in dimensionless space. These results provide evidence that the scaling prescriptions in \Sec{sec:prescriptions} preserve well the normalized implosion trajectories.

\begin{figure}
	\includegraphics[scale=.43]{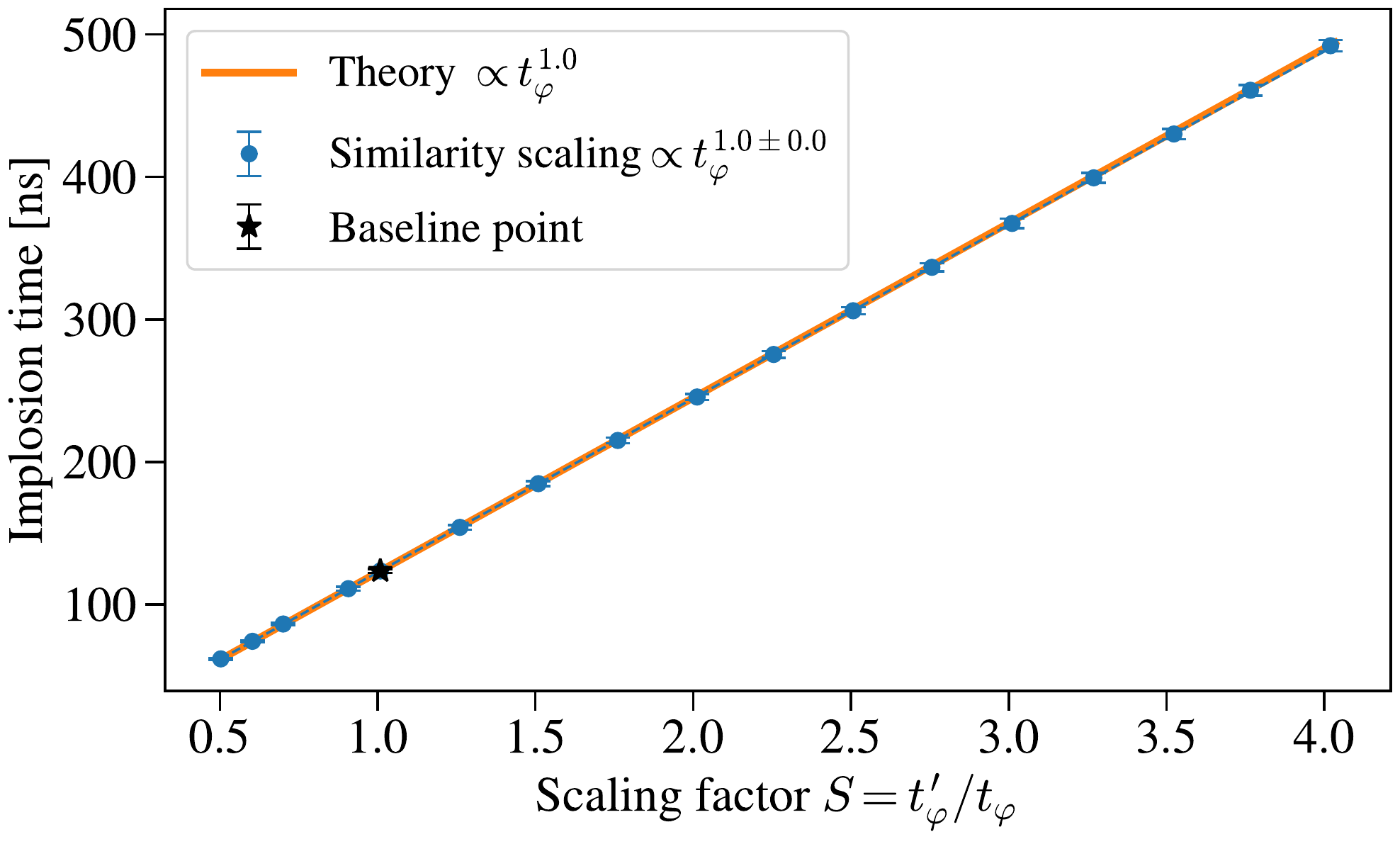}
	\caption{Implosion times\cite{foot:implosion} for the similarity-scaled MagLIF loads.  For reference purposes, the error bars denote the full-width, half-maximum burn time of the neutron-production event in the simulations.  As shown, the measured implosion times follow the expected linear behavior with respect to $S$.}
	\label{fig:implosion_time}
\end{figure}

\begin{figure}
	\includegraphics[scale=.43]{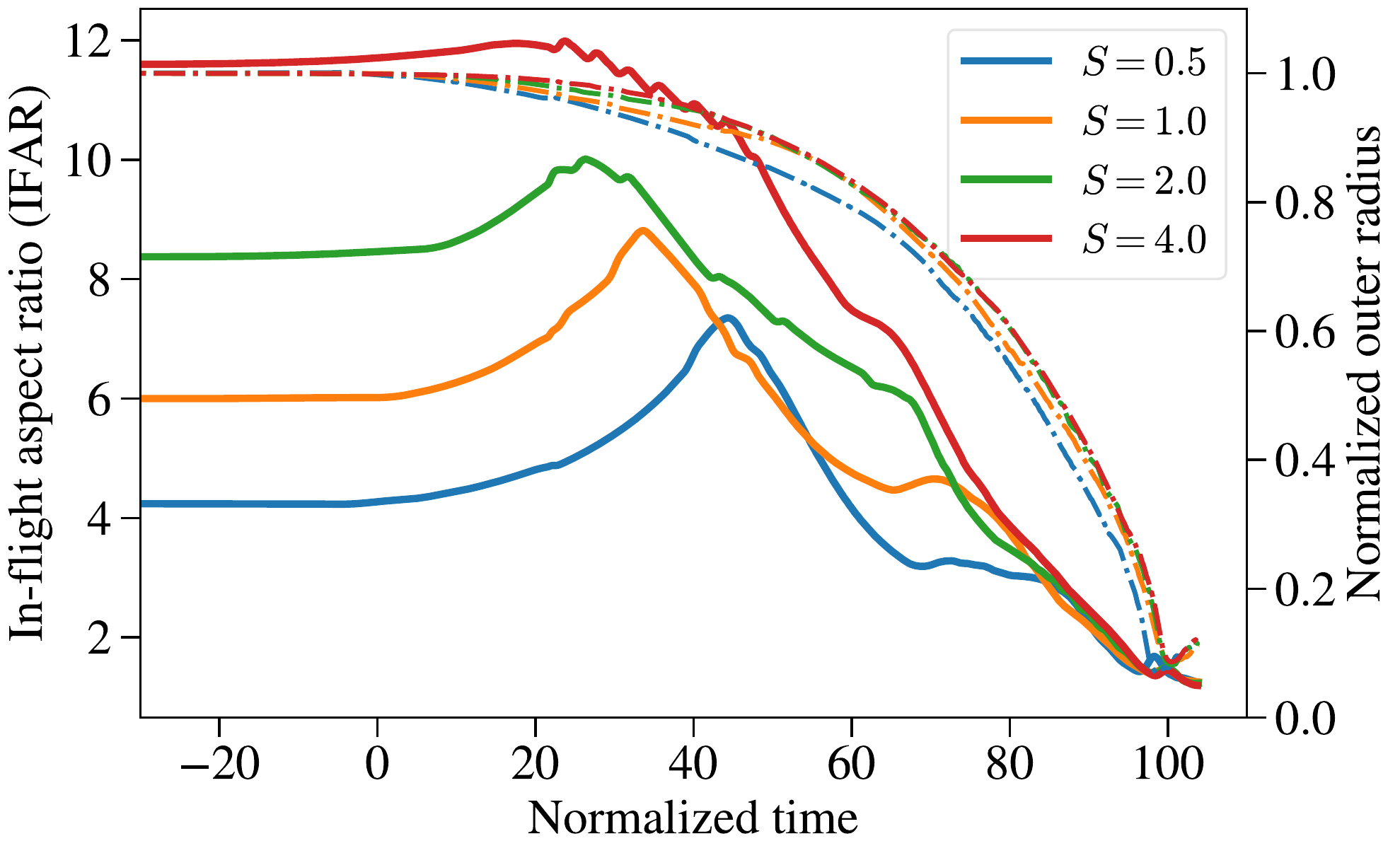}
	\caption{IFAR trajectories (solid lines) for the scaled loads in \Fig{fig:radius_vs_time}.  For reference purposes, the implosion trajectories of the normalized outer radii (dashed lines) are also shown.}
	\label{fig:IFAR_vs_time}
\end{figure}

To be more quantitative, \Fig{fig:implosion_time} shows the measured implosion times\cite{foot:implosion} obtained from \textsc{hydra} calculations for a family of similarity-scaled MagLIF loads.  As expected, the measured implosion times follow the linear scaling with respect to the temporal scaling factor $S$ to a very high degree.  This is indicative that the liners implode in a similar fashion as expected from the scaling theory.

From \Figs{fig:radius_vs_time}--\ref{fig:implosion_time}, we find that the characteristic implosion times, the bang times, and the current-rise times all scale linearly with the characteristic timescale $t_\varphi$ of the driving voltage.  This is a property of similarity scaling. Therefore, it is equivalent to say that, in this study, we are similarity scaling MagLIF loads with respect to the implosion time and the current-rise time.  We shall use these terms below.

The in-flight aspect ratio (IFAR) is a quantity often used to measure the robustness of an imploding ICF shell to the Rayleigh--Taylor instability.\cite{Bose:2017jf}  Larger IFAR values are usually correlated to less stable implosions.  Figure~\ref{fig:IFAR_vs_time} shows the IFAR trajectories calculated using \textsc{hydra} for the scaled MagLIF loads shown in \Fig{fig:radius_vs_time}.  Several points are worth discussing.  First, the IFAR trajectories for the scaled liners evolve in different manners.  Specifically, the IFAR curve for the $S=4.0$ scaled load (denoted by the solid red curve) does not increase substantially from its initial value because it is subject to a weaker magnetic pressure.  In contrast, the faster imploding liners are more strongly shock compressed, so the IFAR increases until shock break out occurs, and the liner begins imploding as a whole.  Second, when increasing the scaling factor $S$, the peak IFAR value for the scaled liners increases.  This means that the slower imploding liners (\eg, the $S=4$ load) are more susceptible to MRT instabilities during the early stages of the implosions.  However, MRT instabilities grow fastest during peak acceleration which occurs later during the implosion as the liner converges towards the axis.  Once the liners have achieved an outer convergence ratio of two ($R_{\rm out}/_{\rm out,0}=0.5$), the IFAR curves have almost coalesced into a single curve.

\begin{figure}
	\includegraphics[scale=.43]{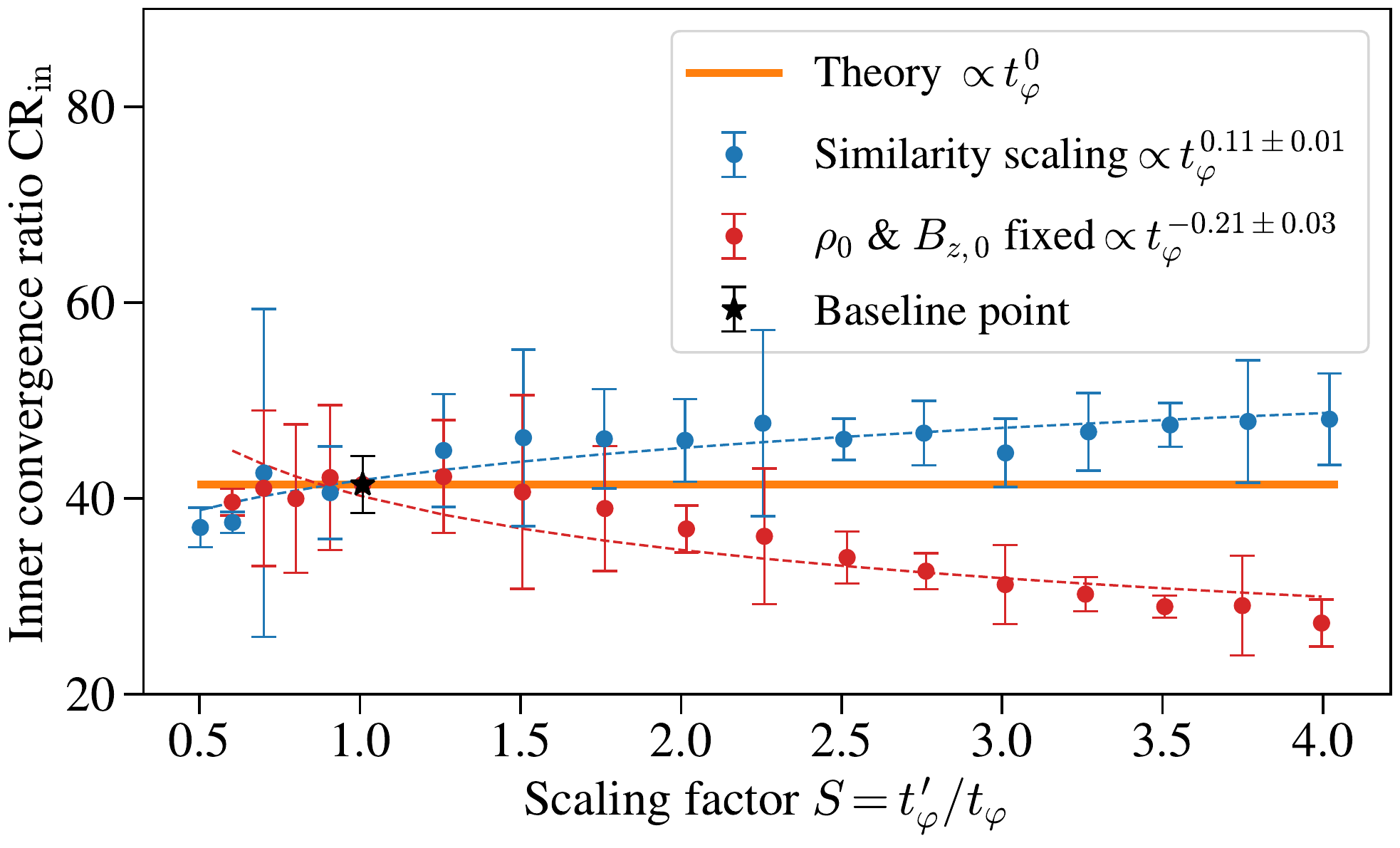}
	\caption{Inner convergence ratio evaluated at peak burn.  The blue points denote simulation results for the similarity-scaled MagLIF loads following the prescriptions given in \Sec{sec:prescriptions}.  The red points denote simulations for scaled MagLIF loads whose radii, length, and preheat are the same as the similarity-scaled liners, but the initial density and initially applied magnetic field are held constant for all loads.  Dashed lines are power-law fits to the simulation data.  The legend shows the fitted scaling exponents.  The error bars correspond to a standard deviation when collecting statistics of the radial position of the inner liner surface along its axial length.  The orange curve denotes the theoretical scaling law.
}
	\label{fig:CR}
\end{figure}

%%%%%%%%%%%%%%%%%%%%%%%%%%%%%%%%%%%%%%%%%%%%%%%%%
%%%%%%%%%%%%%%%%%%%%%%%%%%%%%%%%%%%%%%%%%%%%%%%%%
%%%%%%%%%%%%%%%%%%%%%%%%%%%%%%%%%%%%%%%%%%%%%%%%%
\section{Stagnation conditions}
\label{sec:stagnation}

We now discuss the scaling of the fuel thermodynamic conditions near stagnation.  From Paper~I, we recall that any ``no-$\alpha$" dimensionless dynamical quantity $\smash{\bar{Q}_{\rm no\,\alpha}(\bar{t})}$, \eg, the normalized fuel pressure, can be approximated as a function of the dimensionless parameters defining a MagLIF load:
\begin{equation}
	\bar{Q}_{\rm no\,\alpha} 
		\simeq \mathcal{F}^{(0)}_{\bar{\varphi}_{\rm oc},f} (\bar{t}; c_{1-6},\bar{t}_i, \Pi,\Phi,\Psi,\Upsilon_{\rm rad},\Upsilon_c,\Upsilon_{\rm end}),
	\label{eq:stagnation:general}
\end{equation}
where $\mathcal{F}^{(0)}_{\bar{\varphi}_{\rm oc},f}$ depends on the dimensionless time trace $\bar{\varphi}_{\rm oc}$ of the voltage drive and on the function $f$ parameterizing the behavior of the shunt resistor $R_{\rm loss}$.  $\bar{t}_i$ denotes the dimensionless time parameters, for example $\bar{t}_{\rm loss}$, $\Delta \bar{t}_{\rm loss}$ and $\bar{t}_{\rm preheat}$.  (For further details, see the discussion provided in Sec.~X~A of Paper~I.)  The scaling prescriptions in \Sec{sec:prescriptions} were designed to conserve the dimensionless parameters appearing on the right-hand side of \Eq{eq:stagnation:general} when scaling with respect to $t_\varphi$.  In consequence, the right-hand side remains invariant across timescales for similarity-scaled MagLIF configurations.  Therefore,
\begin{equation}
	\bar{Q}'(\bar{t}) \simeq \bar{Q}(\bar{t}),
	\label{eq:stagnation:qbar}
\end{equation}
where $\bar{Q}$ and $\bar{Q}'$ denote the dimensionless quantities corresponding to a \textit{baseline} and a \textit{scaled} MagLIF configurations, respectively.  (Since $\alpha$ heating is negligible for the MagLIF configurations discussed in this work, we have dropped the ``no$\,\alpha$" subscript.)  Note that \Eq{eq:stagnation:qbar} has already been demonstrated for the particular case of the dimensionless liner implosion trajectories shown in \Fig{fig:radius_norm_vs_time}.

\begin{figure}
	\includegraphics[scale=.43]{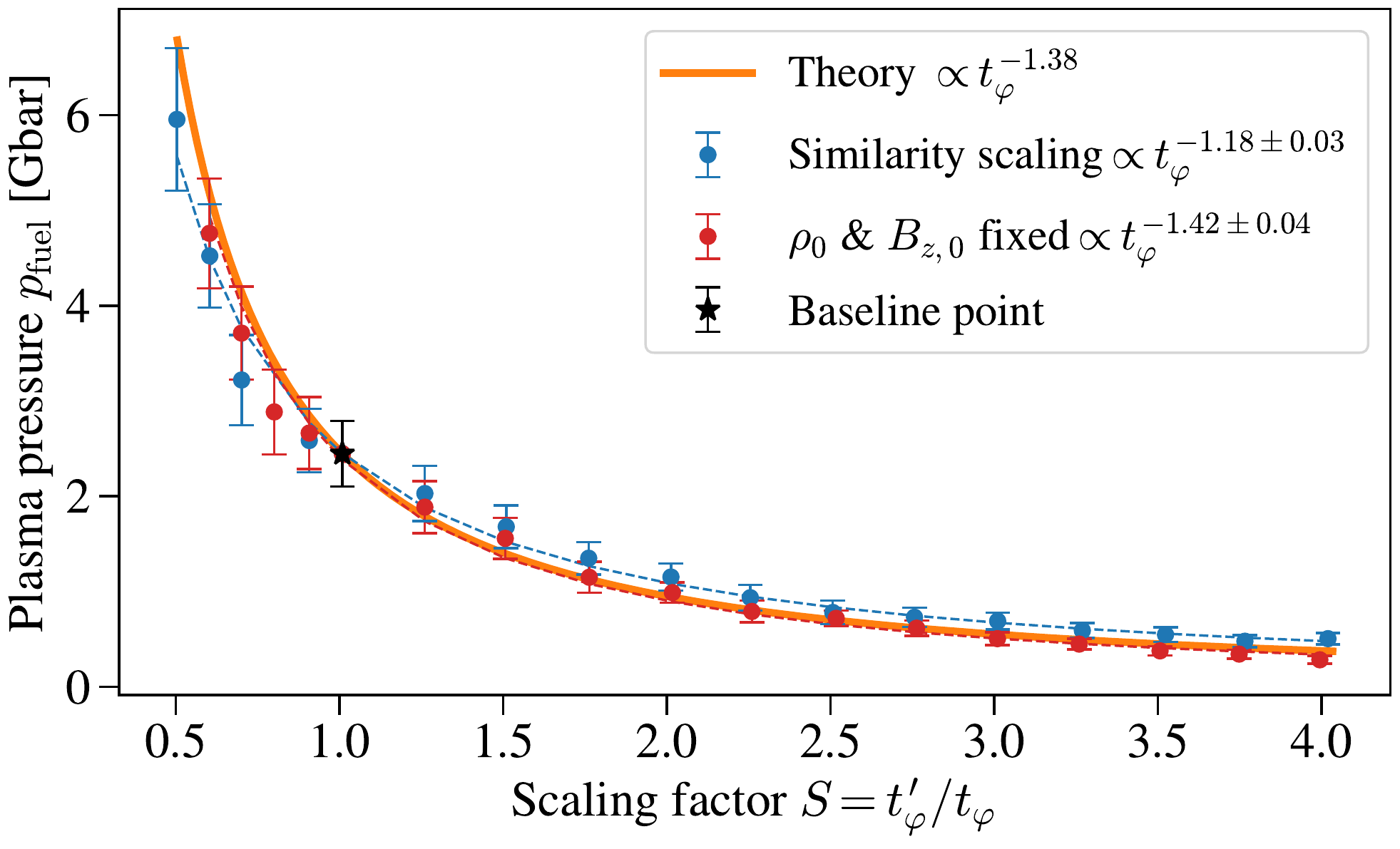}
	\caption{Burn-history averaged plasma pressure.  Blue points denote simulation results for the similarity-scaled MagLIF loads following the prescriptions given in \Sec{sec:prescriptions}.  Red points denote simulation results using the same scaling prescriptions, except for the initial fuel density and magnetic field, which are held constant.  Dashed lines are power-law fits to the simulation data.  The legend shows the fitted scaling exponents.  Error bars denote the burn-weighted standard deviation associated to temporal variations of the plasma pressure near peak burn.  The orange curve denotes the theoretical scaling law given in \Eq{eq:stagnation:pion}.}
	\label{fig:pressure}
\end{figure}

In Paper I, the dimensionless variables $\bar{Q}$ are constructed by normalizing the dimensional quantities $Q$ by functions of the known experimental input parameters defining a MagLIF load.  In other words, $\bar{Q}=Q/Q_0$ where $Q_0$ is a normalization coefficient that only depends on input parameters, \eg, $R_{\rm in,0}$, $\rho_0$, and $E_{\rm preheat}$.  As an example, the fuel pressure $p_{\rm fuel}$ is normalized by the preheat pressure $p_{\rm preheat}\doteq (2/3) E_{\rm preheat} /(\pi R_{\rm in,0}^2 h)$, which is the characteristic fuel pressure at the moment of preheat.  The scaling rules for the dimensional variables $Q$ can be written as follows:
\begin{equation}
	\frac{Q'}{Q} \simeq \frac{Q_0'}{Q_0}.
	\label{eq:stagnation:q}
\end{equation}
Once a baseline quantity $Q$ is known (calculated via simulations or measured in experiments), the corresponding scaled quantity $Q'$ is obtained by multiplying $Q$ by a known function $Q_0'/Q_0$ of the two sets of input parameters for different similarity-scaled MagLIF configurations.  We emphasize that the scaling relations \eq{eq:stagnation:qbar} and \eq{eq:stagnation:q} are only valid for similarity-scaled MagLIF configurations.  In the following, we shall make use of \Eqs{eq:stagnation:qbar} and \eq{eq:stagnation:q} to derive the scaling rules for the fuel thermodynamic conditions near stagnation.

In the absence of non-ideal energy-loss mechanisms, the main factor determining the thermodynamic conditions of the fuel plasma is the inner-convergence ratio $\mathrm{CR}_{\rm in}(t) \doteq R_{\rm in,0} / R_{\rm in}(t)$ of the liner.  Since the inner convergence ratio at peak burn $\mathrm{CR}_{\rm in,bang} \doteq \mathrm{CR}_{\rm in}(t=t_{\rm bang})$ is a dimensionless dynamical quantity, we expect that it should be conserved for the scaled loads; in other words,
\begin{equation}
	\mathrm{CR}_{\rm in,bang}' \simeq \mathrm{CR}_{\rm in,bang}.
\end{equation}
In \Fig{fig:CR}, we plot the $\mathrm{CR}_{\rm in,bang}$ values measured from simulations for the similarity-scaled MagLIF loads (shown in blue).  As observed, there is roughly a $10\%$ variation in $\mathrm{CR}_{\rm in,bang}$ when compared to the baseline configuration ($S=1$).   There is acceptable agreement with the theory even as the implosion timescales are varied by close to an order of magnitude.  We note, however, that the fitted scaling law $\mathrm{CR}_{\rm in,bang} \propto t_\varphi^{0.11}$ observed in the simulations will be important to take into account when understanding other discrepancies in the scaling of the fuel thermodynamic conditions discussed later.

For the sake of comparison, we also include a second data set in \Fig{fig:CR} (shown in red).  These simulation data points correspond to scaled MagLIF loads that follow all the scaling prescriptions given in \Sec{sec:prescriptions}, except those for the initial fuel density and magnetic field [\Eqs{eq:scaling:rho}~and~\eq{eq:scaling:Bz}].  We instead keep $\rho_0$ and $B_{z,0}$ constant for all scaled loads.  We shall use this second data set to evaluate the consequences on other metrics when not conserving relative radiation losses and thermal-conduction losses.  As discussed later on, maintaining the initial fuel density while increasing the timescales of the implosions leads to enhanced radiation losses.  These losses quench the plasma temperature earlier in the implosion (in normalized time units), and therefore the apparent $\mathrm{CR}_{\rm in,bang}$ decreases as shown in \Fig{fig:CR}.  

\begin{figure}
	\includegraphics[scale=.43]{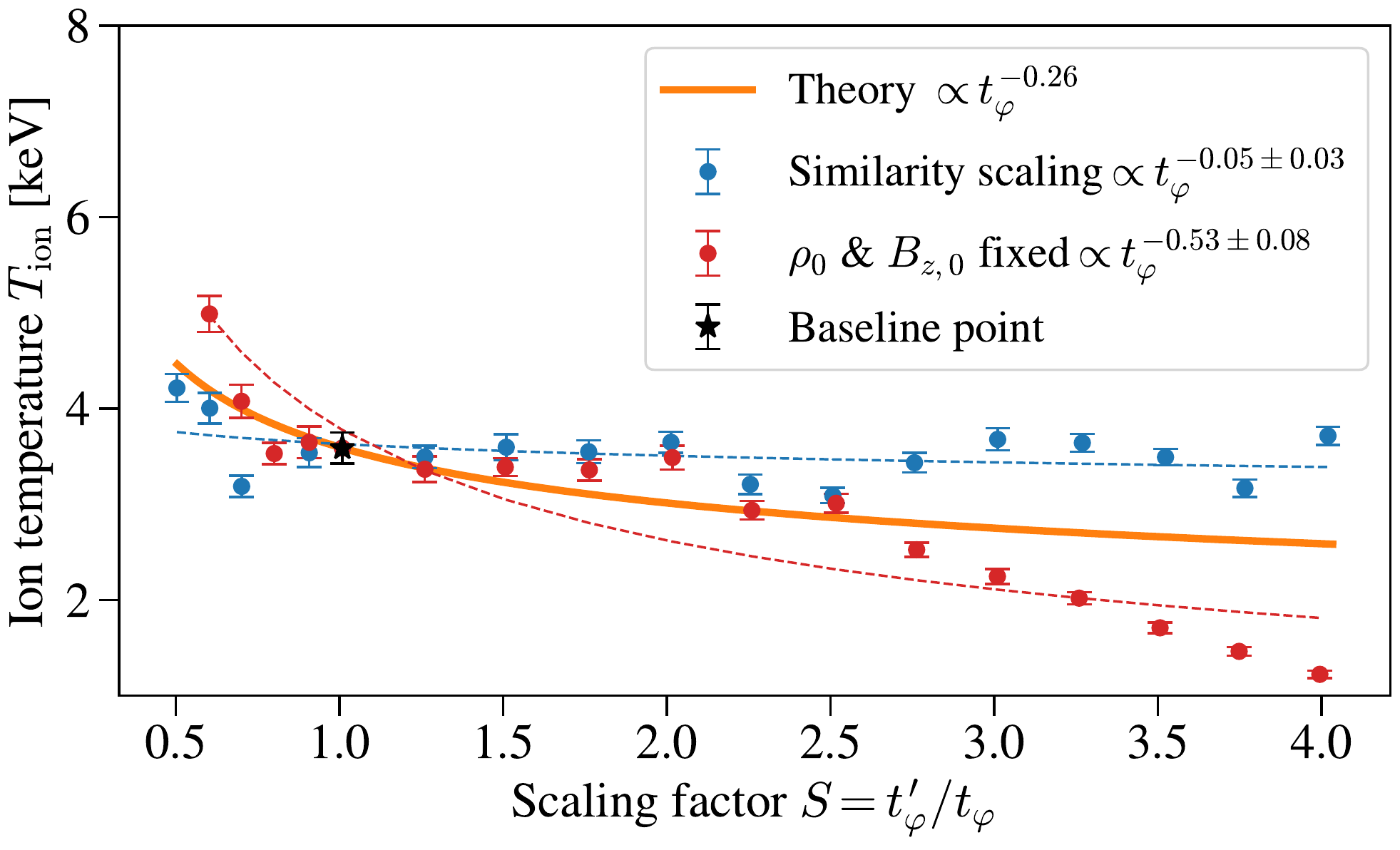}
	\caption{Burn-averaged ion temperature evaluated at peak burn time.  Dashed lines are power-law fits to the simulation data.  The legend shows the fitted scaling exponents.  The orange curve is the theoretical scaling curve given in \Eq{eq:stagnation:tion}.}
	\label{fig:tion}
\end{figure}

In \Fig{fig:pressure}, we plot the burn-history averaged plasma pressure for the scaled MagLIF loads.  As in Paper II, burn-history averaged quantities $\langle Q \rangle_{\rm b.h.}$ are calculated as follows:
\begin{equation}
	\langle Q \rangle_{\rm b.h.} 
		\doteq \frac{ \iint_{V_{\rm fuel}(t)} n_i^2 \langle \sigma v \rangle \, Q \,
				 \mathrm{d}V 
			\, \mathrm{d}t}
	  {\iint_{V_{\rm fuel}(t)} n_i^2 \langle \sigma v \rangle \, \mathrm{d}V 
		    \, \mathrm{d}t},
	\label{eq:stagnation:average}
\end{equation}
where $n_i^2 \langle \sigma v \rangle$ is the neutron yield-rate per-unit-volume and $V_{\rm fuel}$ is the volume of the fuel plasma.  As shown in \Fig{fig:pressure}, the burn-history averaged plasma pressure decreases when increasing the implosion timescale.  In Paper~I, the plasma pressure is normalized by the characteristic pressure $p_{\rm preheat}\doteq (2/3) E_{\rm preheat} /(\pi R_{\rm in,0}^2 h)$ achieved in the fuel by the preheat.  Upon substituting \Eqs{eq:numerical:R}~and~\eq{eq:numerical:Epreheat_h} into $p_{\rm preheat}$ and using \Eq{eq:stagnation:q}, we obtain the following scaling rule for the fuel pressure:
\begin{equation}
	\frac{p_{\rm fuel}'}{p_{\rm fuel}} 
			\simeq \frac{p_{\rm preheat}'}{p_{\rm preheat}} 
			= \left( \frac{R_{\rm in,0}}{R_{\rm in,0}'}\right)^2 
			\simeq \left(\frac{t_\varphi'}{t_\varphi}\right)^{-1.38} .
	\label{eq:stagnation:pion}
\end{equation}
As shown in \Fig{fig:pressure}, the scaling curve reproduces the decrease of the plasma pressure, and the deduced scaling exponents are also in close agreement.  From a physics perspective, this result can be understood as follows.  MagLIF loads imploded using shorter timescales achieve higher pressures because their assembled fuel columns at peak burn are more compact, and the energy density is therefore larger.  The inverse occurs when dilating the characteristic timescales of the implosions.

The weaker scaling law observed for the plasma pressure in \Fig{fig:pressure} can be partially explained by the  discrepancy in the scaling law of $\mathrm{CR}_{\rm in,bang}$ shown in \Fig{fig:CR}.  In the absence of non-ideal energy losses, the fuel pressure increases as $\smash{p_{\rm fuel} \propto \mathrm{CR}_{\rm in}^{10/3}}$ when adiabatically compressed.  If we modify the scaling relation \eq{eq:stagnation:pion} to
\begin{equation}
	\frac{p_{\rm fuel}'}{p_{\rm fuel}} 
			\simeq \frac{p_{\rm preheat}'}{p_{\rm preheat}}  \left( \frac{\mathrm{CR}_{\rm in,bang}'}{\mathrm{CR}_{\rm in,bang}}\right)^{10/3}
\end{equation}
and substitute the measured scaling law $\mathrm{CR}_{\rm in,bang} \propto t_\varphi^{0.11}$ obtained from the simulations, we find that $\smash{p_{\rm fuel} \propto t_\varphi^{-1.01}}$.  The scaling law becomes weaker, and the agreement between the scaling exponents is marginally improved.  It is likely that there are other effects at play that may bring closer agreement between the theory and simulations.

We now discuss the scaling of the burn-history averaged fuel temperature $\langle T \rangle_{\rm b.h.}$.  In analogy to the fuel pressure, the fuel temperature is normalized by the preheat temperature $k_B \smash{T_{\rm preheat} \doteq p_{\rm preheat}/(2\rho_0/m_i)}$ introduced in Paper~I.  Therefore, the scaling law for the fuel temperature is given by
\begin{equation}
	\frac{T_{\rm fuel}'}{T_{\rm fuel}} 
			\simeq \frac{T_{\rm preheat}'}{T_{\rm preheat}} 
			= 	\frac{p_{\rm preheat}'}{p_{\rm preheat}} 
				\frac{\rho_0}{\rho_0'} 
			\simeq  \left(\frac{t_\varphi'}{t_\varphi}\right)^{-0.26},
	\label{eq:stagnation:tion}
\end{equation}
where we used the scaling rules in \Eqs{eq:numerical:rho} and \eq{eq:stagnation:pion}.  As for the plasma pressure, ion temperatures are  expected to decrease when increasing the implosion timescales, yet the power-law dependence is weaker.  In \Fig{fig:tion}, we compare \Eq{eq:stagnation:tion} to the simulation results.  In agreement with the theory, simulations show a slight decreasing trend in $\langle T \rangle_{\rm b.h.}$ for the similarity-scaled loads.  However, the measured scaling law from the simulations is weaker than that in \Eq{eq:stagnation:tion}.  In a similar manner to the discussion above, when assuming adiabatic compression of the fuel $\smash{(T_{\rm fuel} \propto \mathrm{CR}_{\rm in}^{4/3})}$ and modifying \Eq{eq:stagnation:tion} to include the weak $\smash{\mathrm{CR}_{\rm in,bang} \propto t_\varphi^{0.11}}$ scaling law shown in \Fig{fig:CR}, we obtain $\smash{T_{\rm fuel} \propto t_\varphi^{-0.11}}$, which agrees better with the measured scaling law in \Fig{fig:tion}.  In this particular example, including the weak scaling rule for $\smash{\mathrm{CR}_{\rm in,bang}}$ has helped understand the discrepancy in the scaling law of the fuel temperature.

It is worth noting that, for the second data set where the initial density and axial magnetic field are held constant, the temperature drops as the temporal scaling factor is increased beyond beyond $S\simeq2.5$.  This sharp decrease is not apparent in \Fig{fig:pressure} for the plasma pressure.

\begin{figure}
	\includegraphics[scale=.43]{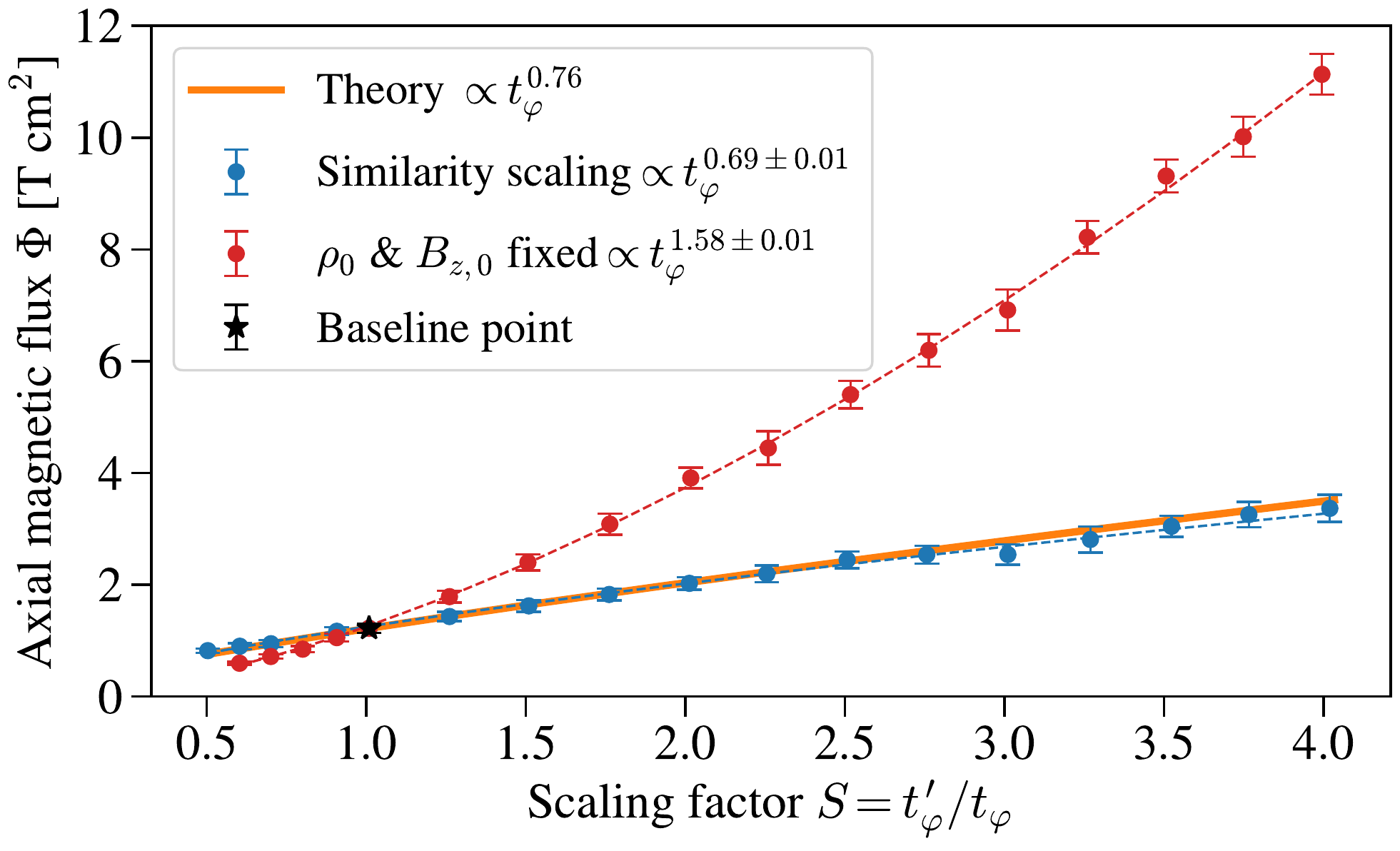}
	\caption{Axial magnetic flux averaged over the neutron yield-rate history.  Error bars denote the burn-weighted standard deviation associated to temporal variations near peak burn.}
	\label{fig:Bflux}
\end{figure}

In \Figs{fig:pressure} and \ref{fig:tion}, the small deviation in the scaling of the inner convergence ratio was shown to affect the measured scaling curves for $p_{\rm fuel}$ and $T_{\rm fuel}$.  To test the scaling theory without the effects of $\smash{\mathrm{CR}_{\rm in,bang}}$, we plot in \Fig{fig:Bflux} the measured scaling laws for the magnetic flux
\begin{equation}
	\Phi_{\rm B}(t) \doteq \frac{1}{h} \int_{V_{\rm fuel}} B_z \, \mathrm{d}V,
\end{equation}
where the volume integral is taken across the volume of the imploding fuel and $h$ is the imploding height of the liner.  In the absence of non-ideal magnetic-field transport effects (\eg, Nernst advection), the magnetic flux is independent of $\mathrm{CR}_{\rm in}$.  The corresponding scaling law is
\begin{equation}
	\frac{\Phi_{\rm B}'}{\Phi_{\rm B}}
		\simeq	\frac{B_{z,0}'}{B_{z,0}} \left( \frac{R_{\rm in,0}'}{R_{\rm in,0}}\right)^2
		\simeq \left(\frac{t_\varphi'}{t_\varphi} \right)^{0.76},
	\label{eq:stagnation:Bzflux}
\end{equation}
which shows good agreement with the simulation data points in \Fig{fig:Bflux} for the similarity-scaled MagLIF loads.

For the second dataset where the initial magnetic field is held constant, the theoretical scaling law is $\Phi_{\rm B} \propto R_{\rm in,0}^2 \propto t_\varphi^{1.38}$.  This scaling law underpredicts the measured scaling law from the simulations.  Two reasons why this might be the case are the following. First, the MagLIF loads considered in the second dataset are not entirely similarity scaled, so there is no reason why the derived scaling laws should hold.  Second, for this particular dataset, non-ideal  magnetic-transport effects, such as Nernst advection, might be changing more drastically for the longer implosions and affect more strongly the compression of the axial magnetic field.

\begin{figure}
	\includegraphics[scale=.43]{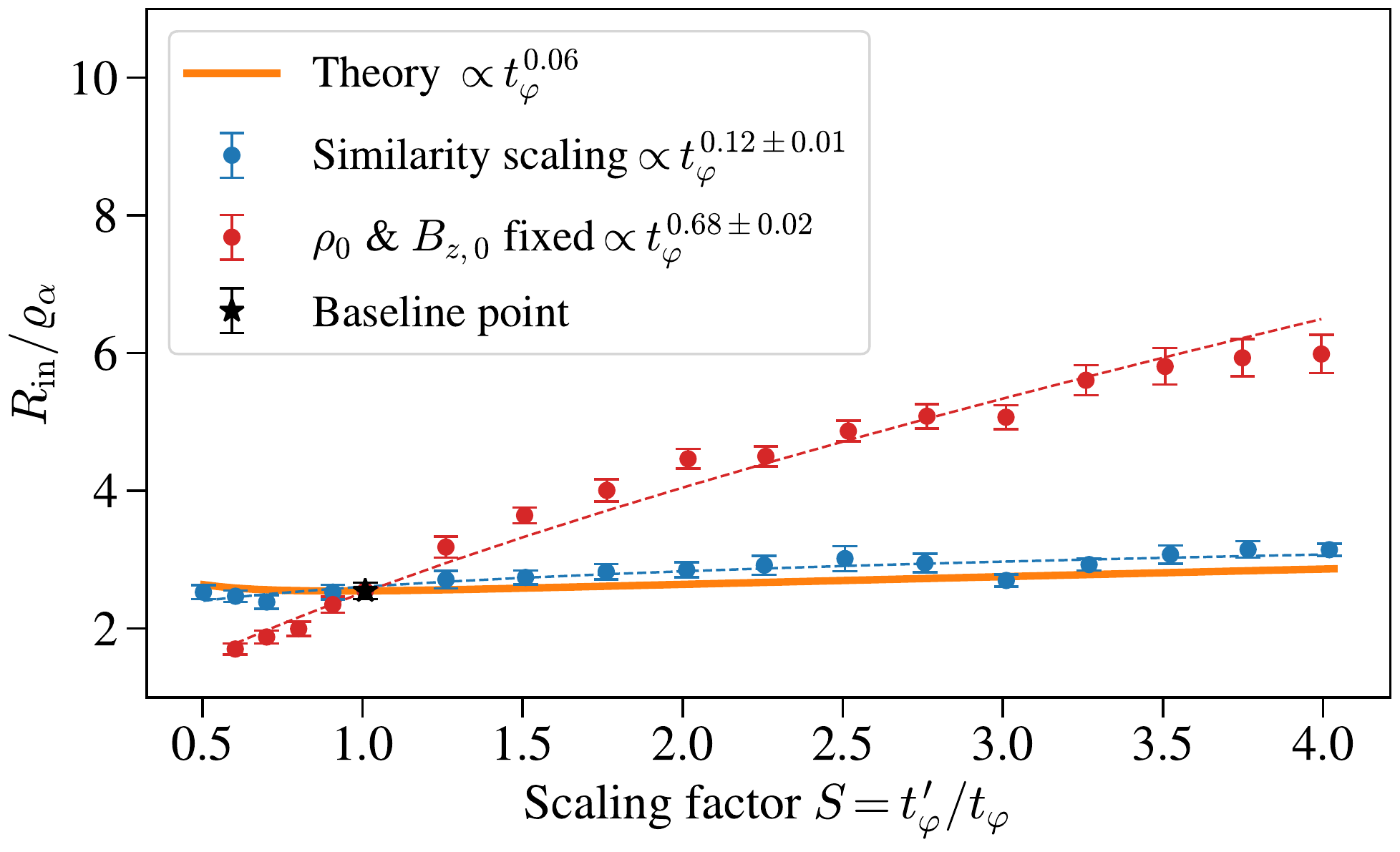}
	\caption{Magnetic-field radius product $\langle B_zR_{\rm in} \rangle$ averaged over the neutron yield-rate history.  Error bars denote the burn-weighted standard deviation associated to temporal variations near peak burn.}
	\label{fig:BzR}
\end{figure}

Another stagnation quantity of interest is the ratio of the fuel column radius $R_{\rm in}$ and the gyroradius $\varrho_\alpha$ of 3.5-MeV $\alpha$ particles.  $R_{\rm in}/\varrho_\alpha$ is proportional to the magnetic-field radius product:
\begin{equation}
	\frac{R_{\rm in}}{\varrho_\alpha}
		\doteq \frac{[B_z(\mathrm{T})] \cdot [R_{\rm in} (\mathrm{cm})]}{26.5}.
\end{equation}  
When $R_{\rm in}/\varrho_\alpha \gg 1$, $\alpha$ particles created by DT fusion reactions are well confined within the fuel column.  The quantity $\smash{\langle B_zR_{\rm in} \rangle}$ is commonly measured in present-day MagLIF implosions.\cite{Schmit:2014fg,Knapp:2015kc,Lewis:2021kz}  When scaling with respect to the implosion time, the scaling rule for $R_{\rm in}/\varrho_\alpha$ is
\begin{equation}
	\frac{\langle R_{\rm in}/\varrho_\alpha \rangle'_{\rm no\,\alpha}}
			{\langle R_{\rm in}/\varrho_\alpha \rangle_{\rm no\,\alpha}}
	=	\frac{\langle B_zR_{\rm in} \rangle'_{\rm no\,\alpha}}
			{\langle B_zR_{\rm in} \rangle_{\rm no\,\alpha}}
		\simeq	\frac{B_{z,0}'}{B_{z,0}}\frac{R_{\rm in,0}'}{R_{\rm in,0}}
		\simeq	\left(\frac{t_\varphi'}{t_\varphi} \right)^{0.07}.
	\label{eq:stagnation:BzR}
\end{equation}
Figure~\ref{fig:BzR} compares the scaling law in \Eq{eq:stagnation:BzR} to the simulation outputs for $R_{\rm in}/\varrho_\alpha$. In simulations, $\langle B_zR_{\rm in} \rangle$ is calculated using
\begin{equation}
	\langle B_zR_{\rm in} \rangle \doteq \frac{\Phi_{\rm B}(t)}{\pi \langle R_{\rm in} \rangle},
\end{equation}
where $\langle R_{\rm in} \rangle$ is the average inner radius along the axial length of the liner.  As shown, both the simulation and theory results show an almost constant scaling curve for $R_{\rm in}/\varrho_\alpha$, meaning that magnetic confinement of $\alpha$ particles does not decrease when going to longer implosion times using the scaling strategy detailed in \Sec{sec:prescriptions}.  Intuitively, this is can be understood by noting that the initial magnetic field $B_{z,0}$ is reduced for longer implosion times, but the initial liner inner radius $R_{\rm in,0}$ increases offsetting the effect.  Of course, the simulations results for the second dataset show an increase in $R_{\rm in}/\varrho_\alpha$ since the initial magnetic field is held fixed across scales while the initial liner radius is increasing for larger $S$.

As our final comparison for this section, we discuss the scaling law for the fuel internal energy $U$.  By similarity, the fuel internal energy scales as the total energy delivered to the fuel during the preheat stage; \ie
\begin{equation}
	\frac{U'}{U}
		\simeq \frac{E_{\rm preheat}'}{E_{\rm preheat}}
		\simeq \left(\frac{t_\varphi'}{t_\varphi} \right)^{0.88}.
	\label{eq:stagnation:U}
\end{equation}
Figure \ref{fig:U} compares \Eq{eq:stagnation:U} with the simulation results.  As shown, there is good agreement between theory and simulations of the similarity-scaled MagLIF loads.  Interestingly, the second simulation data set, where the initial density and magnetic field are held constant, shows a surplus of internal energy within the fuel for longer implosion timescales.  As we shall discuss in \Sec{sec:loss}, this surplus of energy is due to reduced fuel end losses.

%%%%%%%%%%%%%%%%%%%%%%%%%%%%%%%%%%%%%%%%%%%%%%%%%
%%%%%%%%%%%%%%%%%%%%%%%%%%%%%%%%%%%%%%%%%%%%%%%%%
%%%%%%%%%%%%%%%%%%%%%%%%%%%%%%%%%%%%%%%%%%%%%%%%%
\section{Burn-width and conservation of relative losses}
\label{sec:loss}

\begin{figure}
	\includegraphics[scale=.43]{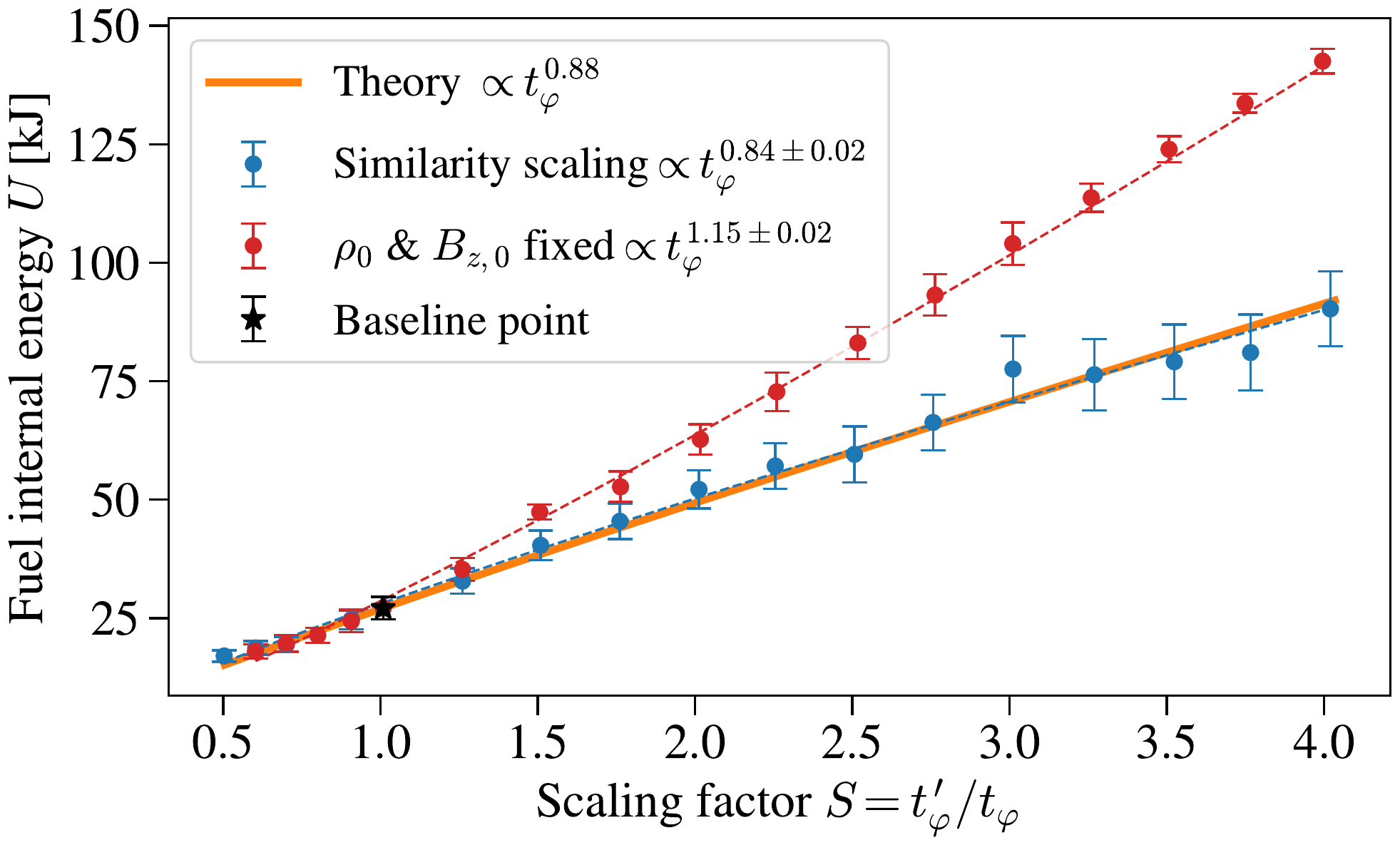}
	\caption{Fuel internal energy averaged over the neutron yield-rate history.  Error bars denote the burn-weighted standard deviation associated to temporal variations near peak burn.}
	\label{fig:U}
\end{figure}

\begin{figure}
	\includegraphics[scale=.43]{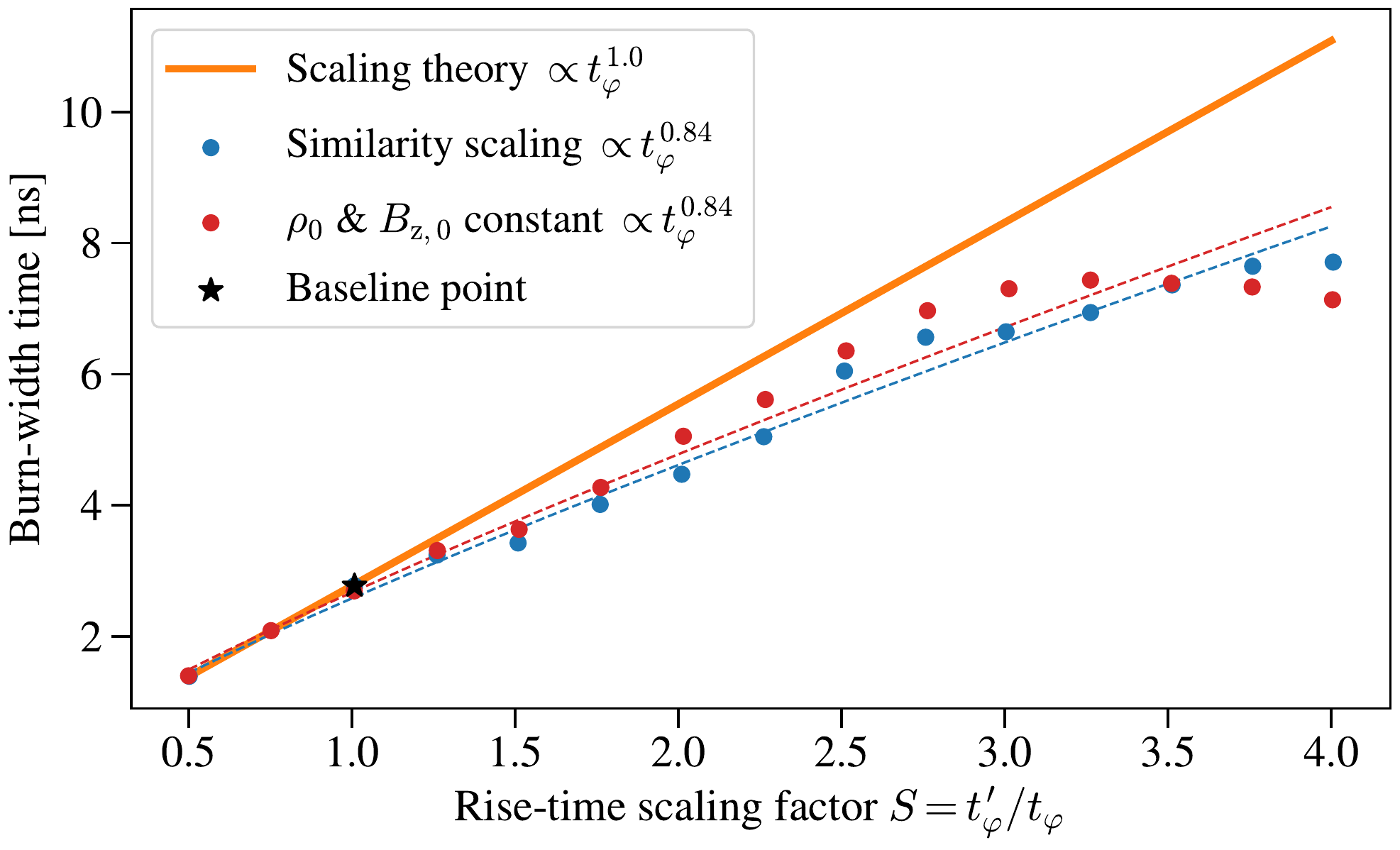}
	\caption{Burn width $\tau_{\rm bw}$ extracted from simulations.  The burn width is calculated by measuring the full-width, half-maximum of the burn-rate curves.  Up to $S=3$, $\tau_{\rm bw}$ shows the expected linear behavior when scaling the voltage rise time but begins to diverge for larger scaling values.}
	\label{fig:tau}
\end{figure}

Based on similarity-scaling arguments, we expect that all timescales involved in a MagLIF implosion scale linearly with the characteristic timescale $t_\varphi$.  In \Fig{fig:tau}, we further test this hypothesis by comparing the measured burn-width time $\tau_{\rm bw}$ extracted from simulations to the linear scaling curve.  For both data sets, $\tau_{\rm bw}$ shows the expected linear dependence up to $S=2.5$.  Beyond this value, the simulation results diverge from the analytical scaling rule.  For $S=4$, the discrepancy between the theory and the simulations is roughly 30\% as the timescales are varied by an order of magnitude.

\begin{figure*}
	\includegraphics[scale=.43]{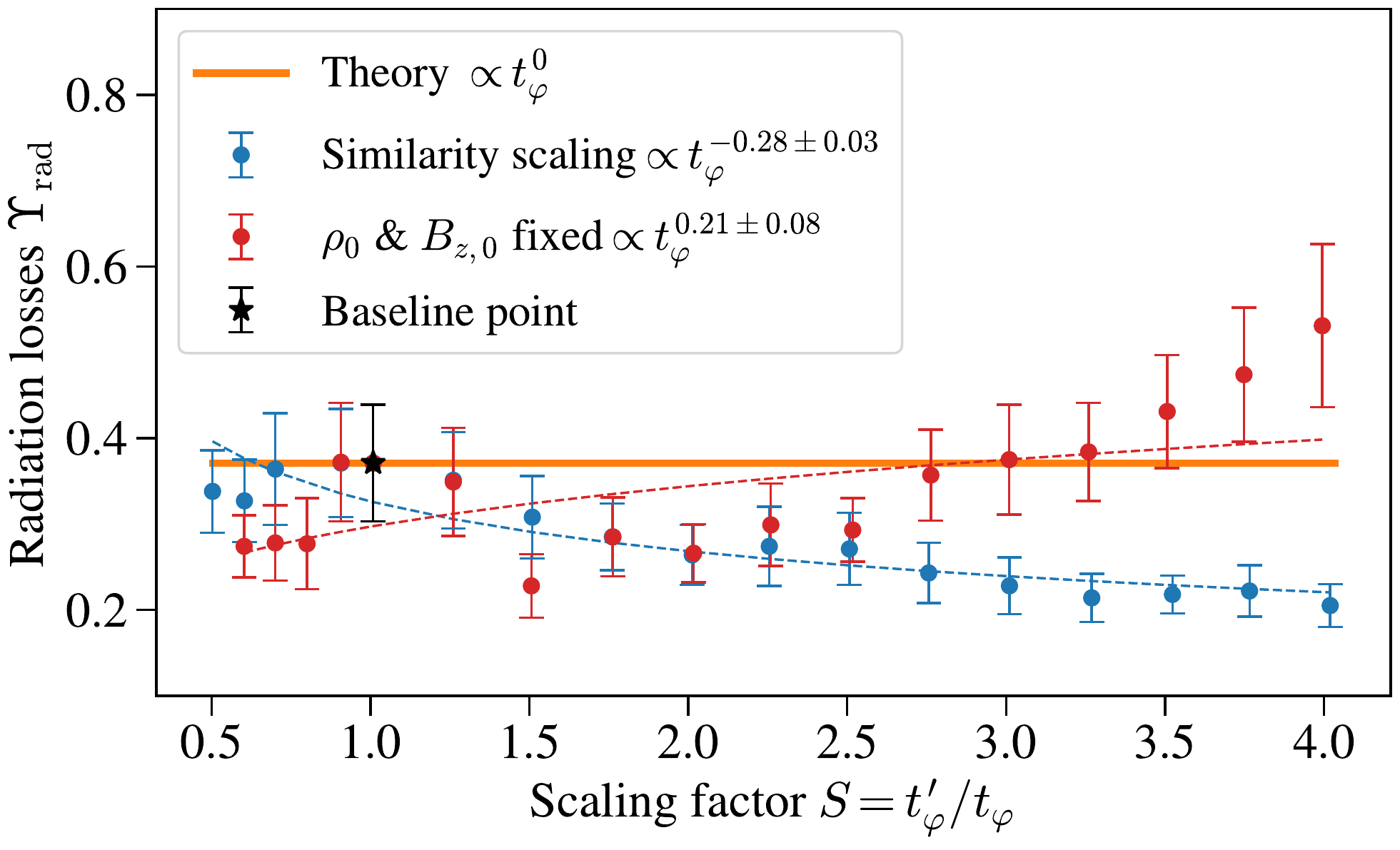}
	\hspace{0.5cm}
	\includegraphics[scale=.43]{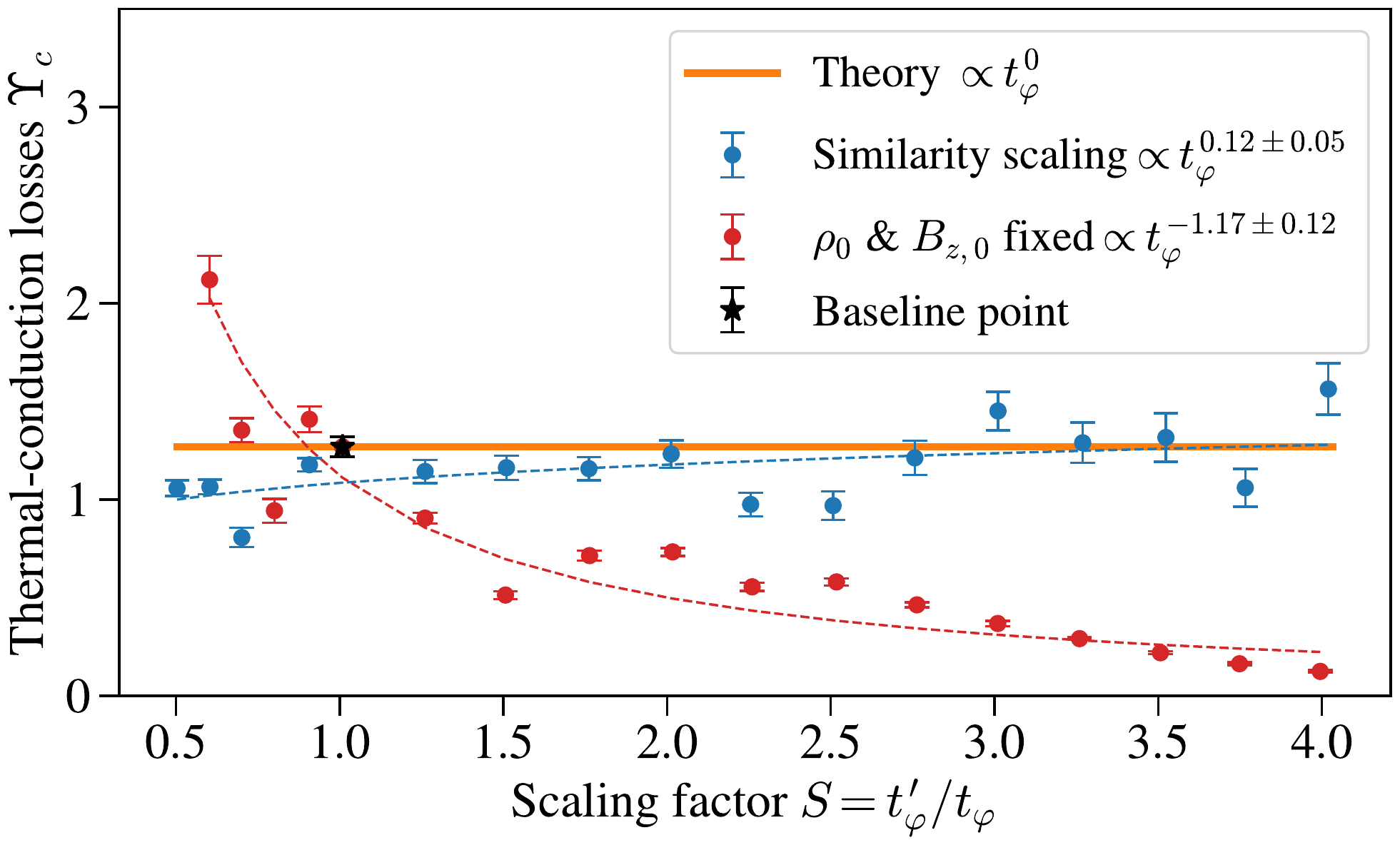}
	\caption{Left: Relative radiation energy losses as characterized by the parameter $\Upsilon_{\rm rad}$ in \Eq{eq:loss:Upsilon_rad}. Right: Relative thermal-conduction energy losses as characterized by the parameter $\Upsilon_c$ in \Eq{eq:loss:Upsilon_c}.  These quantities are evaluated by substituting burn-weighted plasma parameters using \Eq{eq:stagnation:average} representing the plasma conditions near the hot plasma column.  Error bars denote the burn-weighted standard deviation associated with temporal variations near peak burn.}
	\label{fig:Loss}
\end{figure*}

The scaling prescriptions for the initial fuel density and the axial magnetic field intend to conserve relative radiation losses and thermal-conduction losses.  The relative effects of these processes can be measured by the dimensionless parameters $\Upsilon_{\rm rad}$ and $\Upsilon_c$ given in Sec.~V of Paper~I:\cite{foot:losses}
\begin{align}
	\Upsilon_{\rm rad}	
			& \doteq	0.67~
						\frac{\left[\rho ({\rm g/cm^3})\right] 
									\cdot [\tau_{\rm bw}({\rm ns} )] }
						{A \cdot [T (\mathrm{keV})]^{1/2}  },
		\label{eq:loss:Upsilon_rad} \\
 	 \Upsilon_c	
			& \doteq	0.04 ~
						\frac{ [T (\mathrm{keV})] \cdot  [t_{\rm bw}({\rm ns} )]}
						{ \left[ B_z ({\rm T})\right] 	\cdot [R_{\rm in}({\rm cm})]^2 },
		\label{eq:loss:Upsilon_c}
\end{align}
where $A=2$ for DD.  In \Fig{fig:Loss} (left), the parameter $\Upsilon_{\rm rad}$ is evaluated using the burn-history averaged stagnation quantities discussed in \Sec{sec:stagnation}.  For the similarity-scaled MagLIF loads, relative radiation losses decrease when increasing the implosion timescales, and the measured scaling law from the simulations is $\Upsilon_{\rm rad}\propto t_\varphi^{-0.28}$.  This measured scaling law can be explained from the previously discussed scaling trends for the fuel pressure, ion temperature, and burn-width time.  When substituting the fuel density $\rho$ in favor of the fuel pressure $p_{\rm fuel}$, we obtain $\Upsilon_{\rm rad}\propto p_{\rm fuel} \tau_{\rm bw}/T^{3/2}$.  Based on the simulation results in \Figs{fig:pressure}, \ref{fig:tion}, and \ref{fig:tau}, we have $p_{\rm fuel} \propto t_\varphi^{-1.18}$, $T\propto t_\varphi^{0.05}$, and $\tau_{\rm bw}\propto t_\varphi^{0.84}$.  When combined, these scaling laws lead to $\Upsilon_{\rm rad} \propto t_\varphi^{-0.27}$, which agrees with the fitted scaling law in \Fig{fig:Loss}.  It is worth noting that, for the second data set, $\Upsilon_{\rm rad}$ increases considerably for $S\geq 3$, which is expected due to the relatively large initial fuel densities for these configurations.  The increase in $\Upsilon_{\rm rad}$ correlates with the decrease in $\langle T_{\rm ion} \rangle_{\rm b.h.}$ in \Fig{fig:tion}.

\begin{figure}
	\includegraphics[scale=.43]{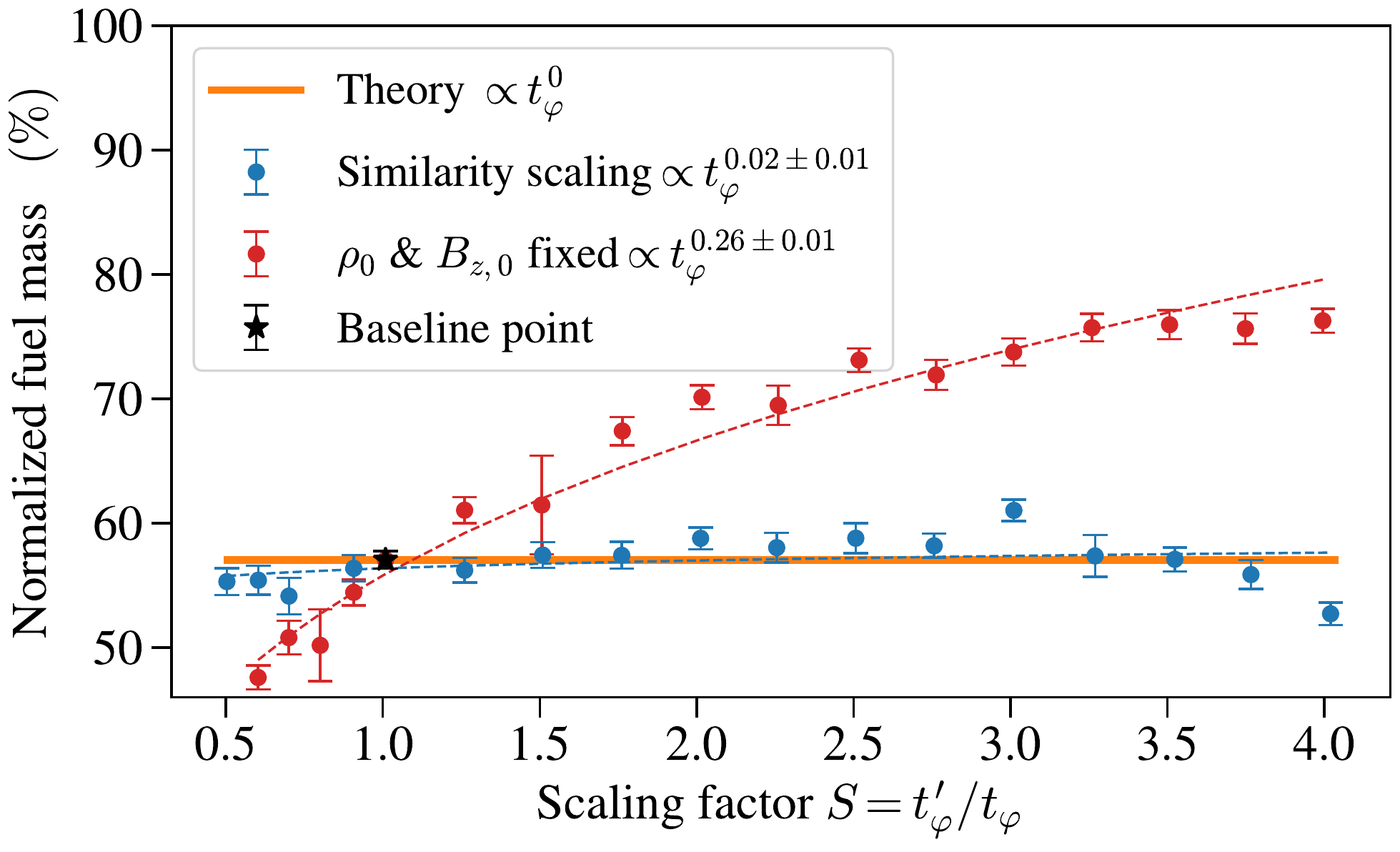}
	\caption{Burn-history averaged normalized fuel mass $\bar{m}_{\rm fuel}(t) \doteq m_{\rm fuel}(t)/m_{\rm fuel}(0)$.  Error bars denote the burn-weighted standard deviation associated to temporal variations near peak burn.}
	\label{fig:NormMass}
\end{figure}

Regarding thermal-conduction losses, \Fig{fig:Loss} (right) shows that the dimensionless parameter $\Upsilon_c$ is overall well conserved when increasing the implosion timescales.  According to the two metrics $\Upsilon_{\rm rad}$ and $\Upsilon_c$, thermal-conduction losses are larger than radiation losses.  The second data set, where the initial density and magnetic field are held constant, shows a sharp decrease in $\Upsilon_c$.  This is mainly due to the sharp decrease in the ion temperature near peak burn, as shown in \Fig{fig:tion}.

\begin{figure}
	\includegraphics[scale=0.43]{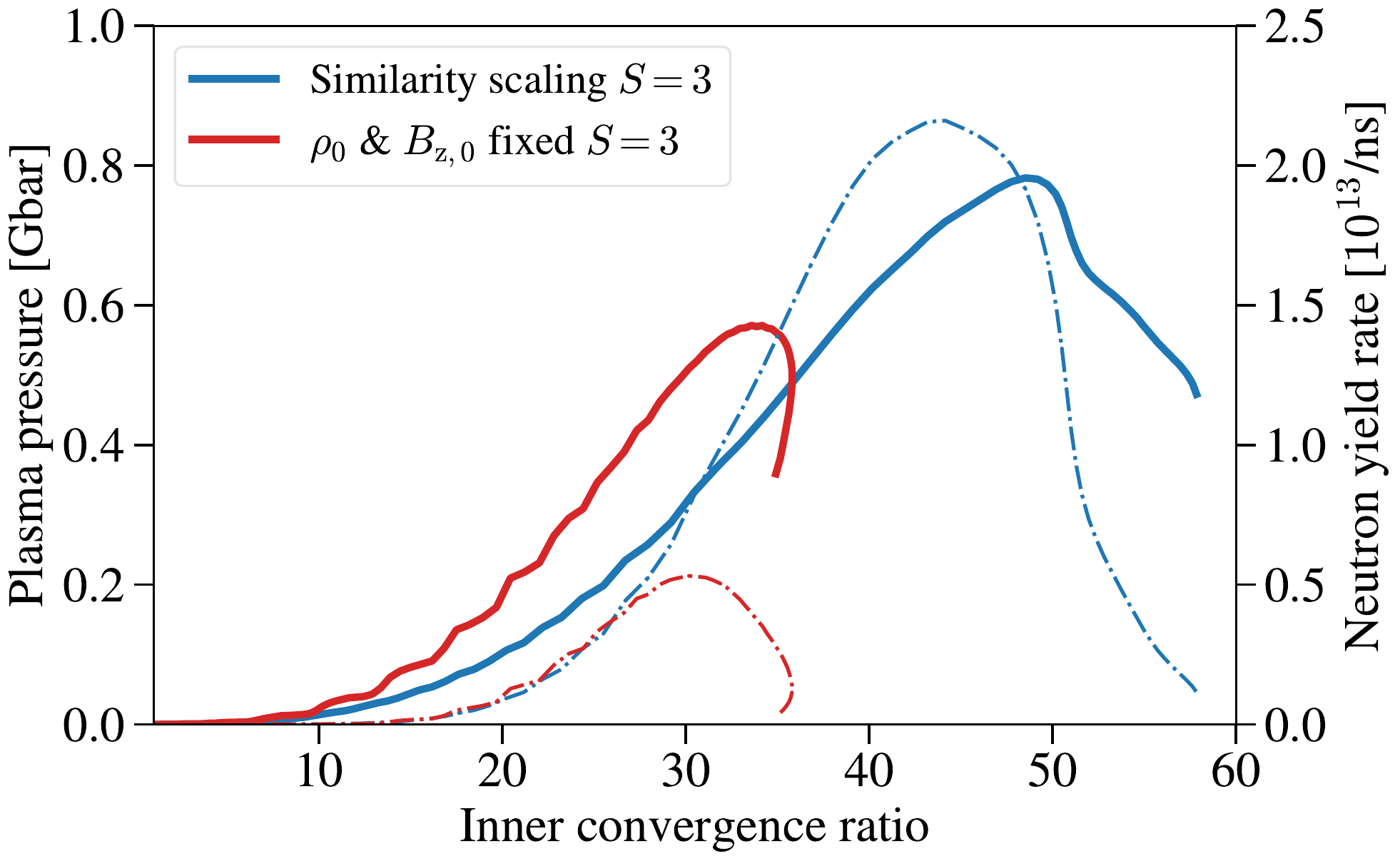}
	\caption{Burn-averaged plasma pressure $p(t)$ (solid lines) and neutron yield rate $\dot{Y}(t)$ (dashed lines) plotted versus inner convergence ratio $\mathrm{CR}_{\rm in}(t)$ for the two $S=3$ scaled loads discussed in this paper.  The second configuration compresses the fuel in a more adiabatic manner due to reduced end losses, yet enhanced radiation losses truncate the neutron pulse earlier in the implosion.  The similarity-scaled configuration has more end losses and compresses the fuel less efficiently, but enhanced fusion yield is achieved with this configuration.}
	\label{fig:pressure_time}
\end{figure}

\begin{figure*}
	\includegraphics[scale=.43]{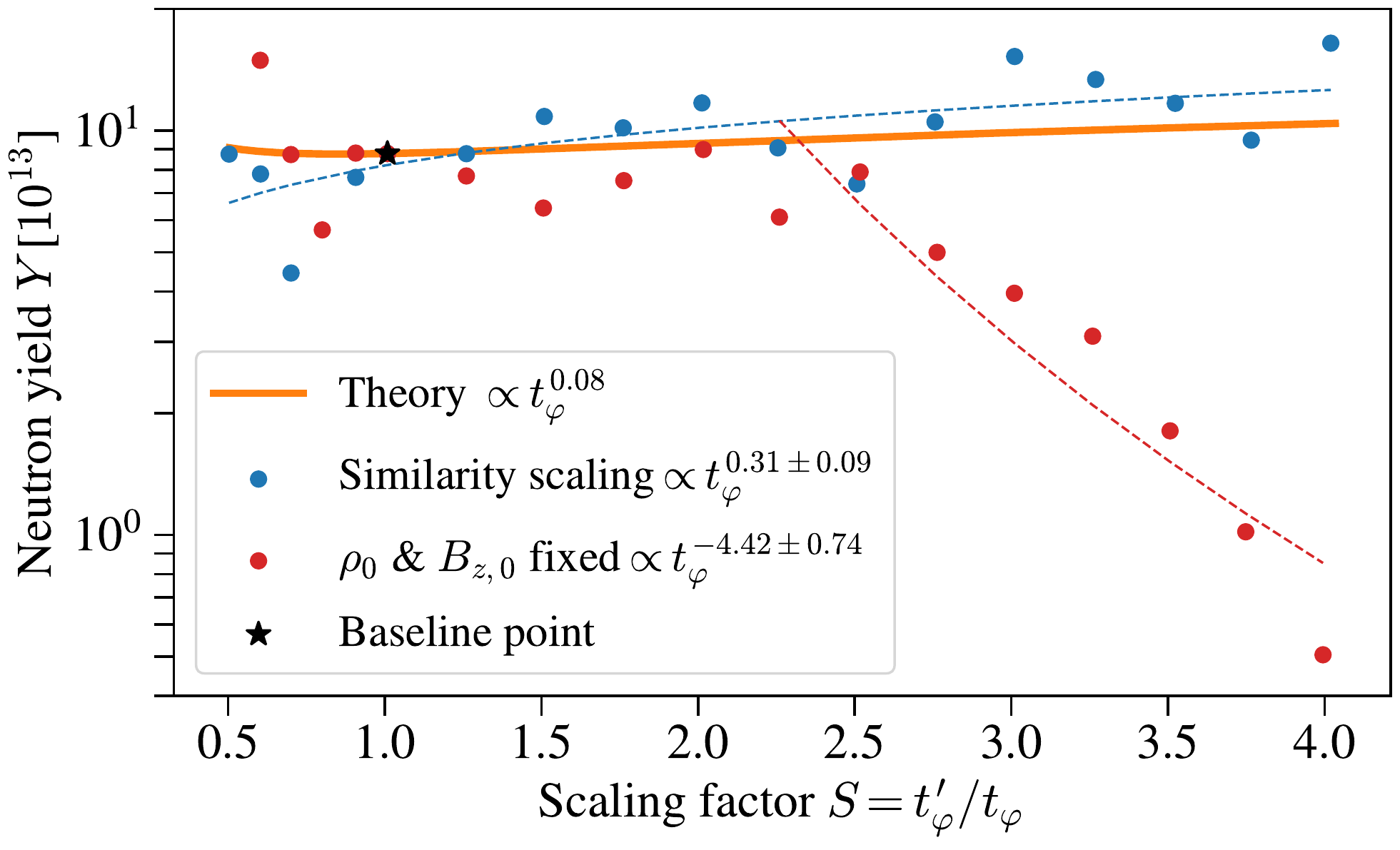}
	\hspace{0.5cm}
	\includegraphics[scale=.43]{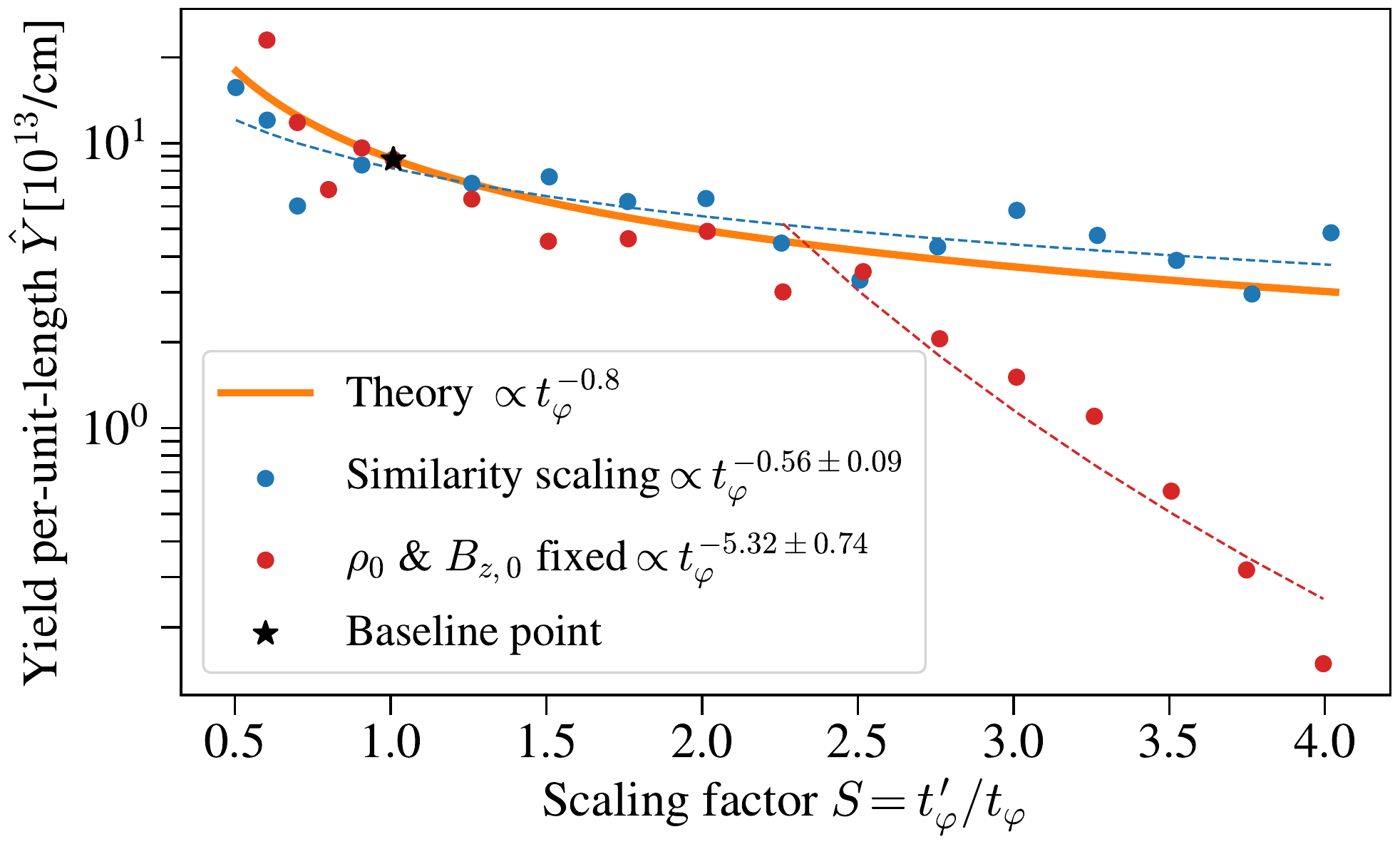}
	\caption{Neutron yield (left) and neutron yield per-unit-length (right) obtained from \textsc{hydra} simulations.  The blue points correspond to the similarity-scaled MagLIF configurations.  The red points denote the scaled MagLIF configurations that followed all the scaling prescriptions except for the initial fuel density and axial magnetic field, which were instead held constant across scales.  Dashed lines are power-law fits to the simulation data.  The legend shows the fitted scaling exponents.  The orange curves are the theoretical scaling laws.}
	\label{fig:yield}
\end{figure*}

As discussed in \Sec{sec:prescriptions}, the liner height scales almost linearly with the implosion timescale in order to conserve relative end-flow energy losses and fuel-mass losses.  To verify this scaling rule, in simulations we tallied the total fuel mass inventory $m_{\rm fuel}(t)$ located within the imploding regions of the studied MagLIF configurations.  Based on similarity-scaling arguments, it is expected that the normalized fuel mass $\bar{m}_{\rm fuel}(t) \doteq m_{\rm fuel}(t)/m_{\rm fuel}(0)$ will remain invariant.  Figure~\ref{fig:NormMass} shows the normalized fuel mass evaluated near peak burn.  For the similarity-scaled loads (shown in blue), the normalized fuel inventory is approximately 57\% and remains constant across timescales.  Surprisingly, the second simulation data set shows that the normalized fuel inventory grows when increasing the implosion timescales.  To make sense of this trend, when modeling the fuel loss from the liner region as a rarefraction wave (see Sec.~VI of Paper~I), we find that the rate of fuel losses depends on the square root of the characteristic fuel temperature.  When keeping the initial density constant across timescales, the characteristic fuel temperature decreases for longer implosions (see \Fig{fig:tion}).  Therefore, less fuel escapes the liner cavity, and more fuel mass remains in the imploding region.

The observed behavior of fuel losses in \Fig{fig:NormMass} helps explain some trends of the scaling laws previously discussed.  For example, the second simulation data set in \Fig{fig:pressure} shows a similar scaling law for the fuel pressure even though the measured inner-convergence ratios $\mathrm{CR}_{\rm in,bang}$ in \Fig{fig:CR} are smaller than those of the similarity-scaled loads.  To understand this observation, in \Fig{fig:pressure_time}, we plot the total pressure within the fuel versus $\mathrm{CR}_{\rm in}$ for the two $S=3$ loads discussed.  Due to reduced end losses, the fuel in the second load shown in \Fig{fig:pressure_time} is more adiabatically compressed; \ie the fuel reaches higher pressures at a given $\mathrm{CR}_{\rm in}$ as compared to its similarity-scaled counterpart.  However, due to the larger initial fuel density, radiation losses (as measured by the relative radiation-loss rate $\tau_{\rm E,rad}^{-1} \doteq P_{\rm rad}/U$ introduced in Paper I) are much stronger for the second load and quench the neutron pulse at an earlier stage in the implosion.  This early termination of the neutron yield leads to the reduced $\mathrm{CR}_{\rm in,bang}$ shown in \Fig{fig:CR} for the second simulation data set.  In contrast, because of the relatively smaller initial fuel densities, the fuel compression of the similarity-scaled loads is less efficient.  Nevertheless, the higher fuel temperatures and reduced radiation losses allow the fuel to converge to higher $\mathrm{CR}_{\rm in,bang}$, reach an equivalent pressure near peak burn, and produce greater neutron yield.

Another interesting observation is that the scaling $m_{\rm fuel}(t)/m_{\rm fuel}(0) \propto t_\varphi^{0.26}$ for the second data set accounts for the ``excess" scaling exponent in the fuel internal energy shown in \Fig{fig:U}.  In other words, reduced end losses are leading to the larger-than-expected scaling in the fuel internal energy $U$ for the second simulation data set with initial fuel density and magnetic field constants.

As a side note, the remarks given in the previous paragraphs show how similarity-scaling theory can be useful even when not exactly reproducing the results generated by more sophisticated multi-physics codes, such as \textsc{hydra}.  Similarity-scaling theory provides easy-to-derive, back-of-the-envelop scaling relations that serve as a benchmark to which simulations results can be compared to.  When deviations in the scaling curves are detected, one can then try to explain those differences in terms of other observed trends and simplified physical models or pictures.  This type of exercises often lead to new physical insights about the functioning of MagLIF that would otherwise be overlooked.

%%%%%%%%%%%%%%%%%%%%%%%%%%%%%%%%%%%%%%%%%%%%%%%%%
%%%%%%%%%%%%%%%%%%%%%%%%%%%%%%%%%%%%%%%%%%%%%%%%%
%%%%%%%%%%%%%%%%%%%%%%%%%%%%%%%%%%%%%%%%%%%%%%%%%
\section{Scaling of MagLIF performance}
\label{sec:performance}

Let us now discuss the scaling of the fusion neutron yield $Y$ when varying the implosion timescale.  The neutron yield follows the scaling of the characteristic yield number $Y_{\rm ref}$ introduced in Paper I.  $Y_{\rm ref}$ obeys the following scaling rule:
\begin{equation}
	\frac{Y'}{Y}
		= 	\left( \frac{\rho_0'}{\rho_0} \right)^2
			 \left( \frac{T_{\rm preheat}'}{T_{\rm preheat}} \right)^{3.6}
			 \left( \frac{R_{\rm in,0}'}{R_{\rm in,0}} \right)^2
			 \frac{h'}{h}
			 \frac{t_\varphi'}{t_\varphi}
\end{equation}  
In the above, we fitted the Bosch--Hale\citep{Bosch:1992aa} expression for the DD reactivity $\langle \sigma v \rangle_{\rm DD}$ to a power law such that $\langle \sigma v \rangle_{\rm DD} \propto T_{\rm ion}^{3.6}$ within the 2 and 5 keV range.  After substituting \Eqs{eq:numerical:R}--\eq{eq:numerical:rho} and \Eq{eq:stagnation:tion}, we obtain the scaling rule for the neutron yield:
%2*(-1.12)+3.6*(-0.26)+2*(0.69)+0.88 + 1 = -0.01
\begin{equation}
	\frac{Y'}{Y}	\simeq	\left( \frac{t_\varphi'}{t_\varphi} \right)^{0.08}.
	\label{eq:performance:yield}
\end{equation}
	
This calculation suggests that the neutron yield will have a weak dependence on the implosion timescales.  Figure~\ref{fig:yield}~(left) compares the scaling law in \Eq{eq:performance:yield} to the neutron yields obtained from simulations.  Regarding the similarity-scaled MagLIF loads (blue points), the simulation results show a $\sim$50\% increase in yield when increasing the implosion timescales.  However, the dependence is rather weak.  Figure~\ref{fig:yield}~(right) shows the yield per-unit-length $\smash{\widehat{Y} \doteq Y/h}$.  According to this metric, MagLIF loads with shorter implosion times perform better since the achieved pressures and ion temperatures are higher.  In summary, similarity-scaled MagLIF loads imploding in longer timescales show a slightly better neutron yield.  However, the enhancement in yield is achieved by the larger fuel volumes of those liners and the longer burn-width times.  The minor enhancement comes at the cost of higher requirements for the preheat energy.

For the second simulation data set in \Fig{fig:yield}, the MagLIF loads tend to perform worse than their similarity-scaled counterparts when increasing the timescale of the implosions.  The drop in performance is due to the sharp temperature decrease shown in \Fig{fig:tion}, which strongly reduces the reactivity.  It is worth noting that keeping the density and magnetic field constant when decreasing the implosion timescales leads to similar performance as the similarity-scaled loads.

\begin{figure}
	\includegraphics[scale=0.43]{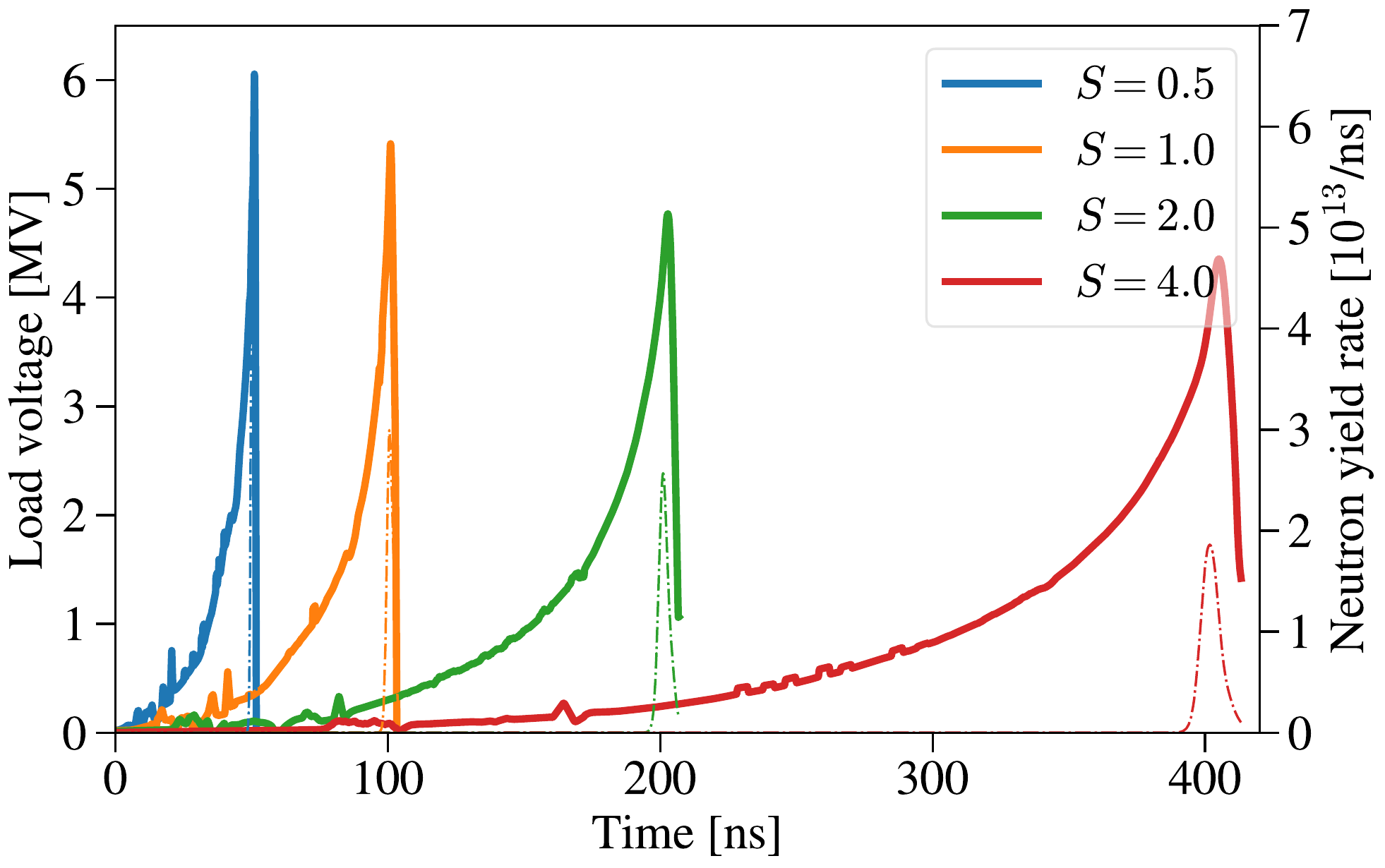}
	\includegraphics[scale=0.43]{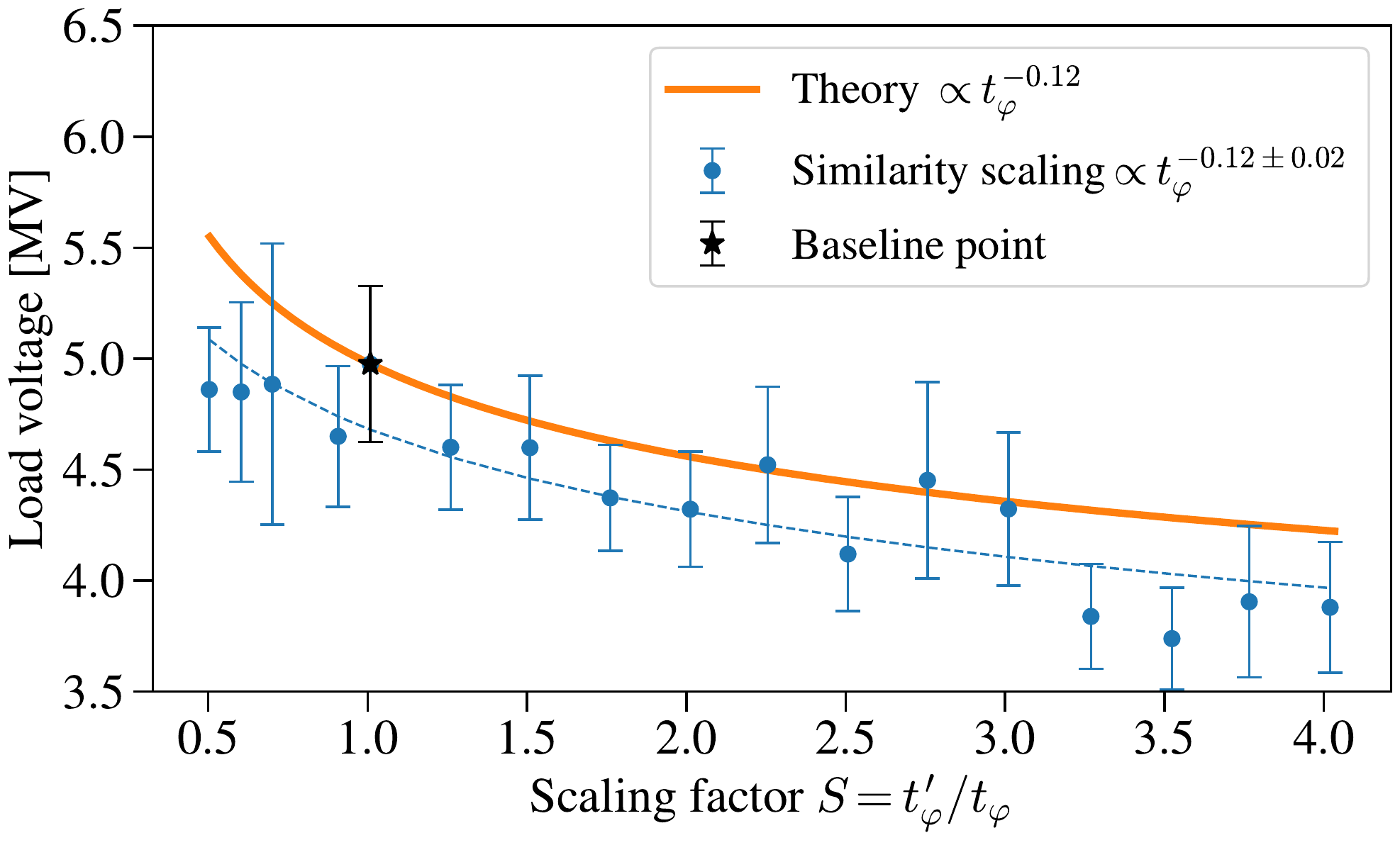}
	\caption{Top: Electrical load voltage\cite{foot:load_voltage} (solid lines) and neutron yield rate (dashed lines) plotted versus time for a family of similarity-scaled MagLIF loads.  Bottom: Electrical load voltage measured near peak burn.}
	\label{fig:voltage}
\end{figure}

\begin{figure}
	\includegraphics[scale=0.43]{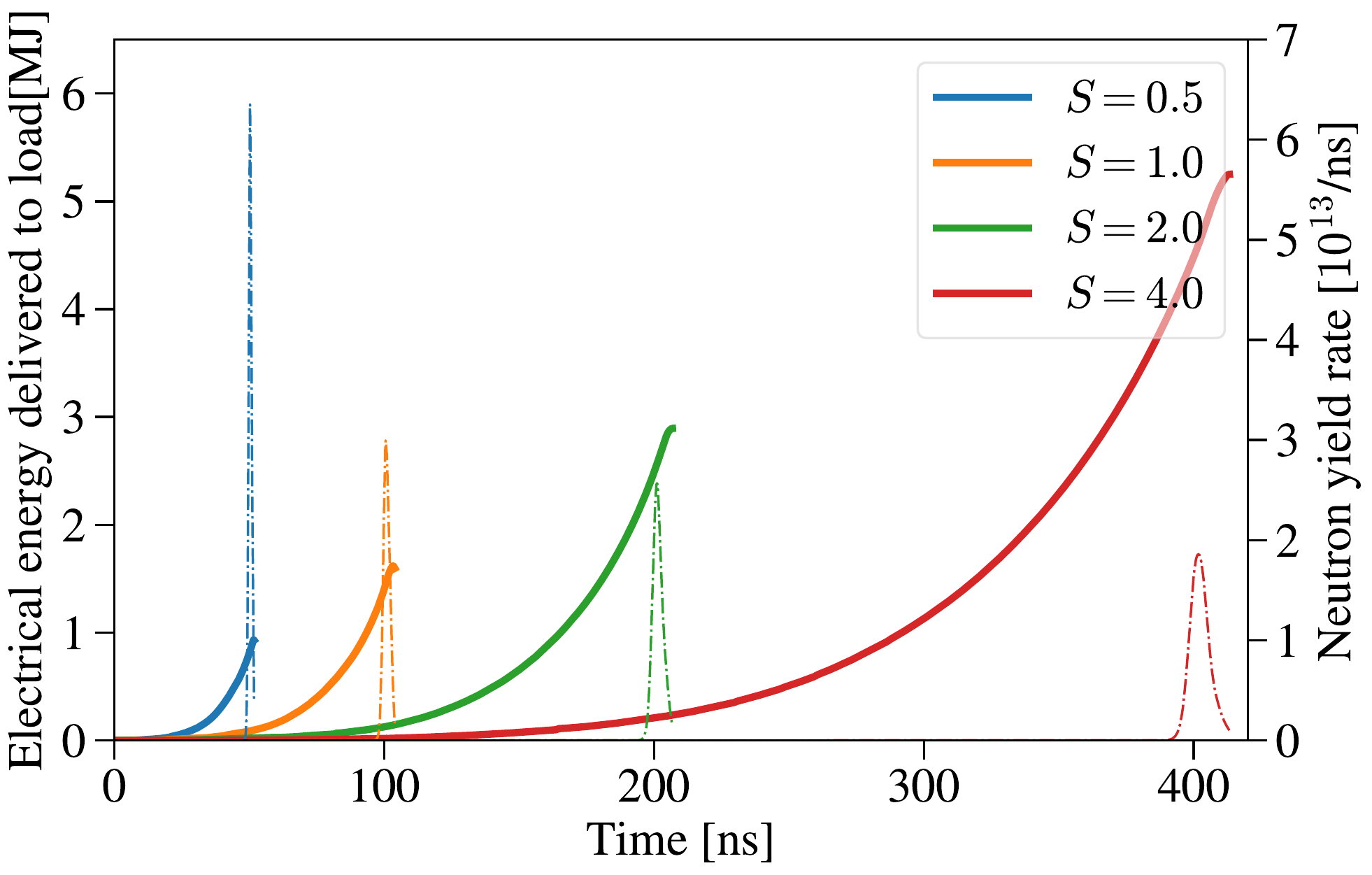}
	\includegraphics[scale=0.43]{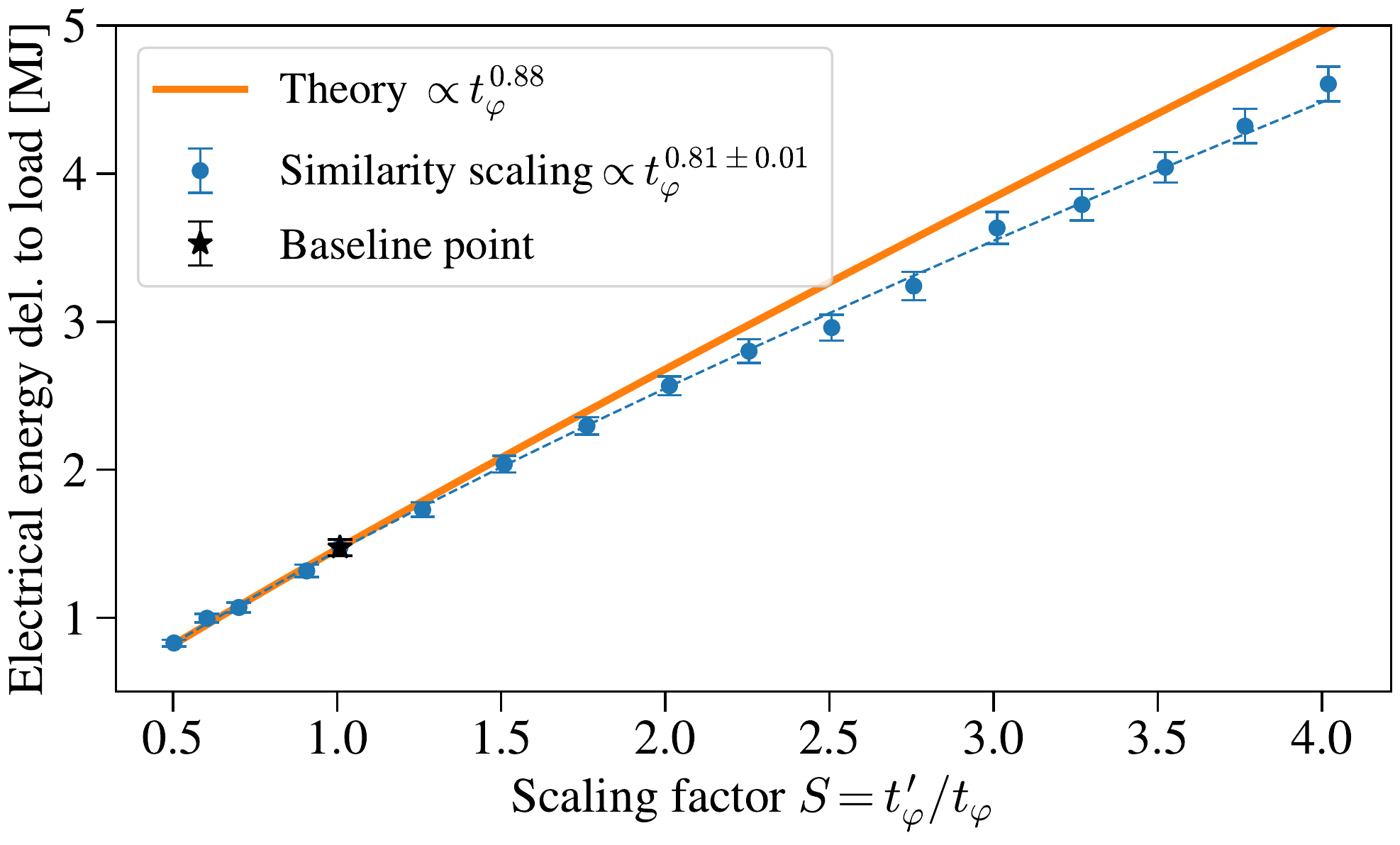}
	\caption{Top: Delivered electrical energy (solid lines) and neutron yield rate (dashed lines) plotted versus time for a family of similarity-scaled MagLIF loads.  Bottom: Electrical energy $E_{\rm load}$ delivered to the load measured near peak burn.}
	\label{fig:energy}
\end{figure}

%%%%%%%%%%%%%%%%%%%%%%%%%%%%%%%%%%%%%%%%%%%%%%%%%
%%%%%%%%%%%%%%%%%%%%%%%%%%%%%%%%%%%%%%%%%%%%%%%%%
%%%%%%%%%%%%%%%%%%%%%%%%%%%%%%%%%%%%%%%%%%%%%%%%%
\section{Discussion}
\label{sec:discussion}

In \Sec{sec:performance}, we showed that varying the characteristic timescale $t_\varphi$ of the voltage drive leads to similar performance for the similarity-scaled MagLIF loads.  One question that can be posed is the following:  from an energy-delivery standpoint, is there an advantage when increasing the current-rise time for MagLIF?  In what follows, we argue that slower implosions do not significantly decrease the voltage at the load region but can increase both preheat-energy and electrical-energy requirements for the MagLIF load.  

It is commonly assumed that, when increasing the current-rise time, the electrical voltage at the load decreases due to the smaller $\mathrm{d}I/\mathrm{d}t$.  In \Fig{fig:voltage} (top), we show the time evolution for the electrical voltages measured at the load region; \ie 
\begin{equation}
	\varphi_{\rm load}(t) \doteq \frac{\mathrm{d}}{\mathrm{d}t} (L_{\rm load} I_l)  + R_{\rm load} I_l,
\end{equation}
where $I_l(t)$ is the current delivered to the load region,
\begin{equation}
	L_{\rm load} \simeq \frac{\mu_0h}{2 \pi} \ln \left( \frac{R_{\rm out,0}}{R_{\rm out}} \right) 
\end{equation}
is the load inductance (neglecting the contribution from magnetic diffusion into the liner), and $R_{\rm load}(t) I_l$ is the resistive voltage drop across the liner.  As shown, the peak voltages only decrease slightly when increasing the implosion timescale by 4x.  The peak voltages do not decrease significantly because the MagLIF liners become longer to mitigate end losses.  Upon assuming that the load voltage is primarily due to the dynamic-inductance term $[\varphi_{\rm load} \simeq \mathrm{d}(L_{\rm load} I_l)/\mathrm{d}t]$, the scaling rule for $\varphi_{\rm load}$ is given by \Eq{eq:scaling:varphi}:
\begin{equation}
	\frac{\varphi_{\rm load}'}{\varphi_{\rm load}}
		\simeq	\frac{\varphi_0'}{\varphi_0}
		=	\frac{h'}{h} \frac{t_\varphi}{t_\varphi'} 
		\simeq \left( \frac{t_\varphi'}{t_\varphi}  \right)^{-0.12}.
	\label{eq:performance:varphi} 
\end{equation}
In \Fig{fig:voltage} (bottom), we show the burn-history averaged electrical voltage at the load for the similarity-scaled MagLIF loads.  As expected from the theory, the peak voltage decreases rather weakly when increasing the implosion timescale.  This behavior is a consequence of MagLIF being an open-ended liner.  For other z-pinch loads, such as wire arrays,\cite{Cuneo:2006hm,Jones:2008ds,Cuneo:2012kv} end effects are less of an issue so the load height does not have to be varied significantly.  Therefore, the voltage near the load wire arrays decreases more markedly with longer implosion times.  In addition, it is worth noting that the electrical power delivered to the load
\begin{equation}
	P_{\rm load} (t) 
		\doteq 	I_l
		\frac{\mathrm{d}}{\mathrm{d}t} \left( L_{\rm load} I_l \right)
\end{equation}
scales identically to $\varphi_0$.

Electrical voltages and peak power do not vary significantly when increasing the implosion timescales.  However, the total electrical energy $E_{\rm load} \doteq \int P_{\rm load} \, \mathrm{d}t$ delivered to the load increases with longer timescales.  Upon using simple dimensional analysis, we find that the electrical energy follows the scaling rule:
\begin{equation}
	\frac{E_{\rm load}'}{E_{\rm load}}
		= 	\frac{h'}{h}
		\simeq \left( \frac{t_\varphi'}{t_\varphi}  \right)^{0.88},
	\label{eq:discussion:Eload}
\end{equation}
which is precisely the scaling rule for the preheat energy $E_{\rm preheat}$ in \Eq{eq:numerical:Epreheat_h}.  In \Fig{fig:energy} (top), we show the time traces for the electrical energy delivered to the similarity-scaled loads.  $E_{\rm load}(t)$ accounts for the mechanical work done on the liner as well as the magnetic energy needed to fill the vacuum region created by the displaced liner.  As shown from the time traces, the delivered electrical energies grow considerably when increasing the implosion times.  The measured scaling curve for the energy delivered up to peak burn is shown in \Fig{fig:energy} (bottom).  The scaling law measured from the simulations follows closely that in \Eq{eq:discussion:Eload}.  These results show that the peak voltage and electrical power delivered to a MagLIF load only decrease slightly when increasing the implosion timescales for MagLIF.  In contrast, the total electrical energy required to drive the implosion and the total preheat energy increase almost linearly with the implosion timescale.  

%  When hydroscaling to higher currents, the current-rise time scales linearly with the peak current 

%to scale MagLIF to higher currents might not be very favorable from the load-physics perspective.

%, where the current rise time increases linearly with the peak current. 

As a final comment for this section, we note that this work does not completely exclude other scaling alternatives that allow for longer implosion times with shorter MagLIF liners.  For example, one possibility is to not scale the laser spot size $R_{\rm spot}$ and the inner radius $R_{\rm cushion}$ of the cushions linearly with the initial inner radius $R_{\rm in,0}$ of the liner, as suggested in \Sec{sec:prescriptions} when invoking geometric scaling.  If a weaker scaling prescription were adopted for $R_{\rm spot}$ and $R_{\rm cushion}$, the resulting smaller cross sections for end flows when scaling to longer implosions could reduce end losses and allow for a weaker scaling of the imploding height $h$ of the liner.  Such alternate scaling strategy could lead to the more attractive scaling laws $\phi\propto P_{\rm load} \propto  1/t_\varphi$ for the circuit voltages and electrical powers associated to longer rise times.  Nevertheless, such scaling prescription would lead to larger laser intensities, which could enhance laser-plasma instabilities and increase the difficulty to deposit the required preheat energy within the relative dilute gas-fill densities for more slowly imploding liners.  This effect could be compensated by dilating the preheat laser pulse, utilizing an expanding beam, or fielding a different wavelength laser.  This and other alternatives to reduce end losses and decrease the scaling requirements for the liner imploding height $h$ are left for future work.

%%%%%%%%%%%%%%%%%%%%%%%%%%%%%%%%%%%%%%%%%%%%%%%%%
%%%%%%%%%%%%%%%%%%%%%%%%%%%%%%%%%%%%%%%%%%%%%%%%%
%%%%%%%%%%%%%%%%%%%%%%%%%%%%%%%%%%%%%%%%%%%%%%%%%
\section{Conclusions}
\label{sec:conclusions}

We applied the similarity scaling framework developed in \Refa{foot:Ruiz_framework} to analytically investigate the performance of MagLIF loads when varying the characteristic rise time of the driving electrical current.  The present theory provides the scaling prescriptions of the experimental input parameters of MagLIF loads when scaling with respect to the rise time.  The theory also gives estimates of the scaling rules for the stagnation conditions and various performance metrics.  We tested the scaling theory against \textsc{hydra} simulations and found good agreement.    

Our findings highlight certain aspects concerning MagLIF-implosion physics that need to be considered when evaluating future designs of more powerful and energetic pulsed-power generators with longer current-rise times.  When considering longer current pulses, our study has identified the following key changes on the physics and requirements of MagLIF implosions.  (i)~When increasing the implosion time of a MagLIF load, the liner axial length must increase to mitigate end-flow mass and energy losses.  (ii)~Because MagLIF liners become axially longer, the requirements for the electrical voltage and power delivered to the load region do not decrease significantly when increasing the current-rise time. In contrast, the electrical energy required to implode the load increases almost linearly with current-rise time. (iii)~In a similar vein, since the initial fuel volume increases with current-rise time, MagLIF requires a more energetic laser to deliver a similar preheat energy per-unit-length before the implosion begins. (iv)~With slower implosions, the initial fuel density must decrease in order to mitigate radiation losses.  This relaxes the requirement for the externally applied magnetic field but may complicate the preheat-energy delivery due to reduced laser absorption via inverse Bremsstrahlung. (v)~When imploding MagLIF loads using longer current-rise times, the fuel stagnation pressure and yield per-unit-length tend to decrease. These results suggest that driving MagLIF loads with higher peak currents and longer current pulses using ``hydroscaling"-type strategies (as originally advocated for laser-driven ICF in \Refa{Nora:2014iq} and discussed previously in \Refa{Schmit:2020jd} for MagLIF) may not be as attractive when compared to a current-scaling strategy with fixed current-rise time (discussed in Paper II). Nevertheless, a systematic study of hydroscaling applied to MagLIF may be interesting, along with a detailed comparison of the results to the current-scaling study presented in Paper~II.  This may be subject for future work.

Setting the MagLIF-load physics aside, from the pulsed-power perspective, a machine with a longer current-rise time is less complex and less expensive to build than one with a shorter current pulse.  It may also be less ``risky" since the electrical voltages at the electrodes are smaller and electrical breakdown may occur later in the current pulse.  These effects may reduce current losses within the magnetically-insulated transmission lines.  However, there is also some risk associated with longer current pulses since there could be additional time for power-loss mechanisms, such as plasma emission, plasma gap closure and uninsulated electron flow, to influence the amount of energy delivered to the load.  For future work, it will be interesting to apply similarity-scaling techniques to study the design space of magnetically-insulated transmission lines and electrical power flow for future pulsed-power generators in order to identify ``conservative" design choices and discover key risks from the pulsed-power perspective.  Such analytical studies should be accompanied by first-principle electromagnetic, particle-in-cell calculations \citep{Welch:2019aa,Bennett:2019aa,Bennett:2021dq} to verify the derived scaling laws.  This will be subject for future work.

One of the authors (D.~E.~Ruiz) was supported in part by Sandia National Laboratories (SNL) Laboratory Directed Research and Development (LDRD) Program, Project 223312.  Sandia National Laboratories is a multimission laboratory managed and operated by National Technology $\&$ Engineering Solutions of Sandia, LLC, a wholly owned subsidiary of Honeywell International Inc., for the U.S. Department of Energy's National Nuclear Security Administration under contract DE-NA0003525.  This paper describes objective technical results and analysis. Any subjective views or opinions that might be expressed in the paper do not necessarily represent the views of the U.S. Department of Energy or the United States Government.

%%%%%%%%%%%%%%%%%%%%%%%%%%%%%%%%%%%%%%%%%%%%%%%%%
%%%%%%%%%%%%%%%%%%%%%%%%%%%%%%%%%%%%%%%%%%%%%%%%%
%%%%%%%%%%%%%%%%%%%%%%%%%%%%%%%%%%%%%%%%%%%%%%%%%

%\bibliographystyle{apsrev-title}
%\bibliography{risetime_scaling,foot}

\begin{thebibliography}{52}
\expandafter\ifx\csname natexlab\endcsname\relax\def\natexlab#1{#1}\fi
\expandafter\ifx\csname bibnamefont\endcsname\relax
  \def\bibnamefont#1{#1}\fi
\expandafter\ifx\csname bibfnamefont\endcsname\relax
  \def\bibfnamefont#1{#1}\fi
\expandafter\ifx\csname citenamefont\endcsname\relax
  \def\citenamefont#1{#1}\fi
\expandafter\ifx\csname url\endcsname\relax
  \def\url#1{\texttt{#1}}\fi
\expandafter\ifx\csname urlprefix\endcsname\relax\def\urlprefix{URL }\fi
\providecommand{\bibinfo}[2]{#2}
\providecommand{\eprint}[2][]{\url{#2}}

\bibitem[{\citenamefont{Atzeni and Meyer-ter Vehn}(2009)}]{Atzeni:2009phys}
\bibinfo{author}{\bibfnamefont{S.}~\bibnamefont{Atzeni}} \bibnamefont{and}
  \bibinfo{author}{\bibfnamefont{J.}~\bibnamefont{Meyer-ter Vehn}},
  \emph{\bibinfo{title}{{The Physics of Inertial Fusion: BeamPlasma
  Interaction, Hydrodynamics, Hot Dense Matter}}}, International Series of
  Monographs on Physics (\bibinfo{publisher}{Oxford University Press Inc.},
  \bibinfo{address}{New York}, \bibinfo{year}{2009}).

\bibitem[{\citenamefont{Lindl}(1998)}]{Lindl:1998bq}
\bibinfo{author}{\bibfnamefont{J.}~\bibnamefont{Lindl}}, ``{Development of the
  indirect-drive approach to inertial confinement fusion and the target physics
  basis for ignition and gain},'' \bibinfo{journal}{Phys. Plasmas}
  \textbf{\bibinfo{volume}{2}}, \bibinfo{pages}{3933} (\bibinfo{year}{1998}).

\bibitem[{Wid()}]{Widner_1977}
\bibinfo{note}{M. M. Widner, Bull. Am. Phys. Soc. \textbf{22}, 1139 (1977).}

\bibitem[{\citenamefont{Lindemuth and Widner}(1981)}]{Lindemuth_1981}
\bibinfo{author}{\bibfnamefont{I.~R.} \bibnamefont{Lindemuth}}
  \bibnamefont{and} \bibinfo{author}{\bibfnamefont{M.~M.}
  \bibnamefont{Widner}}, ``Magnetohydrodynamic behavior of thermonuclear fuel
  in a preconditioned electron beam imploded target,'' \bibinfo{journal}{The
  Physics of Fluids} \textbf{\bibinfo{volume}{24}}, \bibinfo{pages}{746}
  (\bibinfo{year}{1981}).

\bibitem[{\citenamefont{Lindemuth and Kirkpatrick}(1983)}]{Lindemuth:1983aa}
\bibinfo{author}{\bibfnamefont{I.~R.} \bibnamefont{Lindemuth}}
  \bibnamefont{and} \bibinfo{author}{\bibfnamefont{R.~C.}
  \bibnamefont{Kirkpatrick}}, ``{Parameter space for magnetized fuel targets in
  inertial confinement fusion},'' \bibinfo{journal}{Nucl. Fusion}
  \textbf{\bibinfo{volume}{23}}, \bibinfo{pages}{263} (\bibinfo{year}{1983}).

\bibitem[{\citenamefont{Lindemuth}(2015)}]{Lindemuth:2015fu}
\bibinfo{author}{\bibfnamefont{I.~R.} \bibnamefont{Lindemuth}}, ``{The ignition
  design space of magnetized target fusion},'' \bibinfo{journal}{Phys. Plasmas}
  \textbf{\bibinfo{volume}{22}}, \bibinfo{pages}{122712}
  (\bibinfo{year}{2015}).

\bibitem[{\citenamefont{Slutz et~al.}(2010)\citenamefont{Slutz, Herrmann,
  Vesey, Sefkow, Sinars, Rovang, Peterson, and Cuneo}}]{Slutz:2010hd}
\bibinfo{author}{\bibfnamefont{S.~A.} \bibnamefont{Slutz}},
  \bibinfo{author}{\bibfnamefont{M.~C.} \bibnamefont{Herrmann}},
  \bibinfo{author}{\bibfnamefont{R.~A.} \bibnamefont{Vesey}},
  \bibinfo{author}{\bibfnamefont{A.~B.} \bibnamefont{Sefkow}},
  \bibinfo{author}{\bibfnamefont{D.~B.} \bibnamefont{Sinars}},
  \bibinfo{author}{\bibfnamefont{D.~C.} \bibnamefont{Rovang}},
  \bibinfo{author}{\bibfnamefont{K.~J.} \bibnamefont{Peterson}},
  \bibnamefont{and} \bibinfo{author}{\bibfnamefont{M.~E.} \bibnamefont{Cuneo}},
  ``{Pulsed-power-driven cylindrical liner implosions of laser preheated fuel
  magnetized with an axial field},'' \bibinfo{journal}{Phys. Plasmas}
  \textbf{\bibinfo{volume}{17}}, \bibinfo{pages}{056303}
  (\bibinfo{year}{2010}).

\bibitem[{\citenamefont{Gomez et~al.}(2014)\citenamefont{Gomez, Slutz, Sefkow,
  Sinars, Hahn, Hansen, Harding, Knapp, Schmit, Jennings
  et~al.}}]{Gomez:2014eta}
\bibinfo{author}{\bibfnamefont{M.~R.} \bibnamefont{Gomez}},
  \bibinfo{author}{\bibfnamefont{S.~A.} \bibnamefont{Slutz}},
  \bibinfo{author}{\bibfnamefont{A.~B.} \bibnamefont{Sefkow}},
  \bibinfo{author}{\bibfnamefont{D.~B.} \bibnamefont{Sinars}},
  \bibinfo{author}{\bibfnamefont{K.~D.} \bibnamefont{Hahn}},
  \bibinfo{author}{\bibfnamefont{S.~B.} \bibnamefont{Hansen}},
  \bibinfo{author}{\bibfnamefont{E.~C.} \bibnamefont{Harding}},
  \bibinfo{author}{\bibfnamefont{P.~F.} \bibnamefont{Knapp}},
  \bibinfo{author}{\bibfnamefont{P.~F.} \bibnamefont{Schmit}},
  \bibinfo{author}{\bibfnamefont{C.~A.} \bibnamefont{Jennings}},
  \bibnamefont{et~al.}, ``{Experimental demonstration of fusion-relevant
  conditions in magnetized liner inertial fusion},'' \bibinfo{journal}{Phys.
  Rev. Lett.} \textbf{\bibinfo{volume}{113}}, \bibinfo{pages}{155003}
  (\bibinfo{year}{2014}).

\bibitem[{\citenamefont{Gomez et~al.}(2019)\citenamefont{Gomez, Slutz, Knapp,
  Hahn, Weis, Harding, Geissel, Fein, Glinsky, Hansen et~al.}}]{Gomez:2019bg}
\bibinfo{author}{\bibfnamefont{M.~R.} \bibnamefont{Gomez}},
  \bibinfo{author}{\bibfnamefont{S.~A.} \bibnamefont{Slutz}},
  \bibinfo{author}{\bibfnamefont{P.~F.} \bibnamefont{Knapp}},
  \bibinfo{author}{\bibfnamefont{K.~D.} \bibnamefont{Hahn}},
  \bibinfo{author}{\bibfnamefont{M.~R.} \bibnamefont{Weis}},
  \bibinfo{author}{\bibfnamefont{E.~C.} \bibnamefont{Harding}},
  \bibinfo{author}{\bibfnamefont{M.}~\bibnamefont{Geissel}},
  \bibinfo{author}{\bibfnamefont{J.~R.} \bibnamefont{Fein}},
  \bibinfo{author}{\bibfnamefont{M.~E.} \bibnamefont{Glinsky}},
  \bibinfo{author}{\bibfnamefont{S.~B.} \bibnamefont{Hansen}},
  \bibnamefont{et~al.}, ``{Assessing Stagnation Conditions and Identifying
  Trends in Magnetized Liner Inertial Fusion},'' \bibinfo{journal}{IEEE Trans.
  Plasma Sci.} \textbf{\bibinfo{volume}{47}}, \bibinfo{pages}{2081}
  (\bibinfo{year}{2019}).

\bibitem[{\citenamefont{Sefkow et~al.}(2014)\citenamefont{Sefkow, Slutz,
  Koning, Marinak, Peterson, Sinars, and Vesey}}]{Sefkow:2014ik}
\bibinfo{author}{\bibfnamefont{A.~B.} \bibnamefont{Sefkow}},
  \bibinfo{author}{\bibfnamefont{S.~A.} \bibnamefont{Slutz}},
  \bibinfo{author}{\bibfnamefont{J.~M.} \bibnamefont{Koning}},
  \bibinfo{author}{\bibfnamefont{M.~M.} \bibnamefont{Marinak}},
  \bibinfo{author}{\bibfnamefont{K.~J.} \bibnamefont{Peterson}},
  \bibinfo{author}{\bibfnamefont{D.~B.} \bibnamefont{Sinars}},
  \bibnamefont{and} \bibinfo{author}{\bibfnamefont{R.~A.} \bibnamefont{Vesey}},
  ``{Design of magnetized liner inertial fusion experiments using the Z
  facility},'' \bibinfo{journal}{Phys. Plasmas} \textbf{\bibinfo{volume}{21}},
  \bibinfo{pages}{072711} (\bibinfo{year}{2014}).

\bibitem[{\citenamefont{Knapp et~al.}(2019)\citenamefont{Knapp, Gomez, Hansen,
  Glinsky, Jennings, Slutz, Harding, Hahn, Weis, Evans et~al.}}]{Knapp:2019gf}
\bibinfo{author}{\bibfnamefont{P.~F.} \bibnamefont{Knapp}},
  \bibinfo{author}{\bibfnamefont{M.~R.} \bibnamefont{Gomez}},
  \bibinfo{author}{\bibfnamefont{S.~B.} \bibnamefont{Hansen}},
  \bibinfo{author}{\bibfnamefont{M.~E.} \bibnamefont{Glinsky}},
  \bibinfo{author}{\bibfnamefont{C.~A.} \bibnamefont{Jennings}},
  \bibinfo{author}{\bibfnamefont{S.~A.} \bibnamefont{Slutz}},
  \bibinfo{author}{\bibfnamefont{E.~C.} \bibnamefont{Harding}},
  \bibinfo{author}{\bibfnamefont{K.~D.} \bibnamefont{Hahn}},
  \bibinfo{author}{\bibfnamefont{M.~R.} \bibnamefont{Weis}},
  \bibinfo{author}{\bibfnamefont{M.}~\bibnamefont{Evans}},
  \bibnamefont{et~al.}, ``{Origins and effects of mix on magnetized liner
  inertial fusion target performance},'' \bibinfo{journal}{Phys. Plasmas}
  \textbf{\bibinfo{volume}{26}}, \bibinfo{pages}{012704}
  (\bibinfo{year}{2019}).

\bibitem[{\citenamefont{Sinars et~al.}(2020)\citenamefont{Sinars, Sweeney,
  Alexander, Ampleford, Ao, Apruzese, Aragon, Armstrong, Austin, Awe
  et~al.}}]{Sinars:2020bv}
\bibinfo{author}{\bibfnamefont{D.~B.} \bibnamefont{Sinars}},
  \bibinfo{author}{\bibfnamefont{M.~A.} \bibnamefont{Sweeney}},
  \bibinfo{author}{\bibfnamefont{C.~S.} \bibnamefont{Alexander}},
  \bibinfo{author}{\bibfnamefont{D.~J.} \bibnamefont{Ampleford}},
  \bibinfo{author}{\bibfnamefont{T.}~\bibnamefont{Ao}},
  \bibinfo{author}{\bibfnamefont{J.~P.} \bibnamefont{Apruzese}},
  \bibinfo{author}{\bibfnamefont{C.}~\bibnamefont{Aragon}},
  \bibinfo{author}{\bibfnamefont{D.~J.} \bibnamefont{Armstrong}},
  \bibinfo{author}{\bibfnamefont{K.~N.} \bibnamefont{Austin}},
  \bibinfo{author}{\bibfnamefont{T.~J.} \bibnamefont{Awe}},
  \bibnamefont{et~al.}, ``{Review of pulsed power-driven high energy density
  physics research on Z at Sandia},'' \bibinfo{journal}{Phys. Plasmas}
  \textbf{\bibinfo{volume}{27}}, \bibinfo{pages}{070501}
  (\bibinfo{year}{2020}).

\bibitem[{\citenamefont{Yager-Elorriaga
  et~al.}(2022)\citenamefont{Yager-Elorriaga, Gomez, Ruiz, Slutz,
  Harvey-Thompson, Jennings, Knapp, Schmit, Weis, Awe
  et~al.}}]{YagerElorriaga:2022cp}
\bibinfo{author}{\bibfnamefont{D.~A.} \bibnamefont{Yager-Elorriaga}},
  \bibinfo{author}{\bibfnamefont{M.~R.} \bibnamefont{Gomez}},
  \bibinfo{author}{\bibfnamefont{D.~E.} \bibnamefont{Ruiz}},
  \bibinfo{author}{\bibfnamefont{S.~A.} \bibnamefont{Slutz}},
  \bibinfo{author}{\bibfnamefont{A.~J.} \bibnamefont{Harvey-Thompson}},
  \bibinfo{author}{\bibfnamefont{C.~A.} \bibnamefont{Jennings}},
  \bibinfo{author}{\bibfnamefont{P.~F.} \bibnamefont{Knapp}},
  \bibinfo{author}{\bibfnamefont{P.~F.} \bibnamefont{Schmit}},
  \bibinfo{author}{\bibfnamefont{M.~R.} \bibnamefont{Weis}},
  \bibinfo{author}{\bibfnamefont{T.~J.} \bibnamefont{Awe}},
  \bibnamefont{et~al.}, ``{An overview of magneto-inertial fusion on the Z
  machine at Sandia National Laboratories},'' \bibinfo{journal}{Nucl. Fusion}
  \textbf{\bibinfo{volume}{62}}, \bibinfo{pages}{042015}
  (\bibinfo{year}{2022}).

\bibitem[{\citenamefont{Gomez et~al.}(2020)\citenamefont{Gomez, Slutz,
  Jennings, Ampleford, Weis, Myers, Yager-Elorriaga, Hahn, Hansen, Harding
  et~al.}}]{Gomez:2020cd}
\bibinfo{author}{\bibfnamefont{M.~R.} \bibnamefont{Gomez}},
  \bibinfo{author}{\bibfnamefont{S.~A.} \bibnamefont{Slutz}},
  \bibinfo{author}{\bibfnamefont{C.~A.} \bibnamefont{Jennings}},
  \bibinfo{author}{\bibfnamefont{D.~J.} \bibnamefont{Ampleford}},
  \bibinfo{author}{\bibfnamefont{M.~R.} \bibnamefont{Weis}},
  \bibinfo{author}{\bibfnamefont{C.~E.} \bibnamefont{Myers}},
  \bibinfo{author}{\bibfnamefont{D.~A.} \bibnamefont{Yager-Elorriaga}},
  \bibinfo{author}{\bibfnamefont{K.~D.} \bibnamefont{Hahn}},
  \bibinfo{author}{\bibfnamefont{S.~B.} \bibnamefont{Hansen}},
  \bibinfo{author}{\bibfnamefont{E.~C.} \bibnamefont{Harding}},
  \bibnamefont{et~al.}, ``{Performance scaling in magnetized liner inertial
  fusion experiments},'' \bibinfo{journal}{Phys. Rev. Lett.}
  \textbf{\bibinfo{volume}{125}}, \bibinfo{pages}{155002}
  (\bibinfo{year}{2020}).

\bibitem[{\citenamefont{Harvey-Thompson
  et~al.}(2019)\citenamefont{Harvey-Thompson, Geissel, Jennings, Weis, Gomez,
  Fein, Ampleford, Chandler, Glinsky, Hahn et~al.}}]{HarveyThompson:2019ff}
\bibinfo{author}{\bibfnamefont{A.~J.} \bibnamefont{Harvey-Thompson}},
  \bibinfo{author}{\bibfnamefont{M.}~\bibnamefont{Geissel}},
  \bibinfo{author}{\bibfnamefont{C.~A.} \bibnamefont{Jennings}},
  \bibinfo{author}{\bibfnamefont{M.~R.} \bibnamefont{Weis}},
  \bibinfo{author}{\bibfnamefont{M.~R.} \bibnamefont{Gomez}},
  \bibinfo{author}{\bibfnamefont{J.~R.} \bibnamefont{Fein}},
  \bibinfo{author}{\bibfnamefont{D.~J.} \bibnamefont{Ampleford}},
  \bibinfo{author}{\bibfnamefont{G.~A.} \bibnamefont{Chandler}},
  \bibinfo{author}{\bibfnamefont{M.~E.} \bibnamefont{Glinsky}},
  \bibinfo{author}{\bibfnamefont{K.~D.} \bibnamefont{Hahn}},
  \bibnamefont{et~al.}, ``{Constraining preheat energy deposition in MagLIF
  experiments with multi-frame shadowgraphy},'' \bibinfo{journal}{Phys.
  Plasmas} \textbf{\bibinfo{volume}{26}}, \bibinfo{pages}{032707}
  (\bibinfo{year}{2019}).

\bibitem[{\citenamefont{Harvey-Thompson
  et~al.}(2018)\citenamefont{Harvey-Thompson, Weis, Harding, Geissel,
  Ampleford, Chandler, Fein, Glinsky, Gomez, Hahn
  et~al.}}]{HarveyThompson:2018dd}
\bibinfo{author}{\bibfnamefont{A.~J.} \bibnamefont{Harvey-Thompson}},
  \bibinfo{author}{\bibfnamefont{M.~R.} \bibnamefont{Weis}},
  \bibinfo{author}{\bibfnamefont{E.~C.} \bibnamefont{Harding}},
  \bibinfo{author}{\bibfnamefont{M.}~\bibnamefont{Geissel}},
  \bibinfo{author}{\bibfnamefont{D.~J.} \bibnamefont{Ampleford}},
  \bibinfo{author}{\bibfnamefont{G.~A.} \bibnamefont{Chandler}},
  \bibinfo{author}{\bibfnamefont{J.~R.} \bibnamefont{Fein}},
  \bibinfo{author}{\bibfnamefont{M.~E.} \bibnamefont{Glinsky}},
  \bibinfo{author}{\bibfnamefont{M.~R.} \bibnamefont{Gomez}},
  \bibinfo{author}{\bibfnamefont{K.~D.} \bibnamefont{Hahn}},
  \bibnamefont{et~al.}, ``{Diagnosing and mitigating laser preheat induced mix
  in MagLIF},'' \bibinfo{journal}{Phys. Plasmas} \textbf{\bibinfo{volume}{25}},
  \bibinfo{pages}{112705} (\bibinfo{year}{2018}).

\bibitem[{\citenamefont{Weis et~al.}(2021)\citenamefont{Weis, Harvey-Thompson,
  and Ruiz}}]{Weis:2021id}
\bibinfo{author}{\bibfnamefont{M.~R.} \bibnamefont{Weis}},
  \bibinfo{author}{\bibfnamefont{A.~J.} \bibnamefont{Harvey-Thompson}},
  \bibnamefont{and} \bibinfo{author}{\bibfnamefont{D.~E.} \bibnamefont{Ruiz}},
  ``{Scaling laser preheat for MagLIF with the Z-Beamlet laser},''
  \bibinfo{journal}{Phys. Plasmas} \textbf{\bibinfo{volume}{28}},
  \bibinfo{pages}{012705} (\bibinfo{year}{2021}).

\bibitem[{\citenamefont{Schmit et~al.}(2014)\citenamefont{Schmit, Knapp,
  Hansen, Gomez, Hahn, Sinars, Peterson, Slutz, Sefkow, Awe
  et~al.}}]{Schmit:2014fg}
\bibinfo{author}{\bibfnamefont{P.~F.} \bibnamefont{Schmit}},
  \bibinfo{author}{\bibfnamefont{P.~F.} \bibnamefont{Knapp}},
  \bibinfo{author}{\bibfnamefont{S.~B.} \bibnamefont{Hansen}},
  \bibinfo{author}{\bibfnamefont{M.}~\bibnamefont{Gomez}},
  \bibinfo{author}{\bibfnamefont{K.~D.} \bibnamefont{Hahn}},
  \bibinfo{author}{\bibfnamefont{D.~B.} \bibnamefont{Sinars}},
  \bibinfo{author}{\bibfnamefont{K.~J.} \bibnamefont{Peterson}},
  \bibinfo{author}{\bibfnamefont{S.~A.} \bibnamefont{Slutz}},
  \bibinfo{author}{\bibfnamefont{A.~B.} \bibnamefont{Sefkow}},
  \bibinfo{author}{\bibfnamefont{T.~J.} \bibnamefont{Awe}},
  \bibnamefont{et~al.}, ``{Understanding fuel magnetization and mix using
  secondary nuclear reactions in magneto-inertial fusion},''
  \bibinfo{journal}{Phys. Rev. Lett.} \textbf{\bibinfo{volume}{113}},
  \bibinfo{pages}{155004} (\bibinfo{year}{2014}).

\bibitem[{\citenamefont{Knapp et~al.}(2015)\citenamefont{Knapp, Schmit, Hansen,
  Gomez, Hahn, Sinars, Peterson, Slutz, Sefkow, Awe et~al.}}]{Knapp:2015kc}
\bibinfo{author}{\bibfnamefont{P.~F.} \bibnamefont{Knapp}},
  \bibinfo{author}{\bibfnamefont{P.~F.} \bibnamefont{Schmit}},
  \bibinfo{author}{\bibfnamefont{S.~B.} \bibnamefont{Hansen}},
  \bibinfo{author}{\bibfnamefont{M.~R.} \bibnamefont{Gomez}},
  \bibinfo{author}{\bibfnamefont{K.~D.} \bibnamefont{Hahn}},
  \bibinfo{author}{\bibfnamefont{D.~B.} \bibnamefont{Sinars}},
  \bibinfo{author}{\bibfnamefont{K.~J.} \bibnamefont{Peterson}},
  \bibinfo{author}{\bibfnamefont{S.~A.} \bibnamefont{Slutz}},
  \bibinfo{author}{\bibfnamefont{A.~B.} \bibnamefont{Sefkow}},
  \bibinfo{author}{\bibfnamefont{T.~J.} \bibnamefont{Awe}},
  \bibnamefont{et~al.}, ``{Effects of magnetization on fusion product trapping
  and secondary neutron spectraa)},'' \bibinfo{journal}{Phys. Plasmas}
  \textbf{\bibinfo{volume}{22}}, \bibinfo{pages}{056312}
  (\bibinfo{year}{2015}).

\bibitem[{\citenamefont{Lewis et~al.}(2021)\citenamefont{Lewis, Knapp, Slutz,
  Schmit, Chandler, Gomez, Harvey-Thompson, Mangan, Ampleford, and
  Beckwith}}]{Lewis:2021kz}
\bibinfo{author}{\bibfnamefont{W.~E.} \bibnamefont{Lewis}},
  \bibinfo{author}{\bibfnamefont{P.~F.} \bibnamefont{Knapp}},
  \bibinfo{author}{\bibfnamefont{S.~A.} \bibnamefont{Slutz}},
  \bibinfo{author}{\bibfnamefont{P.~F.} \bibnamefont{Schmit}},
  \bibinfo{author}{\bibfnamefont{G.~A.} \bibnamefont{Chandler}},
  \bibinfo{author}{\bibfnamefont{M.~R.} \bibnamefont{Gomez}},
  \bibinfo{author}{\bibfnamefont{A.~J.} \bibnamefont{Harvey-Thompson}},
  \bibinfo{author}{\bibfnamefont{M.~A.} \bibnamefont{Mangan}},
  \bibinfo{author}{\bibfnamefont{D.~J.} \bibnamefont{Ampleford}},
  \bibnamefont{and} \bibinfo{author}{\bibfnamefont{K.}~\bibnamefont{Beckwith}},
  ``{Deep-learning-enabled Bayesian inference of fuel magnetization in
  magnetized liner inertial fusion},'' \bibinfo{journal}{Phys. Plasmas}
  \textbf{\bibinfo{volume}{28}}, \bibinfo{pages}{092701}
  (\bibinfo{year}{2021}).

\bibitem[{\citenamefont{Slutz}(2018)}]{Slutz:2018iq}
\bibinfo{author}{\bibfnamefont{S.~A.} \bibnamefont{Slutz}}, ``{Scaling of
  magnetized inertial fusion with drive current rise-time},''
  \bibinfo{journal}{Phys. Plasmas} \textbf{\bibinfo{volume}{25}},
  \bibinfo{pages}{082707} (\bibinfo{year}{2018}).

\bibitem[{foo({\natexlab{a}})}]{foot:Ruiz_framework}
\bibinfo{note}{D.~E.~Ruiz, P.~F.~Schmit, D.~A.~Yager-Elorriaga, and C.~A.
  Jennings, ``Exploring the parameter space of MagLIF implosions using
  similarity scaling.~~I.~Theoretical framework," (placeholder for reference).}

\bibitem[{\citenamefont{Marinak et~al.}(1996)\citenamefont{Marinak, Tipton,
  Landen, Murphy, Amendt, Haan, Hatchett, Keane, McEachern, and
  Wallace}}]{Marinak:1996fs}
\bibinfo{author}{\bibfnamefont{M.~M.} \bibnamefont{Marinak}},
  \bibinfo{author}{\bibfnamefont{R.~E.} \bibnamefont{Tipton}},
  \bibinfo{author}{\bibfnamefont{O.~L.} \bibnamefont{Landen}},
  \bibinfo{author}{\bibfnamefont{T.~J.} \bibnamefont{Murphy}},
  \bibinfo{author}{\bibfnamefont{P.}~\bibnamefont{Amendt}},
  \bibinfo{author}{\bibfnamefont{S.~W.} \bibnamefont{Haan}},
  \bibinfo{author}{\bibfnamefont{S.~P.} \bibnamefont{Hatchett}},
  \bibinfo{author}{\bibfnamefont{C.~J.} \bibnamefont{Keane}},
  \bibinfo{author}{\bibfnamefont{R.}~\bibnamefont{McEachern}},
  \bibnamefont{and} \bibinfo{author}{\bibfnamefont{R.}~\bibnamefont{Wallace}},
  ``{Three-dimensional simulations of Nova high growth factor capsule implosion
  experiments},'' \bibinfo{journal}{Phys. Plasmas}
  \textbf{\bibinfo{volume}{3}}, \bibinfo{pages}{2070} (\bibinfo{year}{1996}).

\bibitem[{\citenamefont{Koning et~al.}(2009)\citenamefont{Koning, Kerbel, and
  Marinak}}]{Koning:2009}
\bibinfo{author}{\bibfnamefont{J.~M.} \bibnamefont{Koning}},
  \bibinfo{author}{\bibfnamefont{G.~D.} \bibnamefont{Kerbel}},
  \bibnamefont{and} \bibinfo{author}{\bibfnamefont{M.~M.}
  \bibnamefont{Marinak}}, in \emph{\bibinfo{booktitle}{APS Division of Plasma
  Physics Meeting Abstracts}} (\bibinfo{year}{2009}), p.
  \bibinfo{pages}{NP8.101}.

\bibitem[{foo({\natexlab{b}})}]{foot:Ruiz_current}
\bibinfo{note}{D.~E.~Ruiz, P.~F.~Schmit, D.~A.~Yager-Elorriaga, M.~R.~Gomez,
  M.~R.~Weis, C.~A.~Jennings, A.~J.~Harvey-Thompson, P.~F.~Knapp, S.~A.~Slutz,
  D.~J.~Ampleford, K.~Beckwith, and M.~K.~Matzen, ``Exploring the parameter
  space of MagLIF implosions using similarity scaling.~~II.~Current scaling,"
  (placeholder for reference).}

\bibitem[{\citenamefont{Harris}(1962)}]{Harris:1962hu}
\bibinfo{author}{\bibfnamefont{E.~G.} \bibnamefont{Harris}}, ``{Rayleigh-Taylor
  instabilities of a collapsing cylindrical shell in a magnetic field},''
  \bibinfo{journal}{Phys. Fluids} \textbf{\bibinfo{volume}{5}},
  \bibinfo{pages}{1057} (\bibinfo{year}{1962}).

\bibitem[{\citenamefont{Weis et~al.}(2015)\citenamefont{Weis, Zhang, Lau,
  Schmit, Peterson, Hess, and Gilgenbach}}]{Weis:2015hk}
\bibinfo{author}{\bibfnamefont{M.~R.} \bibnamefont{Weis}},
  \bibinfo{author}{\bibfnamefont{P.}~\bibnamefont{Zhang}},
  \bibinfo{author}{\bibfnamefont{Y.~Y.} \bibnamefont{Lau}},
  \bibinfo{author}{\bibfnamefont{P.~F.} \bibnamefont{Schmit}},
  \bibinfo{author}{\bibfnamefont{K.~J.} \bibnamefont{Peterson}},
  \bibinfo{author}{\bibfnamefont{M.}~\bibnamefont{Hess}}, \bibnamefont{and}
  \bibinfo{author}{\bibfnamefont{R.~M.} \bibnamefont{Gilgenbach}}, ``{Coupling
  of sausage, kink, and magneto-Rayleigh-Taylor instabilities in a cylindrical
  liner},'' \bibinfo{journal}{Phys. Plasmas} \textbf{\bibinfo{volume}{22}},
  \bibinfo{pages}{032706} (\bibinfo{year}{2015}).

\bibitem[{\citenamefont{Velikovich and Schmit}(2015)}]{Velikovich:2015jl}
\bibinfo{author}{\bibfnamefont{A.~L.} \bibnamefont{Velikovich}}
  \bibnamefont{and} \bibinfo{author}{\bibfnamefont{P.~F.}
  \bibnamefont{Schmit}}, ``{Bell-Plesset effects in Rayleigh-Taylor instability
  of finite-thickness spherical and cylindrical shells},''
  \bibinfo{journal}{Phys. Plasmas} \textbf{\bibinfo{volume}{22}},
  \bibinfo{pages}{122711} (\bibinfo{year}{2015}).

\bibitem[{\citenamefont{Sinars et~al.}(2010)\citenamefont{Sinars, Slutz,
  Herrmann, McBride, Cuneo, Peterson, Vesey, Nakhleh, Blue, Killebrew
  et~al.}}]{Sinars:2010de}
\bibinfo{author}{\bibfnamefont{D.~B.} \bibnamefont{Sinars}},
  \bibinfo{author}{\bibfnamefont{S.~A.} \bibnamefont{Slutz}},
  \bibinfo{author}{\bibfnamefont{M.~C.} \bibnamefont{Herrmann}},
  \bibinfo{author}{\bibfnamefont{R.~D.} \bibnamefont{McBride}},
  \bibinfo{author}{\bibfnamefont{M.~E.} \bibnamefont{Cuneo}},
  \bibinfo{author}{\bibfnamefont{K.~J.} \bibnamefont{Peterson}},
  \bibinfo{author}{\bibfnamefont{R.~A.} \bibnamefont{Vesey}},
  \bibinfo{author}{\bibfnamefont{C.}~\bibnamefont{Nakhleh}},
  \bibinfo{author}{\bibfnamefont{B.~E.} \bibnamefont{Blue}},
  \bibinfo{author}{\bibfnamefont{K.}~\bibnamefont{Killebrew}},
  \bibnamefont{et~al.}, ``{Measurements of magneto-Rayleigh-Taylor instability
  growth during the implosion of initially solid al tubes driven by the 20-MA,
  100-ns Z Dacility},'' \bibinfo{journal}{Phys. Rev. Lett.}
  \textbf{\bibinfo{volume}{105}}, \bibinfo{pages}{185001}
  (\bibinfo{year}{2010}).

\bibitem[{\citenamefont{McBride et~al.}(2012)\citenamefont{McBride, Slutz,
  Jennings, Sinars, Cuneo, Herrmann, Lemke, Martin, Vesey, Peterson
  et~al.}}]{McBride:2012db}
\bibinfo{author}{\bibfnamefont{R.~D.} \bibnamefont{McBride}},
  \bibinfo{author}{\bibfnamefont{S.~A.} \bibnamefont{Slutz}},
  \bibinfo{author}{\bibfnamefont{C.~A.} \bibnamefont{Jennings}},
  \bibinfo{author}{\bibfnamefont{D.~B.} \bibnamefont{Sinars}},
  \bibinfo{author}{\bibfnamefont{M.~E.} \bibnamefont{Cuneo}},
  \bibinfo{author}{\bibfnamefont{M.~C.} \bibnamefont{Herrmann}},
  \bibinfo{author}{\bibfnamefont{R.~W.} \bibnamefont{Lemke}},
  \bibinfo{author}{\bibfnamefont{M.~R.} \bibnamefont{Martin}},
  \bibinfo{author}{\bibfnamefont{R.~A.} \bibnamefont{Vesey}},
  \bibinfo{author}{\bibfnamefont{K.~J.} \bibnamefont{Peterson}},
  \bibnamefont{et~al.}, ``{Penetrating radiography of imploding and stagnating
  Beryllium liners on the $Z$ Accelerator},'' \bibinfo{journal}{Phys. Rev.
  Lett.} \textbf{\bibinfo{volume}{109}}, \bibinfo{pages}{135004}
  (\bibinfo{year}{2012}).

\bibitem[{\citenamefont{McBride et~al.}(2013)\citenamefont{McBride, Martin,
  Lemke, Greenly, Jennings, Rovang, Sinars, Cuneo, Herrmann, Slutz
  et~al.}}]{McBride:2013gda}
\bibinfo{author}{\bibfnamefont{R.~D.} \bibnamefont{McBride}},
  \bibinfo{author}{\bibfnamefont{M.~R.} \bibnamefont{Martin}},
  \bibinfo{author}{\bibfnamefont{R.~W.} \bibnamefont{Lemke}},
  \bibinfo{author}{\bibfnamefont{J.~B.} \bibnamefont{Greenly}},
  \bibinfo{author}{\bibfnamefont{C.~A.} \bibnamefont{Jennings}},
  \bibinfo{author}{\bibfnamefont{D.~C.} \bibnamefont{Rovang}},
  \bibinfo{author}{\bibfnamefont{D.~B.} \bibnamefont{Sinars}},
  \bibinfo{author}{\bibfnamefont{M.~E.} \bibnamefont{Cuneo}},
  \bibinfo{author}{\bibfnamefont{M.~C.} \bibnamefont{Herrmann}},
  \bibinfo{author}{\bibfnamefont{S.~A.} \bibnamefont{Slutz}},
  \bibnamefont{et~al.}, ``{Beryllium liner implosion experiments on the Z
  accelerator in preparation for magnetized liner inertial fusion},''
  \bibinfo{journal}{Phys. Plasmas} \textbf{\bibinfo{volume}{20}},
  \bibinfo{pages}{056309} (\bibinfo{year}{2013}).

\bibitem[{\citenamefont{Awe et~al.}(2014)\citenamefont{Awe, Jennings, McBride,
  Cuneo, Lamppa, Martin, Rovang, Sinars, Slutz, Owen et~al.}}]{Awe:2014gba}
\bibinfo{author}{\bibfnamefont{T.~J.} \bibnamefont{Awe}},
  \bibinfo{author}{\bibfnamefont{C.~A.} \bibnamefont{Jennings}},
  \bibinfo{author}{\bibfnamefont{R.~D.} \bibnamefont{McBride}},
  \bibinfo{author}{\bibfnamefont{M.~E.} \bibnamefont{Cuneo}},
  \bibinfo{author}{\bibfnamefont{D.~C.} \bibnamefont{Lamppa}},
  \bibinfo{author}{\bibfnamefont{M.~R.} \bibnamefont{Martin}},
  \bibinfo{author}{\bibfnamefont{D.~C.} \bibnamefont{Rovang}},
  \bibinfo{author}{\bibfnamefont{D.~B.} \bibnamefont{Sinars}},
  \bibinfo{author}{\bibfnamefont{S.~A.} \bibnamefont{Slutz}},
  \bibinfo{author}{\bibfnamefont{A.~C.} \bibnamefont{Owen}},
  \bibnamefont{et~al.}, ``{Modified helix-like instability structure on
  imploding z-pinch liners that are pre-imposed with a uniform axial magnetic
  field},'' \bibinfo{journal}{Phys. Plasmas} \textbf{\bibinfo{volume}{21}},
  \bibinfo{pages}{056303} (\bibinfo{year}{2014}).

\bibitem[{\citenamefont{Ruiz et~al.}(2022)\citenamefont{Ruiz, Yager-Elorriaga,
  Peterson, Sinars, Weis, Schroen, Tomlinson, Fein, and
  Beckwith}}]{Ruiz:2022aa}
\bibinfo{author}{\bibfnamefont{D.~E.} \bibnamefont{Ruiz}},
  \bibinfo{author}{\bibfnamefont{D.~A.} \bibnamefont{Yager-Elorriaga}},
  \bibinfo{author}{\bibfnamefont{K.~J.} \bibnamefont{Peterson}},
  \bibinfo{author}{\bibfnamefont{D.~B.} \bibnamefont{Sinars}},
  \bibinfo{author}{\bibfnamefont{M.~R.} \bibnamefont{Weis}},
  \bibinfo{author}{\bibfnamefont{D.~G.} \bibnamefont{Schroen}},
  \bibinfo{author}{\bibfnamefont{K.}~\bibnamefont{Tomlinson}},
  \bibinfo{author}{\bibfnamefont{J.~R.} \bibnamefont{Fein}}, \bibnamefont{and}
  \bibinfo{author}{\bibfnamefont{K.}~\bibnamefont{Beckwith}}, ``Harmonic
  generation and inverse cascade in the z-pinch driven, preseeded multimode,
  magneto-rayleigh-taylor instability,'' \bibinfo{journal}{Phys. Rev. Lett.}
  \textbf{\bibinfo{volume}{128}}, \bibinfo{pages}{255001}
  (\bibinfo{year}{2022}).

\bibitem[{\citenamefont{Ryutov and Dorf}(2014)}]{Ryutov:2014hr}
\bibinfo{author}{\bibfnamefont{D.~D.} \bibnamefont{Ryutov}} \bibnamefont{and}
  \bibinfo{author}{\bibfnamefont{M.~A.} \bibnamefont{Dorf}}, ``{Evolution of
  helical perturbations in a thin-shell model of an imploding liner},''
  \bibinfo{journal}{Phys. Plasmas} \textbf{\bibinfo{volume}{21}},
  \bibinfo{pages}{112704} (\bibinfo{year}{2014}).

\bibitem[{\citenamefont{Schmit and Ruiz}(2020)}]{Schmit:2020jd}
\bibinfo{author}{\bibfnamefont{P.~F.} \bibnamefont{Schmit}} \bibnamefont{and}
  \bibinfo{author}{\bibfnamefont{D.~E.} \bibnamefont{Ruiz}}, ``{A conservative
  approach to scaling magneto- inertial fusion concepts to larger pulsed- power
  drivers},'' \bibinfo{journal}{Phys. Plasmas} \textbf{\bibinfo{volume}{27}},
  \bibinfo{pages}{062707} (\bibinfo{year}{2020}).

\bibitem[{\citenamefont{Vekshtein}(1983)}]{Vekshtein:1983aa}
\bibinfo{author}{\bibfnamefont{G.~E.} \bibnamefont{Vekshtein}}, ``Evolution of
  magnetic field and anomalous thermal losses in a dense plasma,''
  \bibinfo{journal}{Sov. Phys. JETP} \textbf{\bibinfo{volume}{57}},
  \bibinfo{pages}{317} (\bibinfo{year}{1983}).

\bibitem[{\citenamefont{Vekshtein}(1986)}]{Vekshtein:1986aa}
\bibinfo{author}{\bibfnamefont{G.~E.} \bibnamefont{Vekshtein}}, ``Self-similar
  solutions for the compression of a plasma and a magnetic field by a liner,''
  \bibinfo{journal}{Sov. Phys. JETP} \textbf{\bibinfo{volume}{63}},
  \bibinfo{pages}{528} (\bibinfo{year}{1986}).

\bibitem[{\citenamefont{Velikovich et~al.}(2015)\citenamefont{Velikovich,
  Giuliani, and Zalesak}}]{Velikovich:2015gs}
\bibinfo{author}{\bibfnamefont{A.~L.} \bibnamefont{Velikovich}},
  \bibinfo{author}{\bibfnamefont{J.~L.} \bibnamefont{Giuliani}},
  \bibnamefont{and} \bibinfo{author}{\bibfnamefont{S.~T.}
  \bibnamefont{Zalesak}}, ``{Magnetic flux and heat losses by diffusive,
  advective, and Nernst effects in magnetized liner inertial fusion-like
  plasma},'' \bibinfo{journal}{Phys. Plasmas} \textbf{\bibinfo{volume}{22}},
  \bibinfo{pages}{042792} (\bibinfo{year}{2015}).

\bibitem[{foo({\natexlab{c}})}]{foot:cushions}
\bibinfo{note}{The cushions are cylindrical washers placed within the liner
  ends (see \Fig{fig:liners}) that help mitigate the wall
  instability,\cite{McBride:2013gda} which occurs where the liner meets the
  electrode surfaces.}

\bibitem[{foo({\natexlab{d}})}]{foot:preheat}
\bibinfo{note}{In present-day MagLIF experiments on the Z machine, the laser
  pulse length is closer to 5~ns. However, here we choose a 10-ns pulse length
  since a longer pulse length will likely be required for higher preheat
  energies to reduce the laser intensity and the ensuing laser--plasma
  instabilities.}

\bibitem[{\citenamefont{Bose et~al.}(2017)\citenamefont{Bose, Betti, Shvarts,
  and Woo}}]{Bose:2017jf}
\bibinfo{author}{\bibfnamefont{A.}~\bibnamefont{Bose}},
  \bibinfo{author}{\bibfnamefont{R.}~\bibnamefont{Betti}},
  \bibinfo{author}{\bibfnamefont{D.}~\bibnamefont{Shvarts}}, \bibnamefont{and}
  \bibinfo{author}{\bibfnamefont{K.~M.} \bibnamefont{Woo}}, ``{The physics of
  long- and intermediate-wavelength asymmetries of the hot spot: Compression
  hydrodynamics and energetics},'' \bibinfo{journal}{Phys. Plasmas}
  \textbf{\bibinfo{volume}{24}}, \bibinfo{pages}{102704}
  (\bibinfo{year}{2017}).

\bibitem[{foo({\natexlab{e}})}]{foot:implosion}
\bibinfo{note}{The implosion time is defined as the difference between the bang
  time (time at which peak neutron yield rate is achieved) and the
  ``zero-current" time, which is calculated via a linear fit to the rising
  portion of the current traces.}

\bibitem[{foo({\natexlab{f}})}]{foot:losses}
\bibinfo{note}{The numerical coefficients for $\Upsilon_{\rm rad}$ and
  $\Upsilon_{\rm ci}$ are slightly modified to account for DD fuel.}

\bibitem[{\citenamefont{Bosch and Hale}(1992)}]{Bosch:1992aa}
\bibinfo{author}{\bibfnamefont{H.~S.} \bibnamefont{Bosch}} \bibnamefont{and}
  \bibinfo{author}{\bibfnamefont{G.~M.} \bibnamefont{Hale}}, ``{Improved
  formulas for fusion cross-sections and thermal reactivities},''
  \bibinfo{journal}{Nucl. Fusion} \textbf{\bibinfo{volume}{32}},
  \bibinfo{pages}{611} (\bibinfo{year}{1992}).

\bibitem[{foo({\natexlab{g}})}]{foot:load_voltage}
\bibinfo{note}{In \textsc{hydra} calculations, the load voltage $\varphi_{\rm
  load}$ is approximated as the measured voltage at the outer edge of the
  simulation domain, where the electrodes connect to the circuit model, minus
  the voltage drop $\varphi_{\rm void}(t) = L_{\rm
  void}\mathrm{d}I_l/\mathrm{d}t$ associated to the inductance $L_{\rm void}
  \simeq (\mu_0 \Delta h) / (2\pi) \ln(R_{\rm ext}/R_{\rm out,0})$ of the
  initial void region in the simulation. Here $\Delta h$ is the electrode
  spacing, and $R_{\rm ext}$ is the external outer radius of the simulation
  domain.}

\bibitem[{\citenamefont{Cuneo et~al.}(2006)\citenamefont{Cuneo, Sinars,
  Waisman, Bliss, Stygar, Vesey, Lemke, Smith, Rambo, Porter
  et~al.}}]{Cuneo:2006hm}
\bibinfo{author}{\bibfnamefont{M.~E.} \bibnamefont{Cuneo}},
  \bibinfo{author}{\bibfnamefont{D.~B.} \bibnamefont{Sinars}},
  \bibinfo{author}{\bibfnamefont{E.~M.} \bibnamefont{Waisman}},
  \bibinfo{author}{\bibfnamefont{D.~E.} \bibnamefont{Bliss}},
  \bibinfo{author}{\bibfnamefont{W.~A.} \bibnamefont{Stygar}},
  \bibinfo{author}{\bibfnamefont{R.~A.} \bibnamefont{Vesey}},
  \bibinfo{author}{\bibfnamefont{R.~W.} \bibnamefont{Lemke}},
  \bibinfo{author}{\bibfnamefont{I.~C.} \bibnamefont{Smith}},
  \bibinfo{author}{\bibfnamefont{P.~K.} \bibnamefont{Rambo}},
  \bibinfo{author}{\bibfnamefont{J.~L.} \bibnamefont{Porter}},
  \bibnamefont{et~al.}, ``{Compact single and nested tungsten-wire-array
  dynamics at 14{\textendash}19MA and applications to inertial confinement
  fusion},'' \bibinfo{journal}{Phys. Plasmas} \textbf{\bibinfo{volume}{13}},
  \bibinfo{pages}{056318} (\bibinfo{year}{2006}).

\bibitem[{\citenamefont{Jones et~al.}(2008)\citenamefont{Jones, Coverdale,
  Deeney, Sinars, Waisman, Cuneo, Ampleford, LePell, Cochrane, Thornhill
  et~al.}}]{Jones:2008ds}
\bibinfo{author}{\bibfnamefont{B.}~\bibnamefont{Jones}},
  \bibinfo{author}{\bibfnamefont{C.~A.} \bibnamefont{Coverdale}},
  \bibinfo{author}{\bibfnamefont{C.}~\bibnamefont{Deeney}},
  \bibinfo{author}{\bibfnamefont{D.~B.} \bibnamefont{Sinars}},
  \bibinfo{author}{\bibfnamefont{E.~M.} \bibnamefont{Waisman}},
  \bibinfo{author}{\bibfnamefont{M.~E.} \bibnamefont{Cuneo}},
  \bibinfo{author}{\bibfnamefont{D.~J.} \bibnamefont{Ampleford}},
  \bibinfo{author}{\bibfnamefont{P.~D.} \bibnamefont{LePell}},
  \bibinfo{author}{\bibfnamefont{K.~R.} \bibnamefont{Cochrane}},
  \bibinfo{author}{\bibfnamefont{J.~W.} \bibnamefont{Thornhill}},
  \bibnamefont{et~al.}, ``{Implosion dynamics and K-shell x-ray generation in
  large diameter stainless steel wire array Z pinches with various nesting
  configurations},'' \bibinfo{journal}{Phys. Plasmas}
  \textbf{\bibinfo{volume}{15}}, \bibinfo{pages}{122703}
  (\bibinfo{year}{2008}).

\bibitem[{\citenamefont{Cuneo et~al.}(2012)\citenamefont{Cuneo, Herrmann,
  Sinars, Slutz, Stygar, Vesey, Sefkow, Rochau, Chandler, Bailey
  et~al.}}]{Cuneo:2012kv}
\bibinfo{author}{\bibfnamefont{M.~E.} \bibnamefont{Cuneo}},
  \bibinfo{author}{\bibfnamefont{M.~C.} \bibnamefont{Herrmann}},
  \bibinfo{author}{\bibfnamefont{D.~B.} \bibnamefont{Sinars}},
  \bibinfo{author}{\bibfnamefont{S.~A.} \bibnamefont{Slutz}},
  \bibinfo{author}{\bibfnamefont{W.~A.} \bibnamefont{Stygar}},
  \bibinfo{author}{\bibfnamefont{R.~A.} \bibnamefont{Vesey}},
  \bibinfo{author}{\bibfnamefont{A.~B.} \bibnamefont{Sefkow}},
  \bibinfo{author}{\bibfnamefont{G.~A.} \bibnamefont{Rochau}},
  \bibinfo{author}{\bibfnamefont{G.~A.} \bibnamefont{Chandler}},
  \bibinfo{author}{\bibfnamefont{J.~E.} \bibnamefont{Bailey}},
  \bibnamefont{et~al.}, ``{Magnetically driven implosions for inertial
  confinement fusion at sandia national laboratories},'' \bibinfo{journal}{IEEE
  Trans. Plasma Sci.} \textbf{\bibinfo{volume}{40}}, \bibinfo{pages}{3222}
  (\bibinfo{year}{2012}).

\bibitem[{\citenamefont{Nora et~al.}(2014)\citenamefont{Nora, Betti, Anderson,
  Shvydky, Bose, Woo, Christopherson, Marozas, Collins, Radha
  et~al.}}]{Nora:2014iq}
\bibinfo{author}{\bibfnamefont{R.}~\bibnamefont{Nora}},
  \bibinfo{author}{\bibfnamefont{R.}~\bibnamefont{Betti}},
  \bibinfo{author}{\bibfnamefont{K.~S.} \bibnamefont{Anderson}},
  \bibinfo{author}{\bibfnamefont{A.}~\bibnamefont{Shvydky}},
  \bibinfo{author}{\bibfnamefont{A.}~\bibnamefont{Bose}},
  \bibinfo{author}{\bibfnamefont{K.~M.} \bibnamefont{Woo}},
  \bibinfo{author}{\bibfnamefont{A.~R.} \bibnamefont{Christopherson}},
  \bibinfo{author}{\bibfnamefont{J.~A.} \bibnamefont{Marozas}},
  \bibinfo{author}{\bibfnamefont{T.~J.~B.} \bibnamefont{Collins}},
  \bibinfo{author}{\bibfnamefont{P.~B.} \bibnamefont{Radha}},
  \bibnamefont{et~al.}, ``{Theory of hydro-equivalent ignition for inertial
  fusion and its applications to OMEGA and the National Ignition Facility},''
  \bibinfo{journal}{Phys. Plasmas} \textbf{\bibinfo{volume}{21}},
  \bibinfo{pages}{056316} (\bibinfo{year}{2014}).

\bibitem[{\citenamefont{Welch et~al.}(2019)\citenamefont{Welch, Bennett,
  Genoni, Rose, Thoma, Miller, and Stygar}}]{Welch:2019aa}
\bibinfo{author}{\bibfnamefont{D.~R.} \bibnamefont{Welch}},
  \bibinfo{author}{\bibfnamefont{N.}~\bibnamefont{Bennett}},
  \bibinfo{author}{\bibfnamefont{T.~C.} \bibnamefont{Genoni}},
  \bibinfo{author}{\bibfnamefont{D.~V.} \bibnamefont{Rose}},
  \bibinfo{author}{\bibfnamefont{C.}~\bibnamefont{Thoma}},
  \bibinfo{author}{\bibfnamefont{C.}~\bibnamefont{Miller}}, \bibnamefont{and}
  \bibinfo{author}{\bibfnamefont{W.~A.} \bibnamefont{Stygar}}, ``Electrode
  contaminant plasma effects in $1{0}^{7}$-A $z$ pinch accelerators,''
  \bibinfo{journal}{Phys. Rev. Accel. Beams} \textbf{\bibinfo{volume}{22}},
  \bibinfo{pages}{070401} (\bibinfo{year}{2019}).

\bibitem[{\citenamefont{Bennett et~al.}(2019)\citenamefont{Bennett, Welch,
  Jennings, Yu, Hess, Hutsel, Laity, Moore, Rose, Peterson
  et~al.}}]{Bennett:2019aa}
\bibinfo{author}{\bibfnamefont{N.}~\bibnamefont{Bennett}},
  \bibinfo{author}{\bibfnamefont{D.~R.} \bibnamefont{Welch}},
  \bibinfo{author}{\bibfnamefont{C.~A.} \bibnamefont{Jennings}},
  \bibinfo{author}{\bibfnamefont{E.}~\bibnamefont{Yu}},
  \bibinfo{author}{\bibfnamefont{M.~H.} \bibnamefont{Hess}},
  \bibinfo{author}{\bibfnamefont{B.~T.} \bibnamefont{Hutsel}},
  \bibinfo{author}{\bibfnamefont{G.}~\bibnamefont{Laity}},
  \bibinfo{author}{\bibfnamefont{J.~K.} \bibnamefont{Moore}},
  \bibinfo{author}{\bibfnamefont{D.~V.} \bibnamefont{Rose}},
  \bibinfo{author}{\bibfnamefont{K.}~\bibnamefont{Peterson}},
  \bibnamefont{et~al.}, ``Current transport and loss mechanisms in the $z$
  accelerator,'' \bibinfo{journal}{Phys. Rev. Accel. Beams}
  \textbf{\bibinfo{volume}{22}}, \bibinfo{pages}{120401}
  (\bibinfo{year}{2019}).

\bibitem[{\citenamefont{Bennett et~al.}(2021)\citenamefont{Bennett, Welch,
  Laity, Rose, and Cuneo}}]{Bennett:2021dq}
\bibinfo{author}{\bibfnamefont{N.}~\bibnamefont{Bennett}},
  \bibinfo{author}{\bibfnamefont{D.~R.} \bibnamefont{Welch}},
  \bibinfo{author}{\bibfnamefont{G.}~\bibnamefont{Laity}},
  \bibinfo{author}{\bibfnamefont{D.~V.} \bibnamefont{Rose}}, \bibnamefont{and}
  \bibinfo{author}{\bibfnamefont{M.~E.} \bibnamefont{Cuneo}}, ``{Magnetized
  particle transport in multi-MA accelerators},'' \bibinfo{journal}{Phys. Rev.
  Accel. Beams} \textbf{\bibinfo{volume}{24}}, \bibinfo{pages}{060401}
  (\bibinfo{year}{2021}).

\end{thebibliography}

\end{document}